\documentclass[twocolumn]{aastex63}
\usepackage{longtable}
\usepackage{threeparttablex}
\usepackage{lineno}

\graphicspath{{./}{figures/}}

\received{}
\revised{}
\accepted{}
\submitjournal{ApJ}

\shorttitle{The Class I hot corino Ser-emb 11 W}
\shortauthors{Mart\'in-Dom\'enech et al.}

\begin{document}

%\title{Ser-emb 11: A Class I binary protostellar disk candidate source with hot corino chemistry in Serpens}

\title{Hot corino chemistry in the Class I binary
%, protostellar disk candidate 
source Ser-emb 11}

\correspondingauthor{Rafael Mart\'in-Dom\'enech}
\email{rafael.martin\_domenech@cfa.harvard.edu}

\author{Rafael Mart\'in-Dom\'enech}
\affiliation{Center for Astrophysics $|$ Harvard \& Smithsonian\\
60 Garden St., Cambridge, MA 02138, USA}

\author{Jennifer B. Bergner}
\affiliation{University of Chicago Department of the Geophysical Sciences, Chicago, IL 60637, USA}
\affiliation{NHFP Sagan Fellow}

\author{Karin I. \"Oberg}
\affil{Center for Astrophysics $|$ Harvard \& Smithsonian\\
60 Garden St., Cambridge, MA 02138, USA}

\author{John Carpenter}
\affiliation{Joint ALMA Observatory, Alonso de C\'ordova 3107 Vitacura, Santiago, Chile}

\author{Charles J. Law}
\affil{Center for Astrophysics $|$ Harvard \& Smithsonian\\
60 Garden St., Cambridge, MA 02138, USA}

\author{Jane Huang}
\affiliation{Department of Astronomy, 8 University of Michigan, 323 West Hall, 1085 S. University Avenue, Ann Arbor, MI 48109, USA}
\affiliation{NHFP Sagan Fellow}

\author{Jes K. J{\o}rgensen}
\affiliation{Niels Bohr Institute \& Centre for Star and Planet Formation, University of Copenhagen \\
{\O}ster Voldgade 5-7, DK-1350 Copenhagen K}

\author{Kamber Schwarz}
\affiliation{Lunar and Planetary Laboratory, University of Arizona, Tucson, AZ 85721, USA}

\author{David J. Wilner}
\affil{Center for Astrophysics $|$ Harvard \& Smithsonian\\
60 Garden St., Cambridge, MA 02138, USA}

\begin{abstract}

We report the detection of more than 120 emission lines corresponding to 8 complex organic molecules (CH$_3$OH, CH$_3$CH$_2$OH, 
CH$_3$OCH$_3$, CH$_3$OCHO, CH$_3$COCH$_3$, 
NH$_2$CHO, 
CH$_2$DCN, and CH$_3$CH$_2$CN) and 3 isotopologues (CH$_2$DOH, $^{13}$CH$_3$CN, and CH$_3$C$^{15}$N) toward the western component of the Ser-emb 11 binary young stellar object (YSO) using observations with the Atacama Large Millimeter/submillimeter Array at $\sim$1 mm. 
%1mm - 230GHz - Band 6 
%
The complex organic emission was unresolved with a $\sim$0.5$\arcsec$ beam ($\sim$220 au) in a compact region around the central protostar, and  
a population diagram analysis revealed excitation temperatures above 100 K for all COMs, 
indicating the presence of a hot corino. 
%(according to the accepted definition of a hot corino)
%
The estimated column densities were in the range of 10$^{17}$ $-$ 10$^{18}$ cm$^{-2}$ for the O-bearing COMs, and three orders of magnitude lower for the N-bearing species. 
%
%The deuterium fraction measured for CH$_3$OH is four orders of magnitude higher than the elemantal D/H ratio, consistent with the formation of this species at low temperatures.
%
%
%In addition, we observed evidence of spatially resolved redshifted and blueshifted C$^{18}$O emission across Ser-emb 11 W, compatible with the presence of a protostellar disk, %as previously proposed in the literature, 
%although alternative explanations could not be excluded. 
%
%The chemical differentiation observed in the Ser-emb 11 binary system is in line with similar observations toward other multiple sources. 
%The COM emission toward Ser-emb 11 E could be blocked by the dust opacity, as recently observed in the IRAS 4A binary source. 
%
We also report the detection of H$_2$CO and CH$_3$OH emission in %the nearby source ID 68, a very cold YSO that is probably at a very early evolutionary stage. %%ID 68 IS SER-EMB 11!!!
a nearby millimeter source that had not been previously catalogued. 
Ser-emb 11 is classified in the literature as a Class I source near the Class 0/I cutoff. 
The estimated COM relative abundances in Ser-emb 11 W and the other three Class I hot corino sources reported in the literature are consistent with those of Class 0 hot corinos, suggesting a continuity in the chemical composition of hot corinos during protostellar evolution.

\end{abstract}

\keywords{}

\section{Introduction} \label{sec:intro}

%%%%%%%%%%%%%%%1.- Strong opening paragraph focusing on the importance of COMs for the origin of life %%%%%%%%%%%%%%%%%%
Complex (6+ atoms) organic molecules (COMs), the precursors of yet more complex prebiotic species, are thought to be formed in space by energetic processing of interstellar ice mantles and/or by gas-phase reactions following ice sublimation \citep{herbst09}. 
During the star-formation process, COMs
can be incorporated into protostellar and protoplanetary disk structures \citep[][]{oberg20}, 
%On one hand, the icy grains that become incorporated into protostellar disks are expected to be rich in COMs \citep{oberg20}, %p22
%(since the composition of these ice mantles, revealed in hot corinos, is characterized by COM emission). 
%On the other hand, COMs present in the gas accreted by the protostellar disks 
%(previously desorbed from ice mantles) 
%could freeze-out onto the dust grains under particular conditions before they are photodissociated. 
%However, the disk assembly process could alter the composition of the incorporated gas and solids, especially close to the protostar, although chemical reprocessing should be minimal for the material supplied at large distances. This chemical reprocessing could also take place during the subsequent protoplanetary disk stage. 
%Current protoplanetary disk observations do not provide conclusive evidences on how efficient the chemical-dynamical evolution of disks is at resetting the chemistry, but COMs are also observed \citep[see, e.g.,][and references therein]{oberg20}.
%Solar system observations provide evidences for both  chemical reset and chemical inheritance of the material accreted from the solar nebula. 
where they could be preserved in small bodies such as comets \citep[see, e.g.,][]{altwegg16} 
and eventually delivered to nascent planets, perhaps triggering a prebiotic chemistry on their surfaces.

%%%%%%%%%%%%%%%%%%%%%2.- paragraph talking about detection of COMs in the ISM and their origin %%%%%%%%%%%%%%%%%%%%%%%%%%%%
In the interstellar medium (ISM), COMs were first detected in the gas phase of hot (T $\sim$ 300 K) and dense (n$_H$ $\ge$ 10$^6$ cm$^3$) regions around high-mass protostars, so-called hot cores \citep[e.g.,][]{ball70,solomon71,rubin71,blake87,turner91,charnley92}. 
Later on, detection of COMs was also reported toward a number of Class 0 and a few Class I low-mass protostars \citep[e.g.,][]{cazaux03,arce08,oberg10,bianchi19}. 
%and  prestellar cores \citep[][]{marcelino07,vastel14,izaskun16,scibelli20}.
%
%%%%%%%%%%%%%%%%3.- paragraph with locations of COM detections around Class 0 and Class I low-mass protostars %%%%%%%%%%%%%%%
%In Class 0/I sources, COM detection has been associated to outflow shocked regions  \citep{arce08,codella17,lefloch18}, and both the outer \citep{oberg10,jenny17} 
%and innermost \citep{cazaux03,jorgensen05,codella16,jenny19} regions of the envelope surrounding the protostar.
%
%%%%%%%%%%%%%%%%4.- this locations include compact regions called hot corinos, but their origin is unclear (connection with my proposal) %%%%%%%%%%%%%%%%%%%%%%%%%%%%%%%%%%%%%%%%%%%
Those Class 0 and Class I sources where COMs have been detected in a warm (T $\ge$ 100 K) and compact ($\sim$100 au) region around the central protostar are known as hot corinos, in analogy to the hot cores around high-mass protostars. 
%However, it is unclear whether the so-called hot corino chemistry has its origin in the sublimation of ice mantles located in the passively heated inner envelope, the atmosphere of protostellar disks, the accretion shocks at the centrifugal barrier where the infalling-rotating envelope interacts with the protostellar disks, or in outflow shocks at short distances from the central object \citep[see, e.g.,][and references therein]{belloche20}.
%
There is an increased interest in 
%the complex organic content of Class I sources 
Class I hot corinos in particular, 
%(i.e., Class I sources with COM emission detected in a compact and warm region around the central protostar)
%(young protostellar disks are often observed at this stage), 
%since it may reveal the organic reservoir subsequently available to the forming planets. 
%The ice mantle composition at this stage of the star-formation process is shaped by the ice chemistry that takes place in the parental molecular cloud, and the chemical pathways that are activated during the warm-up of the ice mantles prior to their desorption \citep{oberg20}.  
%since recent observations of protoplanetary disks in more evolved Class II/III sources suggest that planet formation likely begins during these early stages of low-mass star formation \citep[][]{kwon09,pagani10,foster13,miotello14}. 
since evidence of early planet formation have been found in Class I and borderline Class I/II sources \citep[][]{alma15,harsono18,sheehan18,tychoniec20,segura20}
%
%%%%%%%%%%%%%%%5.- Connection of hot corinos with protostellar disks %%%%%%%%%%%%%%%%%%%%%%%%%%%%%%%%%%%%%%%%%%%%%%%%%%%%%%%%
%Three of the previously observed hot corino sources were known to harbor a protostellar disk: IRAS 16293-2422A \citep{oya16}, IRAS 4A2 \citep{choi10}, and HH212 \citep{lee17}. 
%In addition to the protostellar disks detected toward IRAS16293-2422A, IRAS 4A2, and HH212 (see above), the detection of compact dust continuum emission toward the VLA4A component in SVS13-A \citep{lefevre17}, Ser-emb 1, Ser-emb 8, Ser-emb 17 \citep{enoch11}, and both the N and S components of L1551-IRS5 \citep{bianchi20}; and B1-c \citep{tychoniec18}, as well as signatures of Keplerian rotation in the COM line emission of BHR71 IRS1 \citep{yang20}, suggests that these sources could also harbor a protostellar disk. 
%
%It was suggested that given the scarcity of hot corino observations, it could be significant that roughly half of them had been detected around protostars that either harbor a protostellar disk or at least are candidates to harbor one. This could indicate an association between both phenomena, 
%although not all detected protostellar disks present complex organic molecule emission \citep[see, e.g.,][]{jenny19,belloche20}.
%However, recent unbiased surveys such as PEACHES have also reported detection of hot corino chemistry in roughly half of the sources in the sample, regardless of the presence of a protostellar disk. 

%%%%%%%%%%%%%%6.- table with hot corinos. %%%%%%%%%%%%%%%%%%%%
Table \ref{tab:hc} lists the Class 0 and Class I sources with reported hot corino chemistry in the literature, according to the above definition (COMs at T $>$ 100 K detected within $\sim$100 au of the central object). 
Eight hot corinos had been detected as of the year 2018. 
This number doubled in the following two years, with the addition of seven confirmed sources, and four hot corino candidates \citep[G211.47-19.27S, G208.68-19.20N1, G210.49-19.79W, G192.12-11.10,][not included in Table \ref{tab:hc}]{hsu20}.  
In addition, there are a number of recent interferometric surveys dedicated to a systematic study of the presence of organic chemistry at planet-forming scales 
%(as well as the specific location and origin of the detected COMs) 
in samples of Class 0/I sources, such as FAUST \citep{bianchi20}, CALYPSO \citep{belloche20}, and PEACHES \citep{yang21}. 
%CALYPSO and PEACHES overlap in some of the targeted sources.
The PEACHES survey recently identified 15 additional Class 0 sources with compact emission of CH$_3$OH and at least one other COM, consistent with the presence of a hot corino %without considering the possible origin of the detected COMs
\citep{yang21}.  
%and seven more that had already been reported as hot corino sources in the literature
However, since the COM excitation temperatures were not constrained in half of these sources, we have not included them in Table \ref{tab:hc}. 
%As mentioned above, the CALYPSO and PEACHES survey pay attention to the possible origin of the detected COMs in the so-called hot corino sources (in general, sources with compact and warm COM emission): either a cannonical hot corino origin (i.e., thermal desorption in the warm, inner envelope), outflows located close to the central protostar, or accretion shocks close to the centrifugal barrier of the potential protostellar disk. 
%In the CALYPSO survey, the four cannonical hot corino sources had been already reported as hot corinos in the literature. 
%The rest of sources with COM emission could have been included in this work as hot corino sources, since it is unclear whether the origin of the hot corinos listed in Table \ref{tab:hc} is cannonical or not. 

%%%%%%%%%%%%%%7.- only 3 toward Class I sources, also interesting %%%%%%%%%%%%%%%%%%%%%%%%%%%%%%%%%%%%%%%%%%%%%%%%
Of the 15 hot corino sources listed in Table \ref{tab:hc}, only three (SVS13-A, Ser-emb 17, and L1551-IRS5) are Class I YSOs (70 K $<$ T$_{bol}$ $<$ 650 K), %\footnote{We note that the three observed Class I hot corino sources were also postulated as protostellar disk candidates \citep{enoch11,bianchi19,bianchi20}.}, 
while the rest are classified as Class 0 (T$_{bol}$ $<$ 70 K). 
Class I sources represent the evolutionary link between the young Class 0 sources, deeply embedded in their parental dense envelopes, and the more evolved Class II/III YSOs, where most of the envelope has dissipated. 
%, and the remaining dust particles and gas-phase molecules are located in a protoplanetary disk. 
It is currently unclear how (and if) the organic chemistry evolves between the Class 0 and Class I evolutionary stages.  
In particular, the small number of detected Class I hot corino sources, along with the large scatter in COM abundances measured toward Class 0 hot corinos  \citep[see, e.g.,][]{jenny19,belloche20}, make it hard to evaluate whether the complex organic chemistry changes during the protostellar phase.

\begin{deluxetable}{cccc}
\caption{Hot corino sources reported in the literature. 
\label{tab:hc}}
\tablehead{Source & Class$^a$ & Binary? & Reference}
\startdata
IRAS 16293 & 0 & y & \citet{cazaux03} \\
IRAS 4A & 0 & y & \citet{bottinelli04} \\
IRAS 2A & 0 & n & \citet{jorgensen05} \\
IRAS 4B & 0 & n & \citet{bottinelli07} \\
HH 212 & 0 & n & \citet{codella16} \\
B335 & 0 & n & \citet{imai16} \\
L483 & 0 & n & \citet{oya17} \\
B1-bS & 0 & n & \citet{lefloch18} \\
SVS13-A & I & y & \citet{bianchi19} \\
Ser-emb 1 & 0 & n & \citet{martin19} \\
Ser-emb 8 & 0 & n & \citet{jenny19} \\
Ser-emb 17 & I & n & \citet{jenny19} \\
L1551-IRS5 & I & y & \citet{bianchi20} \\
BHR 71 IRS1 & 0 & n & \citet{yang20} \\
B1-c & 0 & n & \citet{vanGelder20} \\
\enddata
\tablecomments{$^a$Based on T$_{bol}$ from \citet{andre00,enoch09,pezzuto12,green13,tobin16} }
\end{deluxetable}

%%%%%%%%%%%%%%%%8.- Some are detected in binary sources, and this is interesting. %%%%%%%%%%%%%%%%%%%%%%%%%%%%%%%%%%%%%%%%
In the cases when the hot corino chemistry is detected in binary protostellar systems at either the Class 0 (IRAS 16293-2422, IRAS 4A) or Class I (SVS13-A, and L1551-IRS5) stages, the comparison of the complex organic content observed toward the two components yield mixed results. 
In IRAS 16293-2422 and L1551-IRS5, both components appeared to harbor hot corino chemistry \citep[][]{jorgensen11,bianchi20},  %In L1551-IRS5, the observed COM emission is brighter toward the north component, but a second hot corino could also be associated with the south component \citep{bianchi20}.
while the COM emission detected in IRAS 4A and SVS13-A had been mostly associated with the IRAS 4A2 and VLA4A components, respectively \citep[and ref. therein]{lopezsepulcre16,bianchi19}. 
These findings were somewhat puzzling, since the chemical composition in binary sources might be expected to be similar across the different components, as they should be shaped by the same initial conditions of the parental molecular cloud. %However, recent observations of complex organics seen in absorption at millimeter wavelengths \citep{sahu19} and in emission at centimeter wavelengths \citep{desimone20} suggest that IRAS 4A1 could also harbor a hot corino, obscured by the dust continuum opacity. 
%
%Previous VLA observations toward SVS13-A had revealed that this source was a close binary system formed by the VLA4A and VLA4B components \citep{rodriguez99,anglada00}, with 
It is currently unclear whether this observed chemical differentiation is due to differences in the physical conditions of the individual sources, different evolutionary stages of the individual sources, or differences in the dust opacity toward the two components that could hinder the COM detection in one of them  \citep[][]{sahu19,belloche20,desimone20}. 

%%%%%%%%%%%%%%%%%%%%%8.- This work %%%%%%%%%%%%%%%%%%%%%%%%%%
In this work, we report the serendipitous detection and characterization of hot corino chemistry toward Ser-emb 11 W.
Ser-emb 11 is a binary source where the west component was proposed to harbor a massive disk of 0.13 $M_{\odot}$ \citep{enoch11}. %\citet{enoch11} reported a bolometric luminosity and temperature for Ser-emb 11 of 4.8 $L_{\sun}$ and 77 K, respectively
This source is located in Cluster B of the Serpens Molecular Cloud, a star-forming cloud at a distance of $d$ = 436 $\pm$ 9 pc \citep{gisela18} 
that harbors 
%235 YSOs identified as part of the Spitzer Legacy project "From Molecular Cores to Planet-forming Disks" \citep["c2d",][]{harvey07},
%detected with combined IRAC/MIPS observations 
34 embedded protostars (9 Class 0 and 25 Class I sources) identified in \citet{enoch09}. 
%isolated by combining the "c2d" Spitzer Legacy survey with 1.1 mm Bolocam continuum observations \citep{enoch09} . 
%Note that not all 34 embedded protostars in \citet{enoch09} have an IR counterpart in \citet{harvey07} (e.g., Ser-emb 1). 
%
%\citet{enoch11} reported evidences for a compact disk component around nine of the 34 embedded sources (Ser-emb 1, 4, 5, 6, 7, 8, 11, 15, and 17) based on the emission detected at long \textit{uv} baselines in high-resolution CARMA 230 GHz continuum observations (and therefore coming from dust in compact regions around the central object). 
Ser-emb 11 was classified as a Class I source in \citet{enoch09} according to its bolometric temperature of T$_{bol}$ = 77 K. Since this temperature is close to the Class 0/I cutoff \citep[70 K,][]{andre00}, Ser-emb 11 could also be seen as a transitional source between the Class 0 and Class I stages. 
If we adopt the classification in \citet{enoch09}, Ser-emb 11 W constitutes the fourth hot corino detected toward a Class I source to date. 
%Since the previous three Class I sources with reported hot corino chemistry have bolometric temperatures above 100 K \citep[][]{enoch09,green13,tobin16}, Ser-emb 11 W is the only one to represent the subgroup of Class 0/I transitional sources among the four Class I hot corinos.}
%
We also report the presence of complex organic chemistry toward a nearby millimeter source that had not been previously catalogued. 
%YSO probably associated with the ID 68 infrared source. ID 68 was identified as a very cold YSO, probably at an earlier evolutionary stage than typical Class 0 sources \citep{harvey07}, and had not been previously detected at millimeter/submillimeter wavelengths.

In this paper, the observations are described in Sect. \ref{sec:obs}, along with the calibration and imaging processes. An overview of the Ser-emb 11 source is presented in Sect. \ref{sec:results_overview}. 
The complex molecule detections toward the Ser-emb 11 W component are presented in Sect. \ref{sec:results_hotcorino}, and characterized in Sections  \ref{sec:results_spat}$-$\ref{sec:isot}. 
Sect. \ref{sec:small} is dedicated to the smaller molecules also observed toward this source. 
Sect. \ref{sec:results_rot} presents evidence consistent with the presence of a protostellar disk in Ser-emb 11 W. 
The organic chemistry detected toward the additional millimeter source is presented in Sect. \ref{sec:results_id68}. 
The results are discussed in Sect. \ref{sec:discussion}, 
and the conclusions are summarized in Sect. \ref{sec:conclusions}.

\section{Observations} \label{sec:obs}

\begin{figure}
    \centering
    \includegraphics[width=8cm]{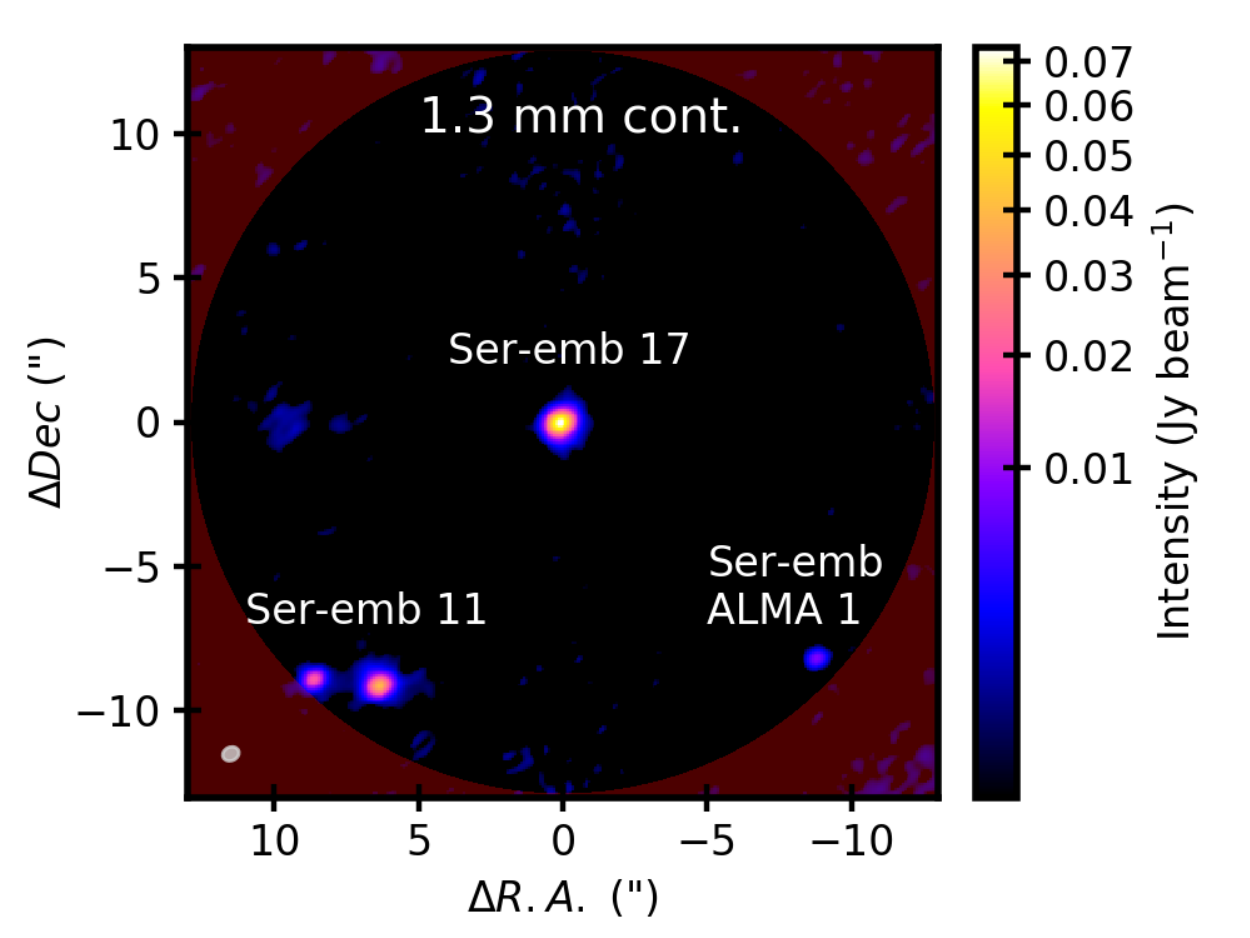}
    \caption{232 GHz continuum emission of the Ser-emb 17 and Ser-emb 11 sources, and an additional millimeter source not classified in the literature that we have named Ser-emb ALMA 1 (Sect. \ref{sec:results_id68}). 
    The map is clipped at $2\sigma$ ($\sigma$ = 0.35 mJy beam$^{-1}$). 
    The size of the synthesized beam is shown on the lower left corner. 
    The three sources were observed within the same field of view, centered at Ser-emb 17. The region of the field of view observed outside the half power beam width of the primary beam is colored in red.}
    \label{fig:fov}
\end{figure}

%The observations analyzed in this paper were 
This work is based on  
observational data from an  ALMA project aiming at systematically characterizing  
the chemistry in disks at different evolutionary stages  (\#2015.1.00964.S). 
This project has previously led to the detection of hot corino chemistry in two Class 0 and one Class I sources in the  Serpens Molecular Cloud \citep[Ser-emb 1, 8, and 17,][]{martin19,jenny19}. 
The observations analyzed in this paper were completed during Cycle 3 using two different Band 6 frequency settings, and  
were centered on Ser-emb 17 
%Class I source in the Serpens Molecular Cloud 
($R.A._{\rm J2000}$ = 18h 29m 06.20s, $Dec._{\rm J2000}$ = +00$^{\circ}$ 30$^{\prime}$ 43.1$^{\prime\prime}$). 
The characterization of the emission detected toward Ser-emb 17 was presented in \citet{jenny19,jenny20}. 
Figure \ref{fig:fov} shows the 1.3 mm continuum emission map of a 26$\arcsec\times$26$\arcsec$ region around Ser-emb 17. 
Emission from the Ser-emb 11 source and another millimeter source that we have named Ser-emb ALMA 1 (Sect. \ref{sec:results_id68}) was also observed within the half power beam width of the primary beam (HPBW $\sim$ 26$\arcsec$). %and is analyzed in this work. 

%The two frequency settings consisted in 
The majority of the data presented here correspond to 
two broad spectral windows of 1.875 GHz bandwidth observed with a native spectral resolution of 1.19$-$1.25 km/s. 
%(the spectral resolution is 2x)
These spectral windows spanned 231.481$-$233.356 GHz and 242.978$-$244.853 GHz, and covered a large number of COM transitions. 
In addition, a series of 14 narrow spectral windows spread across 217$-$233 GHz and 243$-$262 GHz, with bandwidths between 58.6 MHz and 117.2 MHz and a native spectral resolution of 0.16 $-$ 0.20 km/s, were also observed. 
These narrow spectral windows targeted particular transitions of simpler O-, C-, N-, and S-bearing species, namely, SiO, CO, $^{13}$CO, C$^{18}$O, H$_2$CO, C$_2$H, c-C$_3$H$_2$, N$_2$D$^+$, DCN, H$^{13}$CN, HC$^{15}$N,  and CS. 
A list of the observed spectral windows can be found in Table \ref{tab:spw} of Appendix \ref{app-spw}. 
The observations of the first frequency setting (217$-$233 GHz) were carried out on June 05 and 14, 2016, using 42 antennas (longest baseline of 772.8 m) for a total on source time of 19.2 min.  
The second frequency setting (243$-$262 GHz) was observed on May 15 and 16, 2016, using 41 antennas (longest baseline of 640.0 m) for a total on source time of 22.5 min.

\subsection{Calibration of the observed visibilities}\label{sec:obs_cal}
The observed visibilities were initially calibrated by ALMA staff with the Common Astronomy Software Applications (CASA) versions 4.5.3 and 4.7.0, using the sources J1751+0939, J1830+0619 as bandpass and phase calibrators, respectively, and Titan as the absolute flux calibrator. 
%I THINK Jenny did the calibration with the script provided by ALMA, leading to the  calibrated_final.ms and calibrated_source.ms measurement sets that contained the calibrated visibilities for the 5 sources with the two frequency settings, respectively.  
%HOWEVER, I don't know the CASA version she used for this. 
%
%%%%%% selfcal_Ser17_X.py with X=217,230,244,258
In order to work with manageable datasets, the calibrated visibilities centered on the Ser-emb 17 source were split into four separate measurement sets containing the lower and upper sideband visibilities of the two frequency settings. 
%Ser-emb_17_XGHz_cal.ms with X=217,230,244,258
%The continuum data of each measurement set (channel-averaged visibilities of all spectral windows in every sideband after flagging line emission channels) 
%%cal_X_cont.ms with X=217,230,244,258
%was further self-calibrated %\footnote{The self-calibration process only took into account the Ser-emb 17 continuum emission. Including the Ser-emb 11 weaker continuum emission did not change the self-calibration solutions.} 
%%I checked that for the 230GHz continuum. 
%in one round with CASA version 5.4.1. 
Four continuum measurement sets were generated from the channel-averaged visibilities of all spectral windows in each sideband after flagging channels containing line emission. 
One round of phase self-calibration was then performed in CASA 5.4.1 using these continuum datasets. 
For that purpose, the continuum visibilities were imaged using the task \texttt{tclean} with Briggs weighting of the baselines (robustness parameter = 0.5) and the Hogbom deconvolver. 
The polarization-independent (gaintype = T), self-calibration phase solutions were subsequently obtained with the task \texttt{gaincal}, 
%self-calibration solutions in pcal1 table in each folder. 
and applied to the continuum-subtracted, native resolution visibilities. 
Continuum subtraction had been previously performed in the \textit{uv} plane using line-free channels.  
%Ser-emb_17_XGHz_cal.ms.contsub with X=217,230,244,258
The self-calibration process led to an increase of 4$-$12\% in the peak intensity of the continuum images, and a decrease of 6$-$22\% in the corresponding rms. 
Finally, the continuum-subtracted, self-calibrated visibilities corresponding to the different spectral windows were further split into different measurement sets before their imaging. 
%Ser-emb_17_X_line_selfcal.ms.contsub

\subsection{Imaging of the calibrated visibilities}\label{sec:obs_im}

Imaging of the continuum-subtracted, self-calibrated visibilities was performed with CASA version 5.4.2,  %and 5.5.0 
%mainly 5.4.2 in scarfinger, some with 5.5.0 in Mac, but I don't think any of the final images I've used were cleaned in my Mac since they were large cubes that I probably did in sarek or sabriel. 
using the task \texttt{tclean} with Brigs weighting of the baselines (robustness parameter = 0.5)    
down to a $\sim$2$\sigma$ noise threshold, 
with $\sigma$ $\sim$3.0 mJy beam$^{-1}$ for the two 1.875 GHz bandwidth spectral windows, and $\sigma$ $\sim$7.5 mJy beam$^{-1}$ for the narrow spectral windows. 
The corresponding $\sigma$ was estimated as the average rms over ten line-free channels of a 8$\arcsec$ diameter region that included (but was not centered on) the Ser-emb 11 source. 
The two 1.875 GHz bandwidth spectral windows were imaged with a uniform channel width of 0.63 km/s,  
%see sg230ghz\myScriptForImaging_ch3oh.py line 104
while the narrow spectral windows were imaged with a channel width of 0.14 $-$ 0.17 km/s (corresponding to their native channel spacing). 
%see myScriptForImaging_H2CO_03.py, line 188
%
A multiscale deconvolver with default values (\texttt{scales} = 0, 5, 15; \texttt{smallscalebias} = 0.6) was used in order to account for both compact and extended emission. 
During the imaging process, an auto-mask was applied independently to every channel of the self-calibrated visibilities, using the default parameters for auto-masking line emission:  \texttt{sidelobethreshold} = 2.0 $\times$ rms, \texttt{noisethreshold} = 4.25 $\times$ rms,
\texttt{lownoisethreshold} = 1.5 $\times$ rms,
\texttt{minbeamfrac} = 0.3, and 
\texttt{negativethreshold} = 15.0. 
%\texttt{cutthreshold} = 0.4,  and \texttt{growiterations} = 75. 
%%%%%%%%%%%%%%%%%%%%%%%%%%%%%%%%%%%%%%%%%%%%%%
%%%WARNING MESSAGE BECAUSE SOME TABLES ARE OUT-OF-DATE
%2020-10-20 21:04:33	SEVERE	MeasTable::dUTC(Double) (file ../../measures/Measures/MeasTable.cc, line 4396)
%Leap second table TAI_UTC seems out-of-date. 
%Until the table is updated (see the CASA documentation or your system admin),
%times and coordinates derived from UTC could be wrong by 1s or more.
%%%I HAVE CHECKED THAT THIS IS NOT AFFECTING MY IMAGES. THE CONTINUUM IMAGE WITH UPDATED TABLES HAS THE SAME COORDINATES. 
%%%%%%%%%%%%%%%%%%%%%%%%%%%%%%%%%%%%%%%%%%%%%%
The size of the resulting synthesized beam is listed in Table \ref{tab:spw} of Appendix \ref{app-spw}. 
The average synthesized beam for the two 1.875 GHz bandwidth spectral windows was 0.57$\arcsec$ $\times$ 0.47$\arcsec$ (PA $\sim$ -66$^\circ$). 
We applied the primary beam correction to all images. 
This was of particular importance for this work since both analyzed sources were close to the edge of the half power beam width, with primary beam correction factors of $\sim$0.5$-$0.6. 

CASA version 5.5.0 was used to compute the moment 0 maps of individual transitions 
%que es lo que representamos en las distintas figuras del paper
%I had to inspect the images to see what channels I wanted to include in the moment 0 map, so I guess I always did it in my Mac. 
%corresponding to several of the species detected in the different spectral windows, 
integrating the detected emission over the full line width of every emission line. 
The moment 0 rms for each transition was measured in a 8$\arcsec$ diameter region after integration of the same number of line-free channels as the corresponding moment 0 map.  
%This rms is probably underestimated. A better estimation would have been done with a 4" region. In any case, the rms has not been used quantitatively, but only in the figures.
The rms values were in the 5$-$18 mJy beam$^{-1}$ km s$^{-1}$ range in all cases,  
except for the CO 
%and $^{13}$CO 
moment 0 map, with $\sigma\sim$40 mJy beam$^{-1}$ km s$^{-1}$   
%and $\sigma\sim$20 mJy beam$^{-1}$ km s$^{-1}$, respectively.  
This is likely the result of the CO moment 0 map having $\sim$5$\times$ more channels than the rest. 

%CO 8.3 - 26.0 km/s (34.3km/s or 217 channels) 37 mJy!!
%C18O 4.2 - 9.8 km/s (5.6km/s or 35 channels) 11mJy
%13CO 3.4 - 12.3 km/s (8.9km/s or 55 channels) 18mJy
%H2CO 5.2 - 12.8 (7.6km/s or 48 channels) 8.5mJy
%CH3OH 5.7 - 13.3 km/s (7.6km/s or 12 channels) 11mJy
%CH3OH 5.1 - 14.5 km/s (9.4 km/s or 16 channels) 11 mJy
%CH2DOH 6.4 - 12.7 km/s (6.3 km/s or 11 channels) 10mJy
%C2H5OH 6.4 - 11.4 km/s (5km/s or 9 channels) 8mJy
%CH3OCH3 4.5 - 10.1 km/s (5.6km/s or 10 channels) 10 mJy
%CH3OCHO 7.0 - 11.4 km/s (4.4km/s or 8 channels) 9mJy
%NH2CHO 7.0 - 10.8 km/s (3.8km/s or 7 channels) 8.5 mJy
%CH2DCN 6.3 - 12.1 KM/S (5.8KM/S or 10 channels) 10 mJy 
%13CH3CN 7 - 10.2 km/s (3.2 km/s or 6 channels) 7 mJy
%CH3C15N 6.4 - 8.3 (2km/s or 4 channels) 5mJy
%C2H5CN 7.0 - 10.8 km/s (3.8 km/s or 7 channels) 7 mJy

The spectra at the pixels corresponding to the  continuum peaks of the Ser-emb 11 binary source (pixel size = 0.08$\arcsec$) %in the 231.262 GHz $-$ 233.380 GHz and the 242.978 GHz $-$ 244.853 GHz ranges 
were extracted 
%from the FITS files generated by CASA for the corresponding imaged visibilities 
assuming a source velocity of 8.5 km/s as a first approximation.  
%I used MADCUBA to extract the spectra of the two continuum spws and saved them as .fits files. Then, I needed to save them in the same spectra directory in order to use SLIM to identify the different transitions. However, MADCUBA closed and I don't know where I saved the spectra directory and the SLIM product. 
%The three spectra (232 and 2 for the 244 because CASA crashed if I tried to clean the whole spw at once) are saved in the sg_230GHz/ser-emb_17_cont_selfcal folder. 
%I used spectra.py in the two continuum spw folders to extract the spectra and saved them in a .txt file that is in both the spw folder as well as the spectra_ascii_ser-emb_11 folder. In this folder I  did the Gaussian fits. This saves the spectrum in a .txt file. 
%On the other hand, I used spectra_ser_emb11e/w.py to directly extract and plot the spectrum of each source in the serpens folder. 
%and subsequently used to detect and identify molecular transitions. 
%We note that the spectra used for line identification was extracted by the MADCUBA software at the continuum peak of Ser-emb 11 W. 
%The spectra presented in Fig. 3, also used to derive the emission line parameters through Gaussian fitting, was extracted with \texttt{python} at the continuum peak of Ser-emb 11 E and W. 
The rms of the spectra was measured using the Python function \texttt{numpy.std} to calculate the observed standard deviation in six line-free regions of the spectra. %spanned over the 231.262 GHz $-$ 233.380 GHz and the 242.978 GHz $-$ 244.853 GHz ranges.  
An average value of $\sigma\sim$4 mJy beam$^{-1}$ 
%at a spectral resolution of 1.25 km/s 
%for a channel width of of 0.63 km/s 
was found. 
%It was also calculated with the MADCUBA software for ten different line-free regions using the baseline subtraction function, with the same result. 
%This is similar to the channel rms of the imaged visibilities. 

\begin{figure*}[ht!]
\centering
\includegraphics[width=18cm]{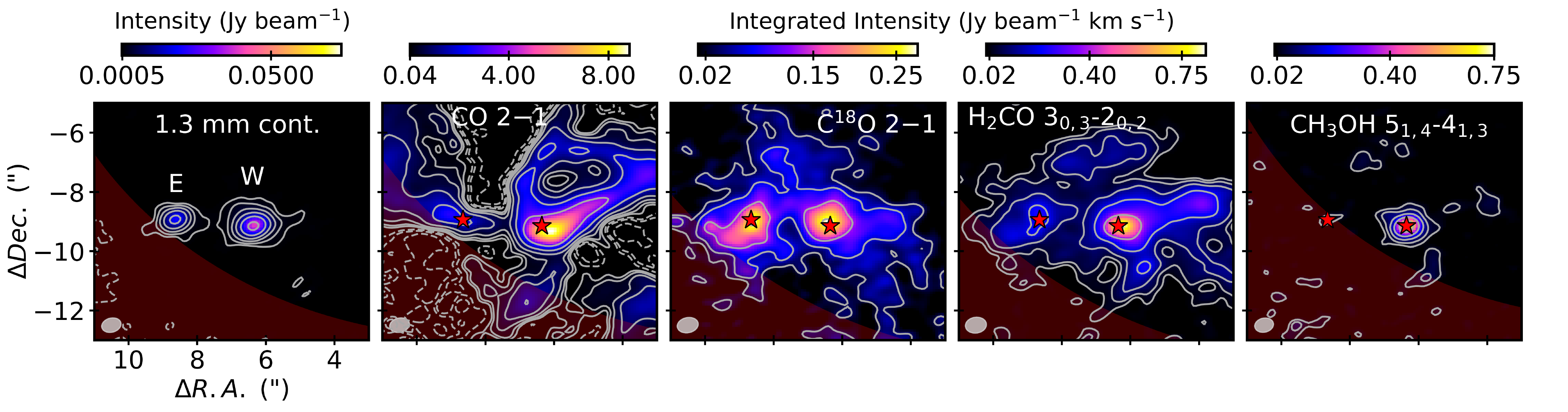}
\caption{
232 GHz continuum, and CO, C$^{18}$O, H$_2$CO, and CH$_3$OH integrated line emission maps toward Ser-emb 11 E and Ser-emb 11 W,  
along with the 3, 6, 12, 24, 48, and 96$\sigma$ contours in white (solid lines), and -3, -6, -12, -24, -48, and -96$\sigma$ contours (dashed lines, see the text). 
%Negative contours help to give a sense of the noise, especifically if the noise is not Gaussian, as it is the case for the extended CO 2$-$1 emission, that is heavily filtered out.  
%Dashed lines in the second panel represent negative contours at the -3 and -24$\sigma$ levels. 
The x and y axes indicate the offset with respect to the center of the observations (Ser-emb 17). 
%(contours for each map using the corresponding value of $\sigma$)
%$\sigma$ = 0.35 mJy beam$^{-1}$ for the continuum. 
%CO 8.3 - 26.0 km/s (34.3km/s or 217 channels) 37 mJy
%C18O 4.2 - 9.8 km/s (5.6km/s or 35 channels) 11mJy
%H2CO 5.2 - 12.8 (7.6km/s or 48 channels) 8.5mJy
%CH3OH 5.7 - 13.3 km/s (7.6km/s or 12 channels) 11mJy
The positions of the continuum emission peaks are marked with a star symbol in each panel. 
Maps are clipped at 1$\sigma$. 
%(each map is clipped at the corresponding value of $\sigma$)
The size of the synthesized beam 
%($\sim$ 0.57$\arcsec$ $\times$ 0.47$\arcsec$) 
is shown on the lower left corner of each panel. 
The region of the field of view observed outside the half power beam width of the primary beam is colored in red.}
\label{overview}
\end{figure*}

\section{Results} \label{sec:results}

\subsection{Overview of the Ser-emb 11 binary  source}\label{sec:results_overview}

Figure \ref{overview} presents an overview of the Ser-emb 11 binary source, as traced by millimeter continuum emission, and CO, C$^{18}$O, H$_2$CO, and CH$_3$OH line emission. 
These species are good tracers of the outflows (CO), envelope (C$^{18}$O), and complex organic chemistry
%wherever the complex organic chemistry is, usually the warm inner envelope or the outflows
(CH$_3$OH) in YSOs. 
H$_2$CO is a COM precursor, and thus traces the complex organic chemistry, but it is also a good tracer of the outflows and protostellar envelopes 
%Since it also traces regions where no other COMs are detected, it is not only a complex organic chemistry tracer. 
\citep{tychoniec21}.
The continuum map shown in the left panel was generated by combining all line-free channels in the 1.875 GHz bandwidth spectral window centered at $\sim$232 GHz. 
The continuum map rms was 0.35 mJy beam$^{-1}$. The continuum emission peaks were used to identify the position of the Ser-emb 11 E and Ser-emb 11 W central protostars, 
marked %with a star symbol 
%The approximate position of the central protostars has been marked 
in every panel of Fig. \ref{overview}.  
%with a dashed green line that represents the 6$\sigma$ contours of the continuum emission. 

The CO J = 2$-$1 moment 0 map is shown in the second panel of Fig. \ref{overview}  
and presents a brighter emission toward the continuum peak of Ser-emb 11 W compared to Ser-emb 11 E. 
%There is also extended emission detected around both components, particularly toward the east of Ser-emb 11 E, and toward the north-west of Ser-emb 11 W. 
The extended emission detected toward the east of Ser-emb 11 E and the north-west of Ser-emb 11 W may be tracing 
%a collimated jet 
outflows 
in the east to west (E-W) %direction, and a wider angle, v-shaped outflow in 
and 
the north-west to south-east (NW-SE) directions, driven by %the 
Ser-emb 11 E and Ser-emb 11 W, 
%central protostars, 
respectively. 
Additional information on the possible outflows is presented in Appendix \ref{app-co-cont}. 
The negative contours in the second panel of Fig. 2 result from incomplete flux recovery and spatial filtering upon the interferometric imaging of extended emission (note that the maximum recoverable scale of the observations was 10$\arcsec$).

The %third panel of Fig. \ref{overview} shows the integrated moment 0 map of the 
C$^{18}$O J = 2$-$1 emission (third panel of Fig. \ref{overview}), a good tracer of the protostar envelopes,   
%In this case, the emission 
presents a similar peak intensity toward the Ser-emb 11 E and W components. 
It also includes a extended component around both protostars, 
suggesting that they have similar gas envelopes. 
%
%The integrated moment 0 map of the $^{13}$CO 2 $-$ 1 emission is not shown in Fig. \ref{overview}, but its spatial distribution was similar to that observed for the C$^{18}$O 2 $-$ 1 emission. 

Line emission from the COM precursor H$_2$CO (fourth panel of Fig. \ref{overview}) 
%is somewhat in between those of the CO and C$^{18}$O 2 $-$ 1 emission lines, presenting extended emission toward both sources, but obscured toward the Ser-emb 11 E component. %compared to the higher intensity observed toward Ser-emb 11 W. 
is detected toward both protostars, and presents also an extended component, but it is a factor of $\sim$4 brighter toward Ser-emb 11 W.  
On the other hand, CH$_3$OH (5$_{1,4}$ $-$ 4$_{1,3}$) emission was only detected in a compact, marginally resolved region toward the continuum peak of Ser-emb 11 W.    
%Therefore, the presence of complex organic chemistry in this binary system seems to be limited to the component postulated as a protostellar disk candidate in \citet{enoch11}. 
This suggests that the western component of the binary system could harbor hot corino chemistry. 
The hot corino nature of Ser-emb 11 W is further evaluated in the following Sections. 

\subsection{Detection of organic molecules toward Ser-emb 11 W}\label{sec:results_hotcorino}

\begin{figure*}%[ht!]
\centering
\includegraphics[width=18cm]{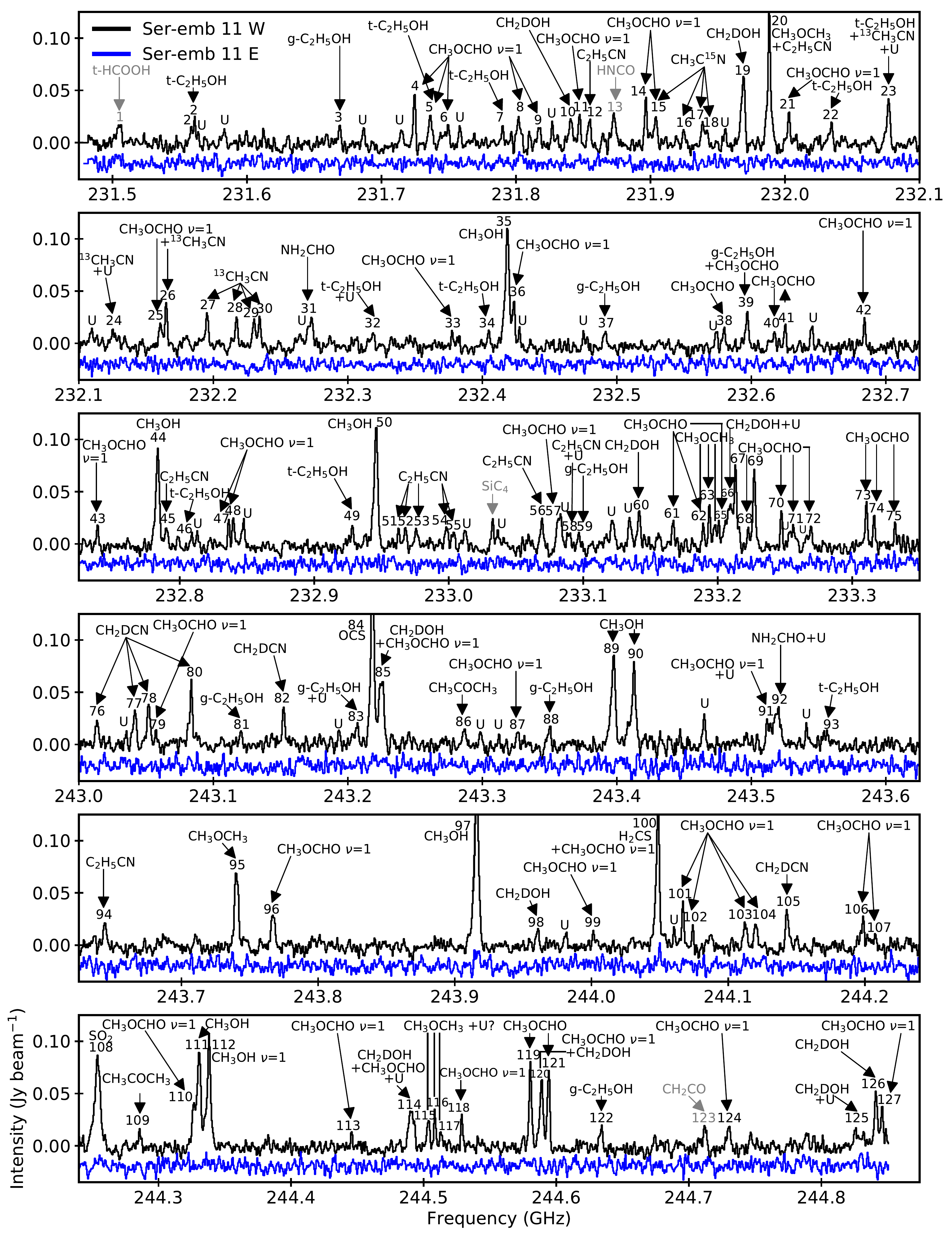}
\caption{Observed spectra toward the continuum peak of Ser-emb 11 E (blue) and Ser-emb 11 W (black) in the 231.262 GHz $-$ 233.380 GHz range (top three panels) and the 242.978 GHz $-$ 244.853 GHz range (bottom three panels). 
The spectra are offset for clarity. 
Identified lines are indicated with a number (see Tables \ref{lines1}$-$\ref{lines6}). 
Tentative detections are indicated in grey. 
Unidentified lines above 4$\sigma$ are marked with a U.
%(above a 3 $\sigma$ level)
}
\label{fig:spec}
\end{figure*}

The spectra at the continuum peak of Ser-emb 11 E and Ser-emb 11 W in the 231.262 GHz $-$ 233.380 GHz range and the 242.978 GHz $-$ 244.853 GHz range are presented in Fig. \ref{fig:spec} (top three panels and bottom three panels, respectively). 
While no transitions were observed toward Ser-emb 11 E in this spectral range, more than 120 lines were identified toward Ser-emb 11 W.  
In order to assign the emission lines, we used the Spectal Line Identification and Modelling (SLIM) module within the MADCUBA\footnote{The MAdrid Data CUBe Analysis package is a software developed at the Center of Astrobiology (Madrid, INTA-CSIC) to visualize and analyze single spectra and data cubes \citep{rivilla16a,rivilla16b}.} package \citep{madcuba}, which makes use of the Jet Propulsion Laboratory \citep[JPL;][]{pick98} and the Cologne Database for Molecular Spectroscopy \citep[CDMS;][]{mull05} spectral catalogs. 
To confirm the assignments, the SLIM module allowed us to simultaneously identify all detectable transitions of a particular species in our spectral range. 
%For every putative assignment, we checked for competing identifications, and looked for all the detectable transitions of that particular molecule in our spectral range, confirming that there were not any missing lines in our observations. 
%Most of the identified species presented more than one emission line in our spectral range, or at least one line intense enough to be unambiguously assigned. 
%For the rest of species with only one observed emission line, the detection was considered tentative. 
%
A list of the identified molecular transitions in Fig. \ref{fig:spec} is presented in Tables \ref{lines1}$-$\ref{lines6} of Appendix \ref{app-lines}. 
%11 distinct species (5 O-bearing COMs, 3 N-bearing COMs, and 3 small S-bearing molecules) were identified in Fig. \ref{fig:spec}, along with 3 additional isotopologues. 
%Another 3 species (2 O-bearing and 1 N-bearing molecule) were also tentatively detected in the same spectral region. 
The detected species presented multiple emission lines in our spectral range that were observed with a signal-to-noise ratio (SNR) higher than 3 ($\sigma$ $\sim$ 4 mJy, Section \ref{sec:obs_im}). Additional tentatively detected lines (observed with SNR $<$ 3) corresponding to these species are also indicated in Fig. \ref{fig:spec} and listed in Appendix \ref{app-lines}, but they were not considered in the following discussion. Species with only one line observed at SNR $>$ 3 were considered tentative detections.%, and are labelled as such in Tables \ref{lines1} and \ref{lines6}. 
%Unidentified SNR $>$ 3 emission lines are marked with a U in Fig. \ref{fig:spec}. 

A total of five O-bearing COMs were detected toward the continuum peak of Ser-emb 11 W: CH$_3$OH, CH$_3$CH$_2$OH, CH$_3$OCH$_3$, CH$_3$OCHO, and CH$_3$COCH$_3$, along with the CH$_2$DOH isotopologue %. The observed emission lines are listed in 
(Tables \ref{lines1}-\ref{lines5}). 
%In addition to CH$_3$OH and CH$_2$DOH (Table \ref{lines1}), we detected another four O-bearing COMs: 
%CH$_3$OH (Table \ref{lines1}), 
%CH$_3$CH$_2$OH (Table \ref{lines1}), 
%CH$_3$COCH$_3$ (Table \ref{lines1}), 
%CH$_3$OCH$_3$ (Table \ref{lines2}), and 
%CH$_3$OCHO (Tables \ref{lines2}$-$\ref{lines5}) 
Two emission lines tentatively assigned to CH$_2$CO and t-HCOOH are also included in Table \ref{lines1}. 
Table \ref{lines6} lists the observed emission lines corresponding to three confirmed N-bearing COMs (NH$_2$CHO, CH$_2$DCN, and CH$_3$CH$_2$CN), and the  $^{13}$CH$_3$CN and CH$_3$C$^{15}$N isotopologues, as well as one tentatively detected N-bearing species (HNCO). 
%
%Another three small (4 atoms or less) S-bearing species were also detected in Fig. \ref{fig:spec}.  %(Table \ref{lines6}):  SO$_2$, OCS, and H$_2$CS. 
%A second SO$_2$ transition, as well as one transition corresponding to the O$^{13}$CS isotopologue were detected in two of the narrow spectral windows presented in Sect. \ref{sec:small}. 
Three emission lines corresponding to small (4 atoms or less) S-bearing species are also identified in Fig. \ref{fig:spec}. They are presented in Sect. \ref{sec:small} along with the additional small molecules detected in dedicated, narrow spectral windows.

\begin{figure*}%[ht!]
\centering
\includegraphics[width=\textwidth]{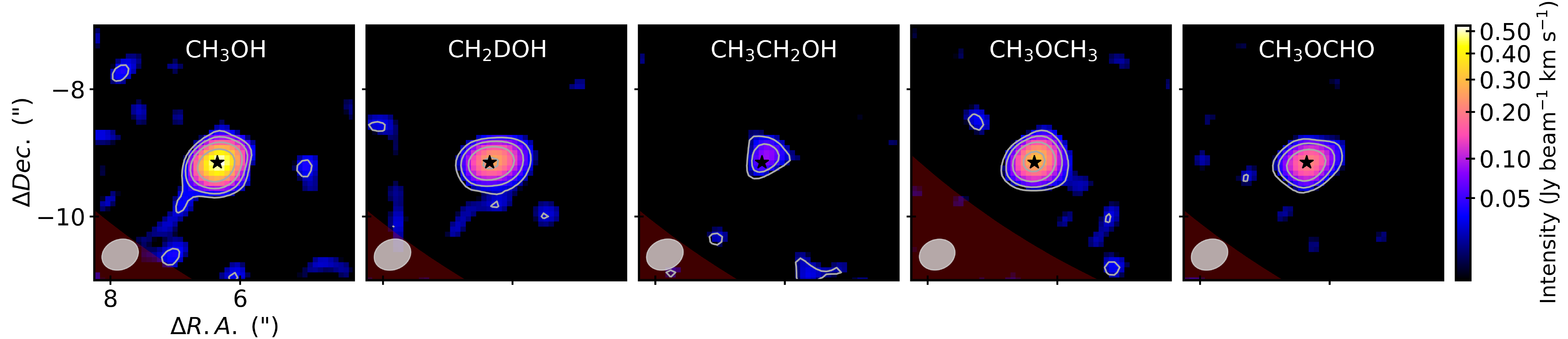}
\includegraphics[width=\textwidth]{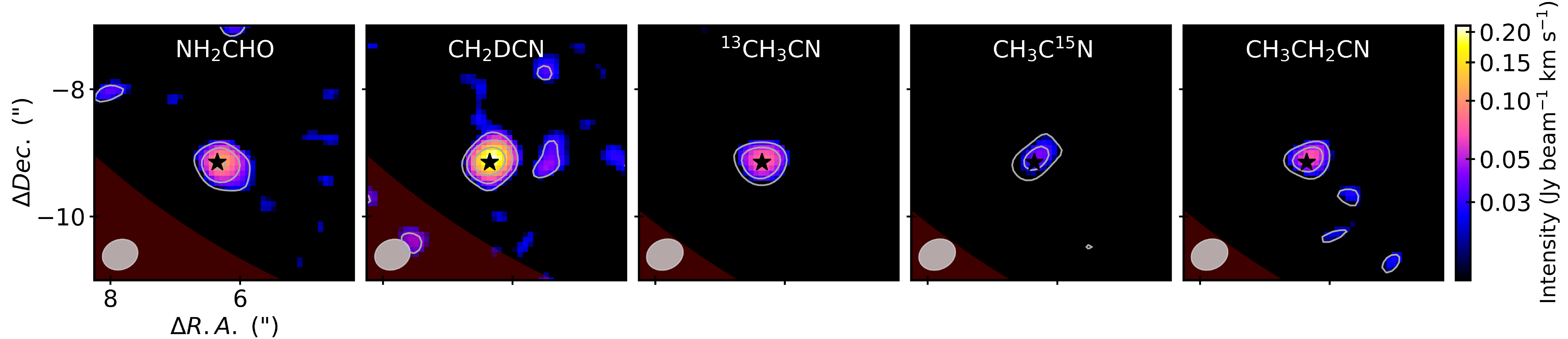}
\caption{
Integrated moment 0 maps toward Ser-emb 11 W, along with the 3, 6, 12, and 24$\sigma$ contours, of the O-bearing (top panels) and N-bearing (bottom panels) COM transitions highlighted in Tables \ref{lines1}-\ref{lines6} of Appendix \ref{app-lines}. 
%(contours for each map using the corresponding value of $\sigma$)
%CH3OH 5.1 - 14.5 km/s (9.4 km/s or 16 channels) 11 mJy
%CH2DOH 6.4 - 12.7 km/s (6.3 km/s or 11 channels) 10mJy
%C2H5OH 6.4 - 11.4 km/s (5km/s or 9 channels) 8mJy
%CH3OCH3 4.5 - 10.1 km/s (5.6km/s or 10 channels) 10 mJy
%CH3OCHO 7.0 - 11.4 km/s (4.4km/s or 8 channels) 9mJy
%%%%%%%%%%%%%%%%%%%%
%NH2CHO 7.0 - 10.8 km/s (3.8km/s or 7 channels) 8.5 mJy
%CH2DCN 6.3 - 12.1 KM/S (5.8KM/S or 10 channels) 10 mJy 
%13CH3CN 7 - 10.2 km/s (3.2 km/s or 6 channels) 7 mJy
%CH3C15N 6.4 - 8.3 (2km/s or 4 channels) 5mJy
%C2H5CN 7.0 - 10.8 km/s (3.8 km/s or 7 channels) 7 mJy
The position of the continuum emission peak is marked with a star symbol in each panel. 
Maps are clipped at 2$\sigma$. 
%(with $\sigma$ = 10 mJy for all panels, since all panels use the same color scale)
%The upper level energy of the transition is indicated in every panel. 
The size of the synthesized beam 
%($\sim$ 0.57$\arcsec$ x 0.47$\arcsec$) 
is shown on the lower left corner of every panel. 
The region of the field of view observed outside the half power beam width of the primary beam is colored in red.
}
\label{overview_com}
\end{figure*}

\subsection{COM spatial distribution in Ser-emb 11 W}\label{sec:results_spat}

In order to explore the spatial distributions of the different O- and N-bearing COMs detected toward Ser-emb 11 W, Fig. \ref{overview_com} presents the moment 0 maps of the strongest observed transitions of every species among those with E$_{up}$ $>$ 75 K (highlighted in Tables \ref{lines1}-\ref{lines6} of Appendix \ref{app-lines}). 
Unblended emission lines were preferentially selected.  
CH$_3$COCH$_3$ is not included in Fig. \ref{overview_com},  since its emission lines were not strong enough to adequately image them. 
%For species with several detected isotopologues, only the most abundant one has been represented. 
%We note that even though two CH$_3$COCH$_3$ lines were unambiguously identified in the extracted spectrum toward the continuum peak of Ser-emb 11 W (see Fig. \ref{fig:spec} and Table \ref{lines1}), this emission did not meet the requirements to be masked according to the imaging process described in Sect. \ref{sec:obs_im}. Therefore, the resulting integrated moment 0 maps were dirty images, and are not included in Fig. \ref{overview_com}. %I could get the moment 0 map with manual masking.
%
In all cases, the emission of the organic molecules appears  unresolved in a compact region around the central protostar.  %that could correspond to a hot corino. %(see also Sect. \ref{sec:results_col}). 

We estimated the size of the emitting region observed for the CH$_3$OH (10$_{3,7}$ $-$ 11$_{2,9}$) line (the strongest unblended COM emission line with E$_{up}$ $>$ 75 K, top left panel of Fig. \ref{overview_com}), using a 2D Gaussian fit in CASA.   
After deconvolving the synthesized beam, the resulting estimated size was $\Omega_{source}$ = 0.17$\arcsec$ $\times$ 0.11$\arcsec$ (74 au $\times$ 48 au). 
We assumed this emission size for all COMs throughout the rest of the paper,  
%We note that the hot corino would thus be unresolved in our observations. 
%This estimated size is the largest source size consistent with the observations, but 
although the actual size could be smaller. 

Assuming that the unresolved COM emission corresponds to a hot corino, and that COMs were sublimated from the ice mantles upon passive heating of the inner envelope, we can calculate the expected size of the hot corino based on the protostellar radial temperature profile:
%The minimum hot corino size would correspond to the distance beyond which the temperature of the gas phase is lower than 100 K (the desorption temperature of the ice mantles). 
%This distance can be calculated with Eq. \ref{eq_temp}:

\begin{equation}
    T(r) = 60 \bigg( \frac{r}{1.34\times10^{4}\rm{au}} \bigg)^{-q} \bigg( \frac{L_{bol}}{10^5L_{\sun}} \bigg)^{q/2} \rm{K}, \label{eq_temp}
\end{equation}

with $q$ = 2/(4+$\beta$) \citep{chandler00}. 
Assuming $\beta$ = 1.5 \citep{jenny19} and $L_{bol}$ = 4.8 $L_{\sun}$ \citep{enoch11}, 
the estimated radial distance at which the temperature drops below 100 K (the desorption temperature of a water-rich ice mantle) is $\sim$24 au. 
This corresponds to a source size (diameter) of $\Omega_{100K}$ $\sim$ 0.11$\arcsec$ $\times$ 0.11$\arcsec$ (diameter), which is only a factor of $\sim$1.5 smaller than our adopted emission region.

\subsection{COM excitation temperatures and column densities in Ser-emb 11 W} \label{sec:results_col}

For those COMs with detected transitions spanning a wide range of upper level energies ($\Delta$E$_{up}$ $\gtrsim$ 150 K), 
%WARNING: CH2DCN with 135K has a good rotational diagram, but C2H5OH with 144K does not. 
a population diagram analysis was used to derive excitation temperatures and column densities from the observed line integrated intensities. 
The analysis was adopted from \citet{goldsmith99} and is described in Appendix \ref{app-rot}. 
The line integrated intensities ($\int{S_{\nu} dv}$) and uncertainties were calculated by fitting a Gaussian to the spectral features observed in 
%the spectrum extracted at the Ser-emb 11 W continuum peak
Fig. \ref{fig:spec}, 
using the Levenberg-Marquardt minimization implementation of \texttt{scipy.optimize.curve\_fit} in Python. 
In the case of multiple blended transitions, a multiple Gaussian fitting approach was adopted. More information about the fitting process can be found in Appendix \ref{app-lines}. 
The estimated position, width, integrated intensities, and uncertainties of the identified emission lines are listed in Tables \ref{lines1}$-$\ref{lines6}. 
The average full width half maximum (FWHM) of the O- and N-bearing COM emission lines was 3.0 $-$ 3.5 km/s. 
CH$_3$OH and NH$_2$CHO presented broader emission lines (FWHM $\sim$4.2 km/s), 
while the CH$_3$OCHO lines were slightly narrower ($\sim$2.7 km/s). 
%
%%As shown in Sect. \ref{sec:results_hotcorino}, the complex organic emission is located in a compact region around the central Ser-emb 11 W protostar. 
%The average size of the synthesized beam in our observations was $\Omega_{beam}$ = 0.57$\arcsec$ $\times$ 0.47$\arcsec$ (Sect. \ref{sec:obs_im}), while the estimated size of the hot corino toward Ser-emb 11 W was $\Omega_{source}$ $\le$ 0.17$\arcsec$ $\times$ 0.11$\arcsec$ 
%%according to the 2D Gaussian fit of the CH$_3$OH 10$_{3,7}$ $-$ 11$_{2,9}$ integrated moment 0 map after deconvolving the beam size 
%(Sect. \ref{sec:results_hotcorino}). 
%We note that this estimated source size 
%%(that was assumed as the emitting region of all detected COMs) 
%is the largest source size consistent with the observations, but the actual size of the emitting region could be smaller. %

\begin{figure*}
\gridline{
\fig{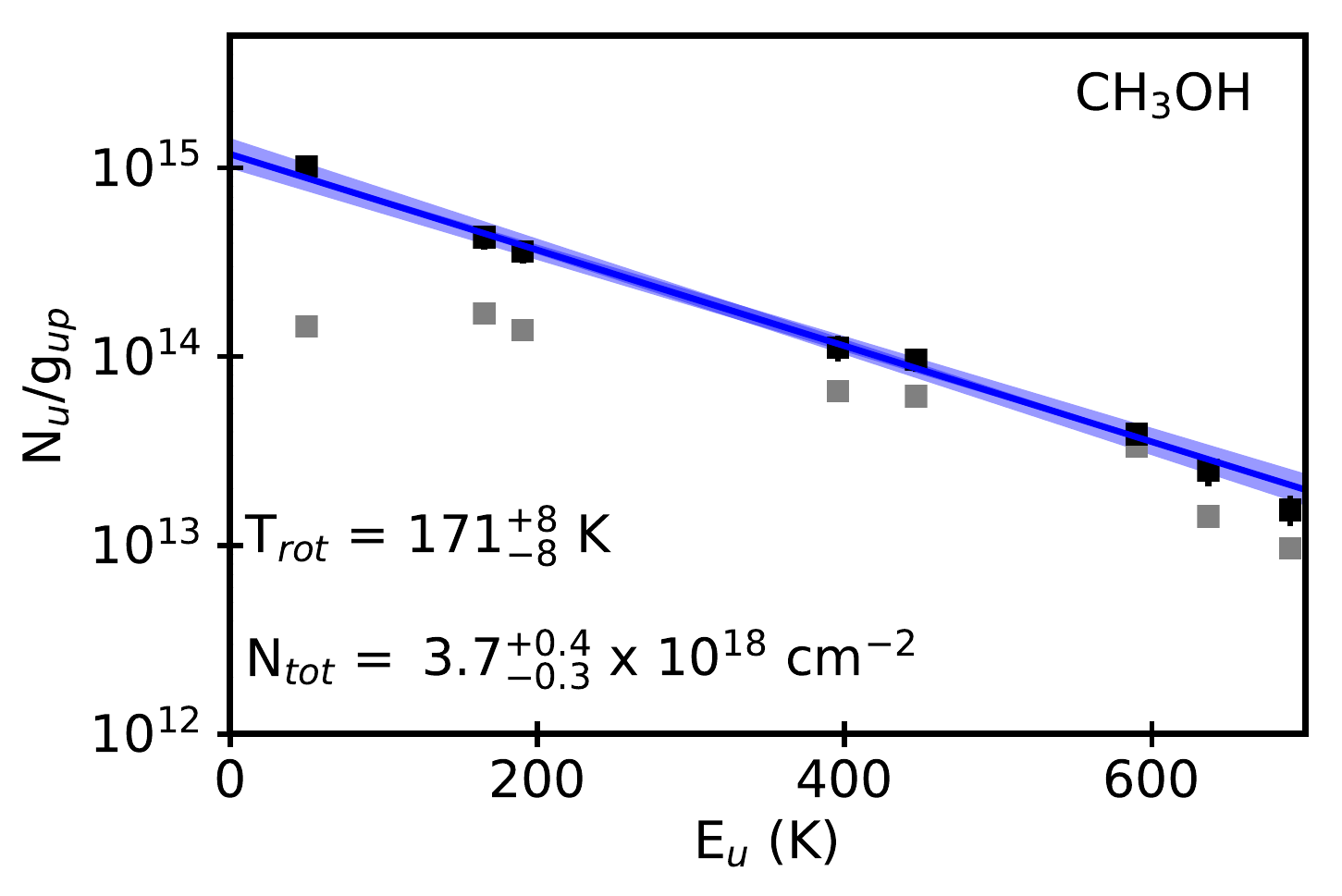}{0.45\textwidth}{}
\fig{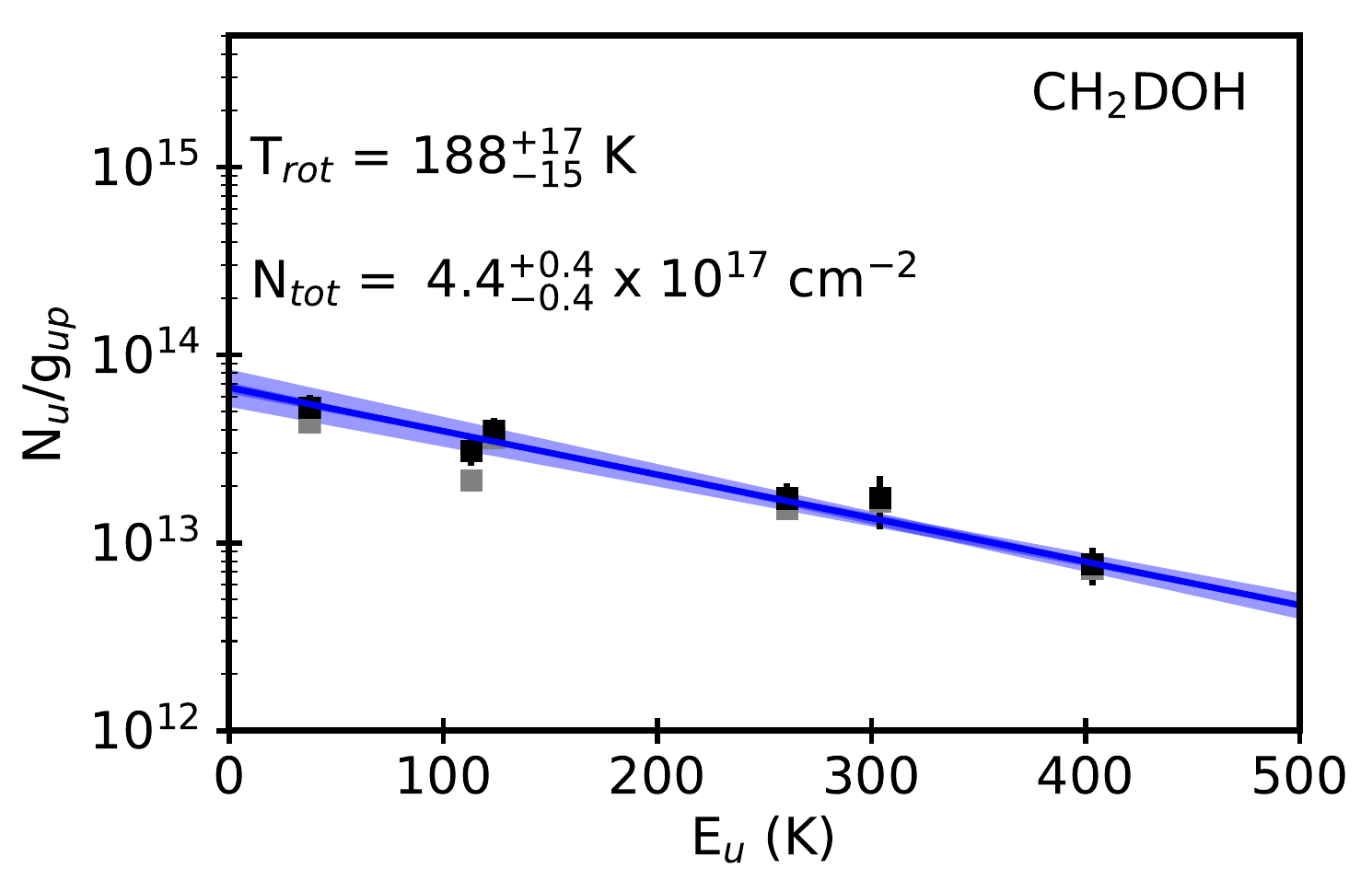}{0.45\textwidth}{}}
\vspace{-7.5mm}
\gridline{
\fig{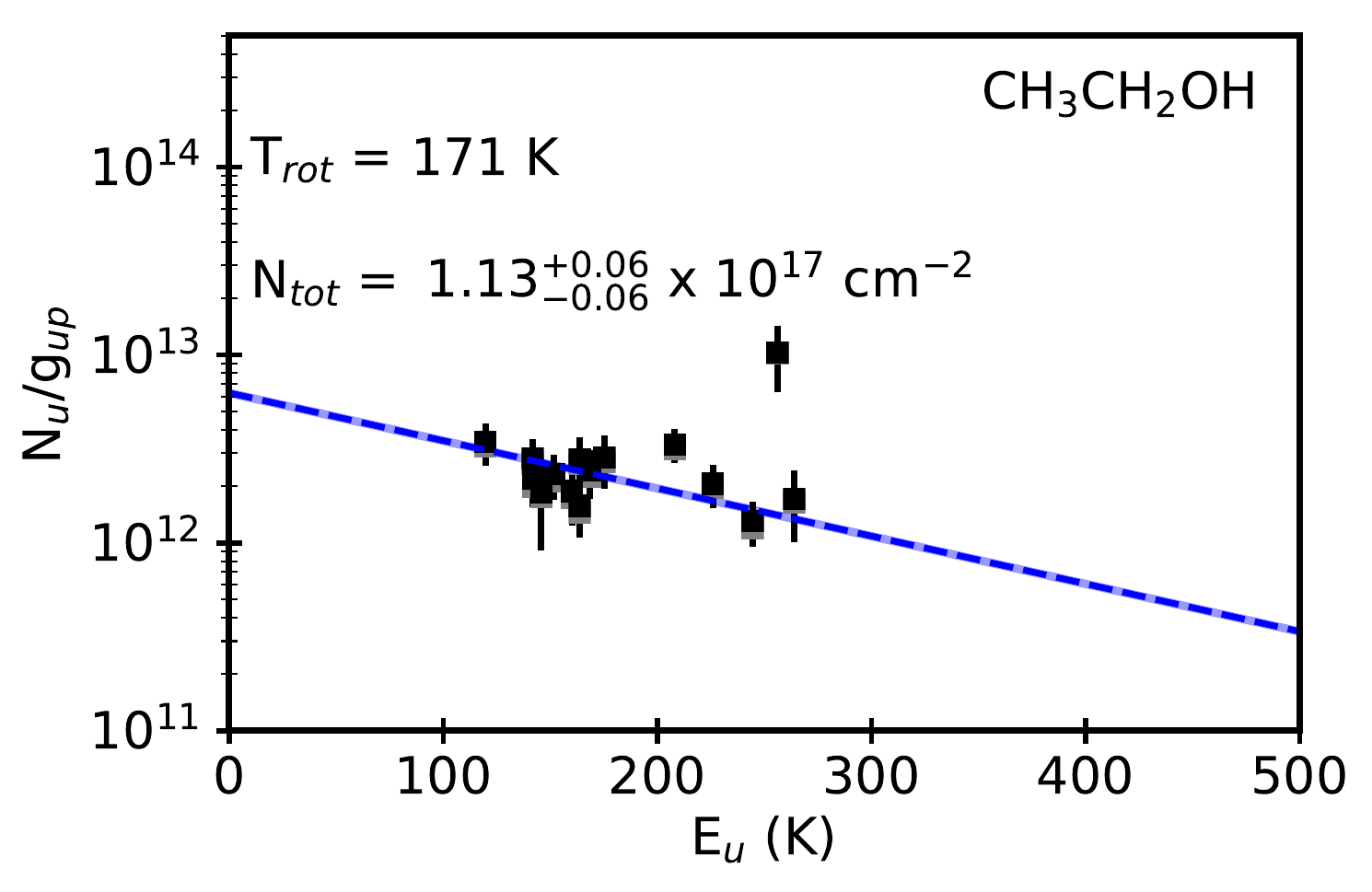}{0.45\textwidth}{}
\fig{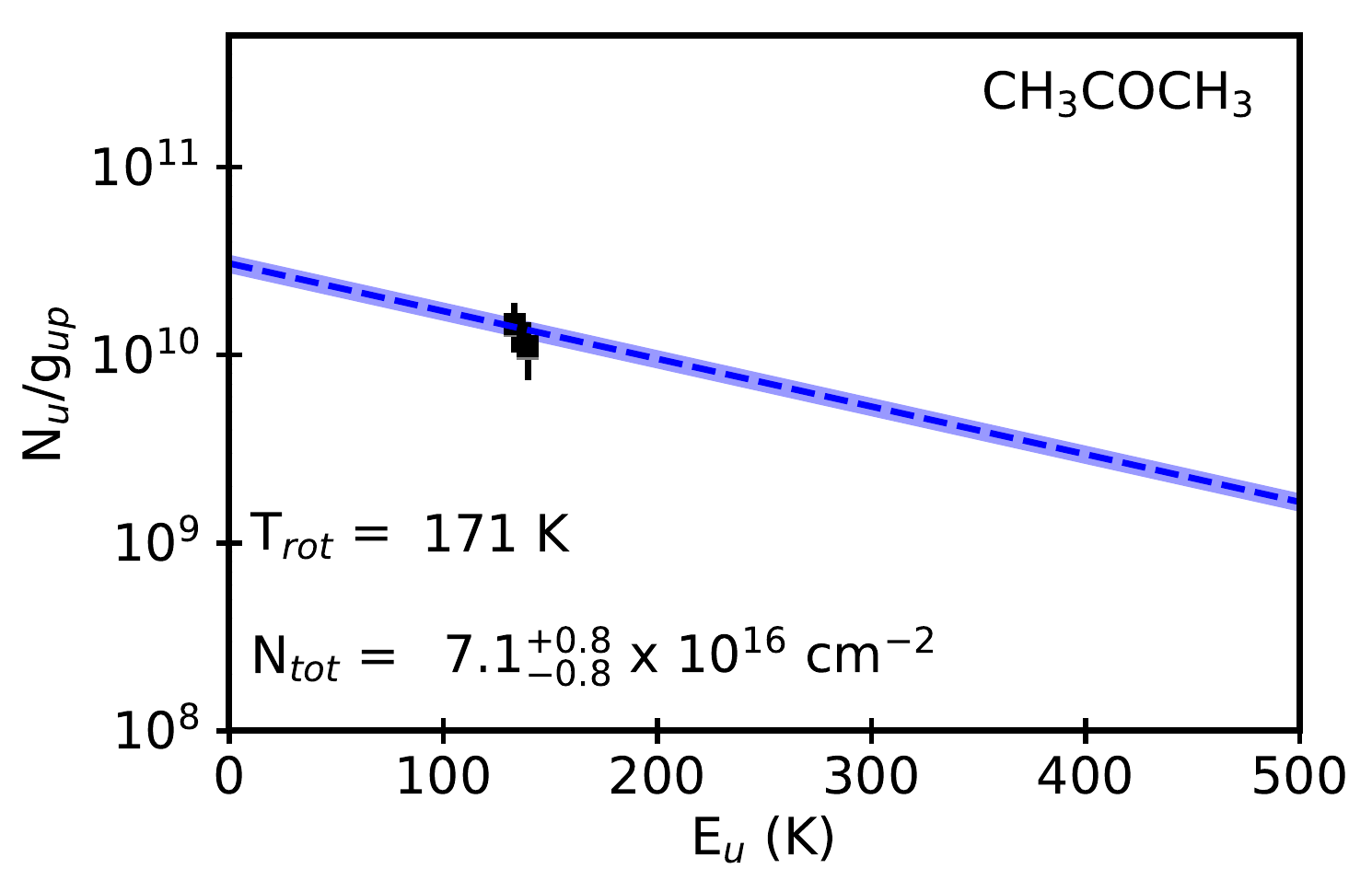}{0.45\textwidth}{}}
\vspace{-7.5mm}
\gridline{
\fig{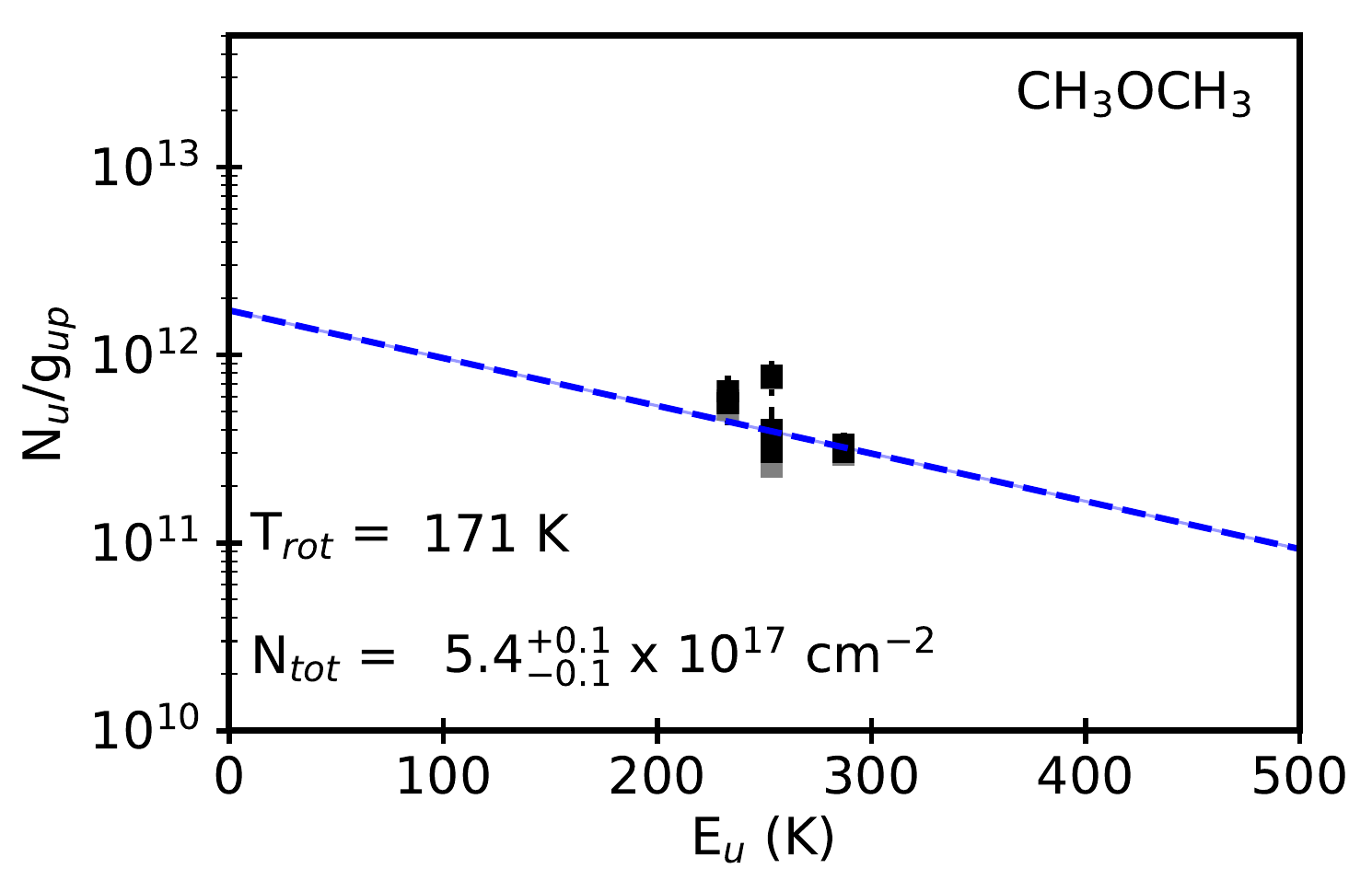}{0.45\textwidth}{}
\fig{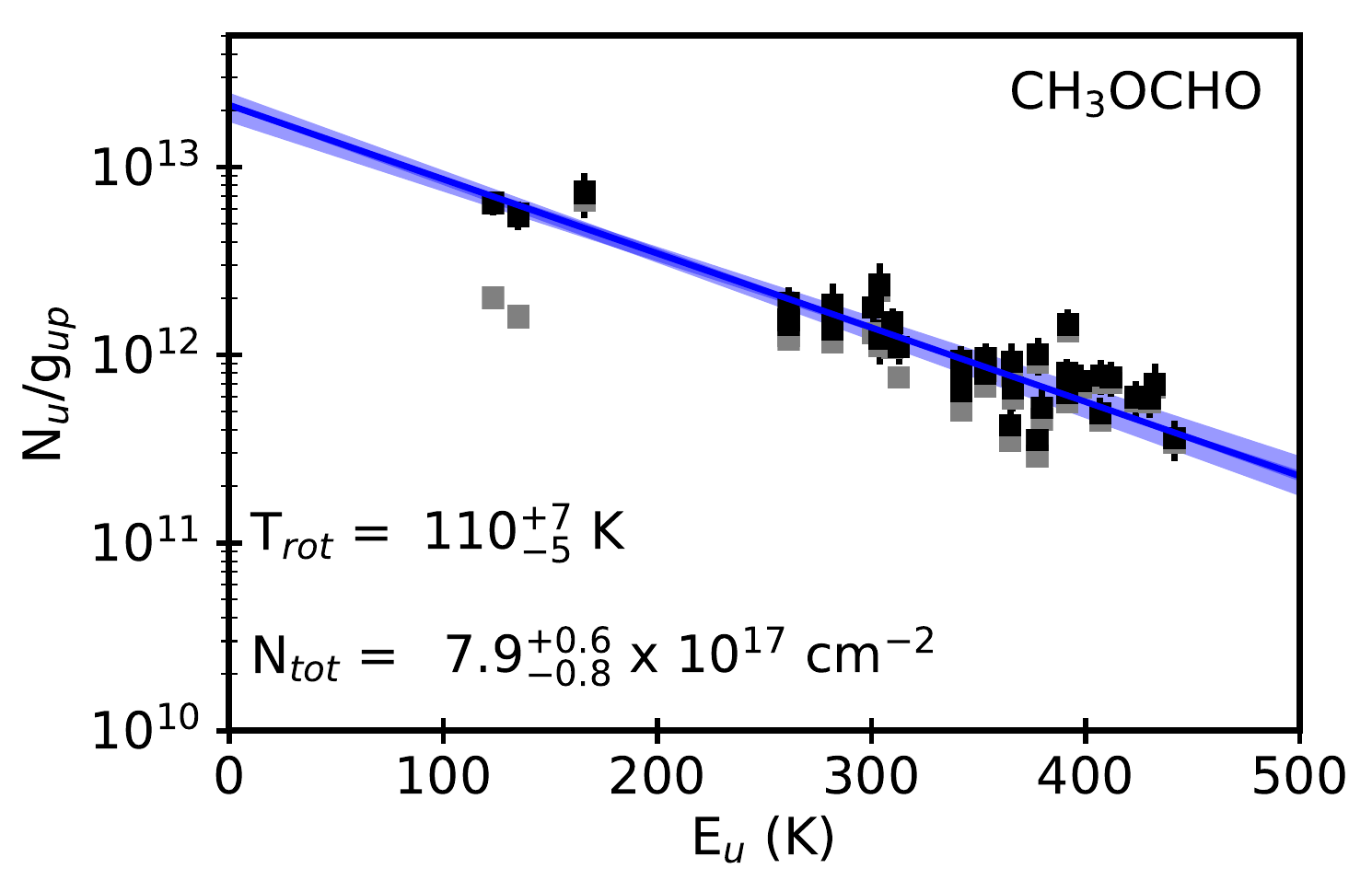}{0.45\textwidth}{}
}
\vspace{-7.5mm}
\caption{Population diagrams for the detected O-bearing COMs toward Ser-emb 11 W (black squares).
%The squares correspond to the different emission lines observed toward the continuum peak of Ser-emb 11 W. 
The plotted 1$\sigma$ error bars include a 10\% absolute calibration uncertainty,  
%to the integrated flux density errors provided by the Gaussian fit to account for calibration errors, 
since lines observed with two different frequency settings were used.  
%The observed values were not corrected from the optical depth of the emission lines. 
The upper level populations are corrected according to Eq. \ref{nucorr}. 
The same data points but without the optical depth correction ($C_{\tau}$) are shown as grey squares to highlight the effects of the line optical depths. 
The MCMC best-fit to the population diagrams is shown in blue, along with the 1$\sigma$ confidence region (shaded zone). 
%The best exponential fit to the data according to Eq. \ref{rot} is represented by a solid line, and t
The estimated best-fit rotational temperature and column density are indicated in every panel. 
%Only the transitions with $E_u$ $>$ 200 K were taken into account in the case of CH$_3$OCHO (see the text). 
%A maximum source size of $\Omega_{source}$ = 0.17$\arcsec$ $\times$ 0.11$\arcsec$ was assumed (see the text). 
%We note that the optical depth of most of the observed CH$_3$OH lines was $\tau$ $>$ 0.3. Therefore, the optically thin line assumption (Sect. \ref{app-rot}) may not be valid in this case, and the derived values should be corrected (see the text). The observed CH$_3$CH$_2$OH and CH$_3$COCH$_3$ transitions did not span a wide enough range of E$_{up}$, and the exponential fit could not be performed. 
The CH$_3$OH rotational temperature was adopted in order to estimate the column densities 
of CH$_3$CH$_2$OH and CH$_3$COCH$_3$. 
%(see the text). 
In these cases the 
%dashed line represents the exponential fit according to Eq. \ref{rot} using the $T_{rot}$ and $N_T$ values reported in Table \ref{abundances}.
MCMC best-fit is represented with dashed lines. 
}
\label{orot}
\end{figure*}

\begin{figure*}
\gridline{
\fig{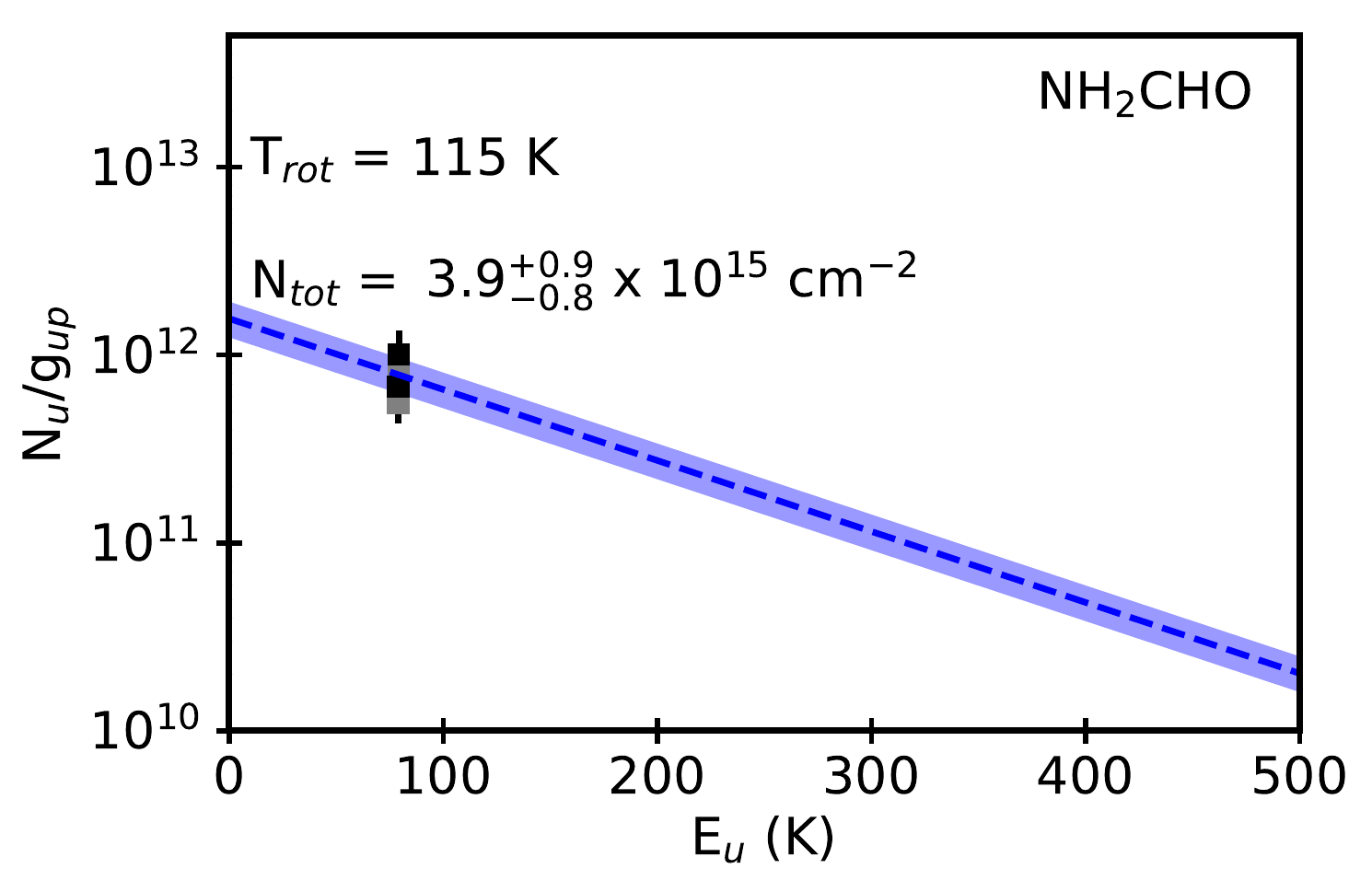}{0.45\textwidth}{}}
\vspace{-7.5mm}
\gridline{
\fig{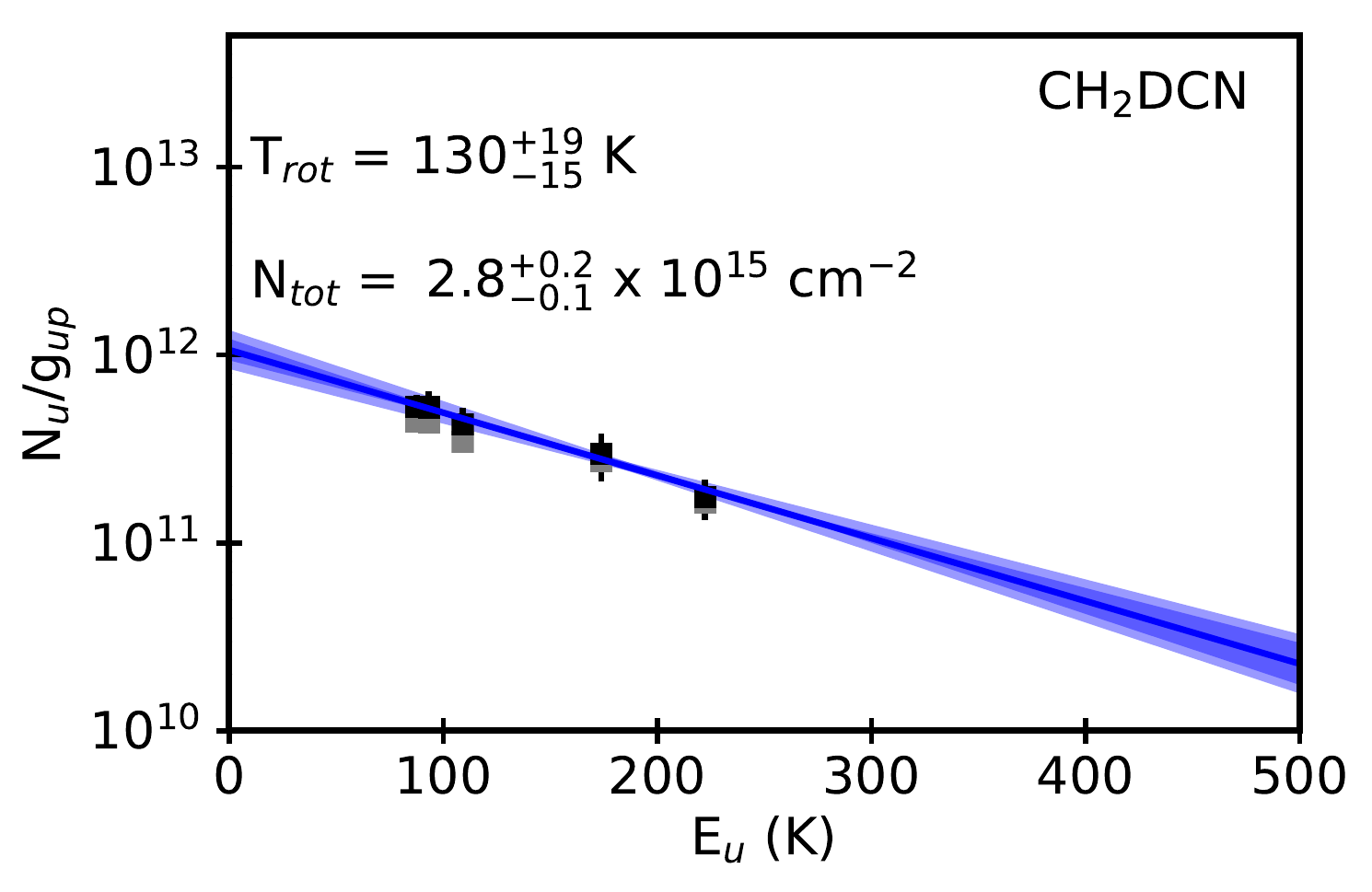}{0.45\textwidth}{}
\fig{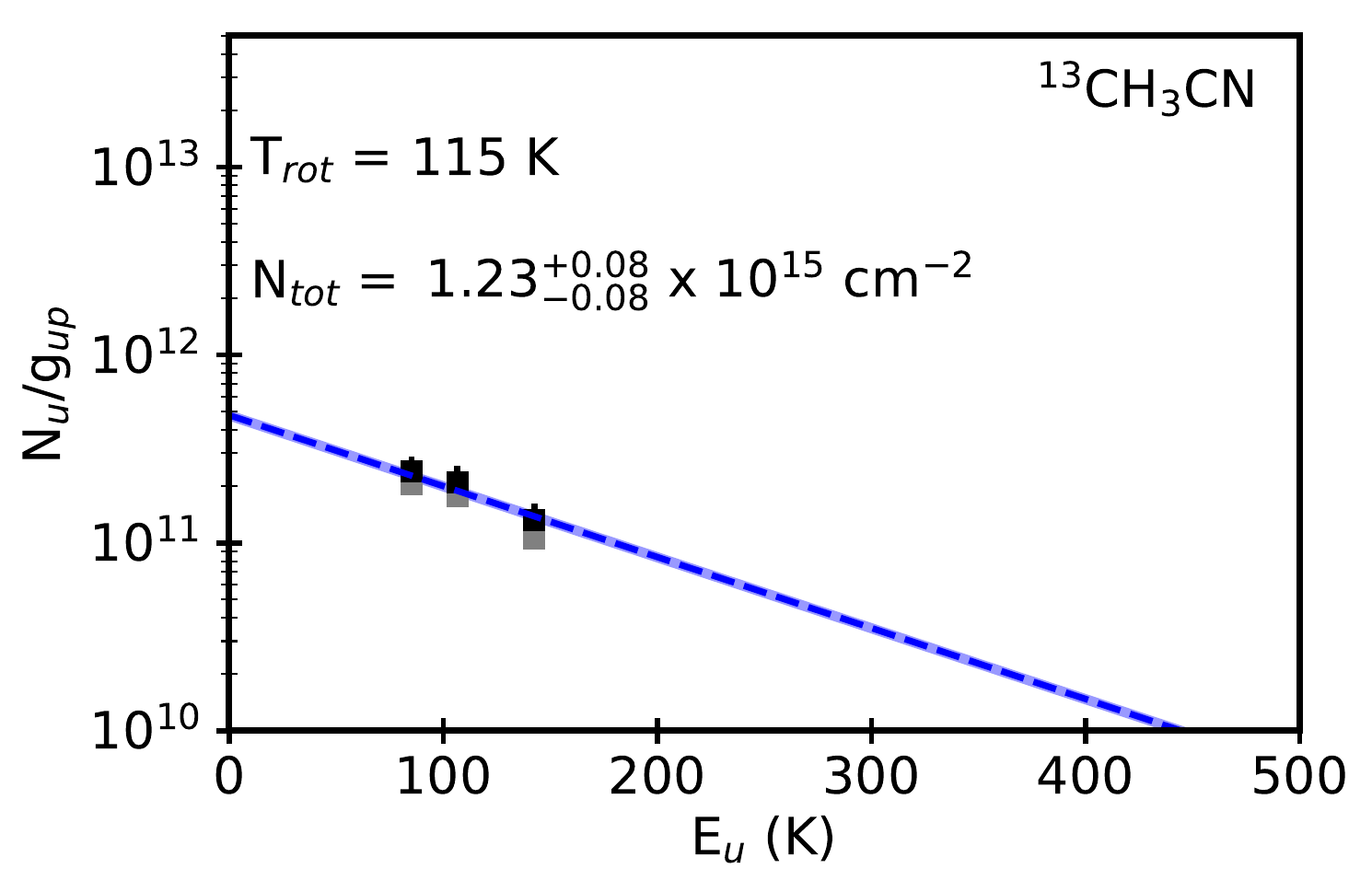}{0.45\textwidth}{}}
\vspace{-7.5mm}
\gridline{
\fig{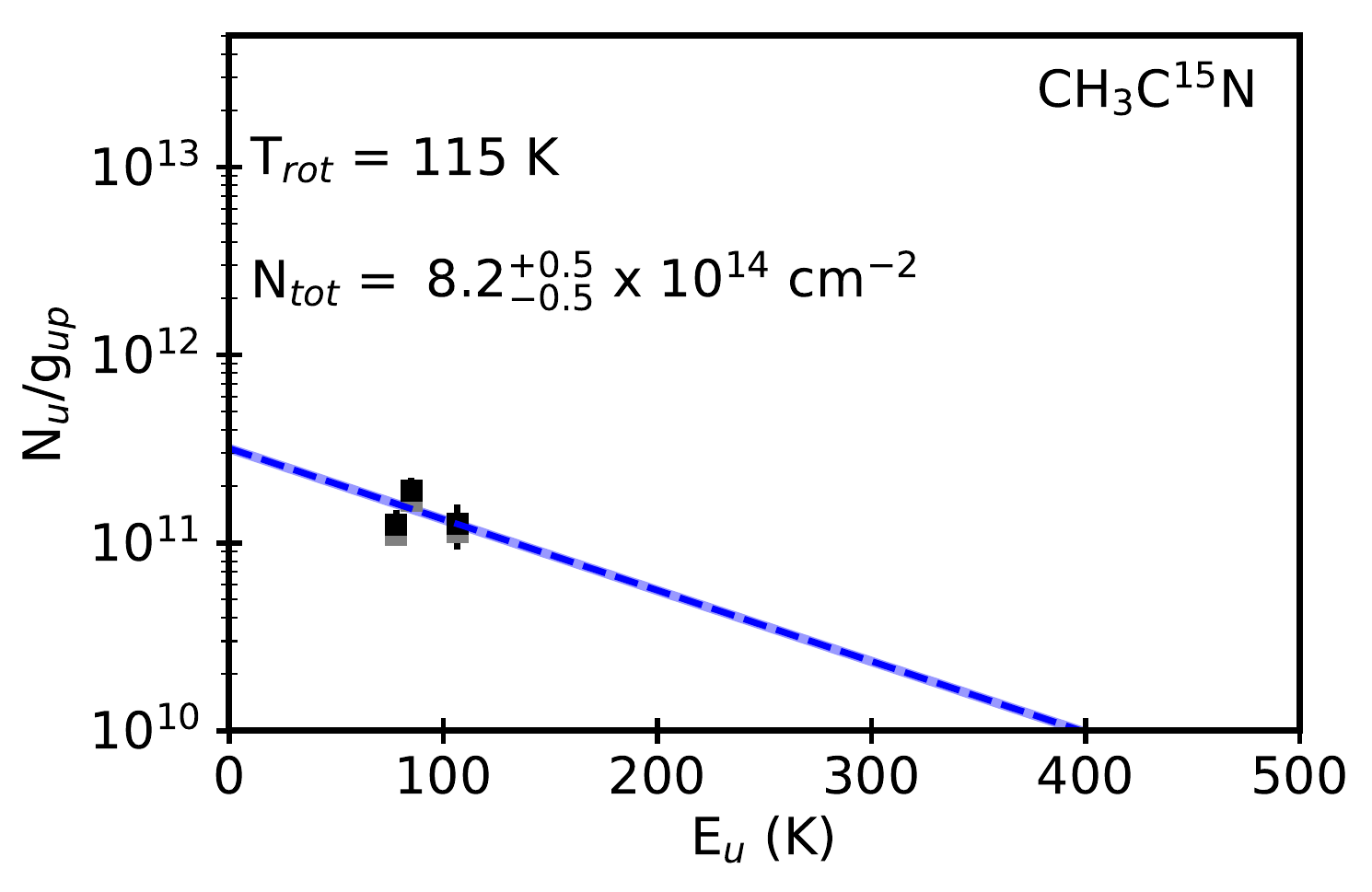}{0.45\textwidth}{}
\fig{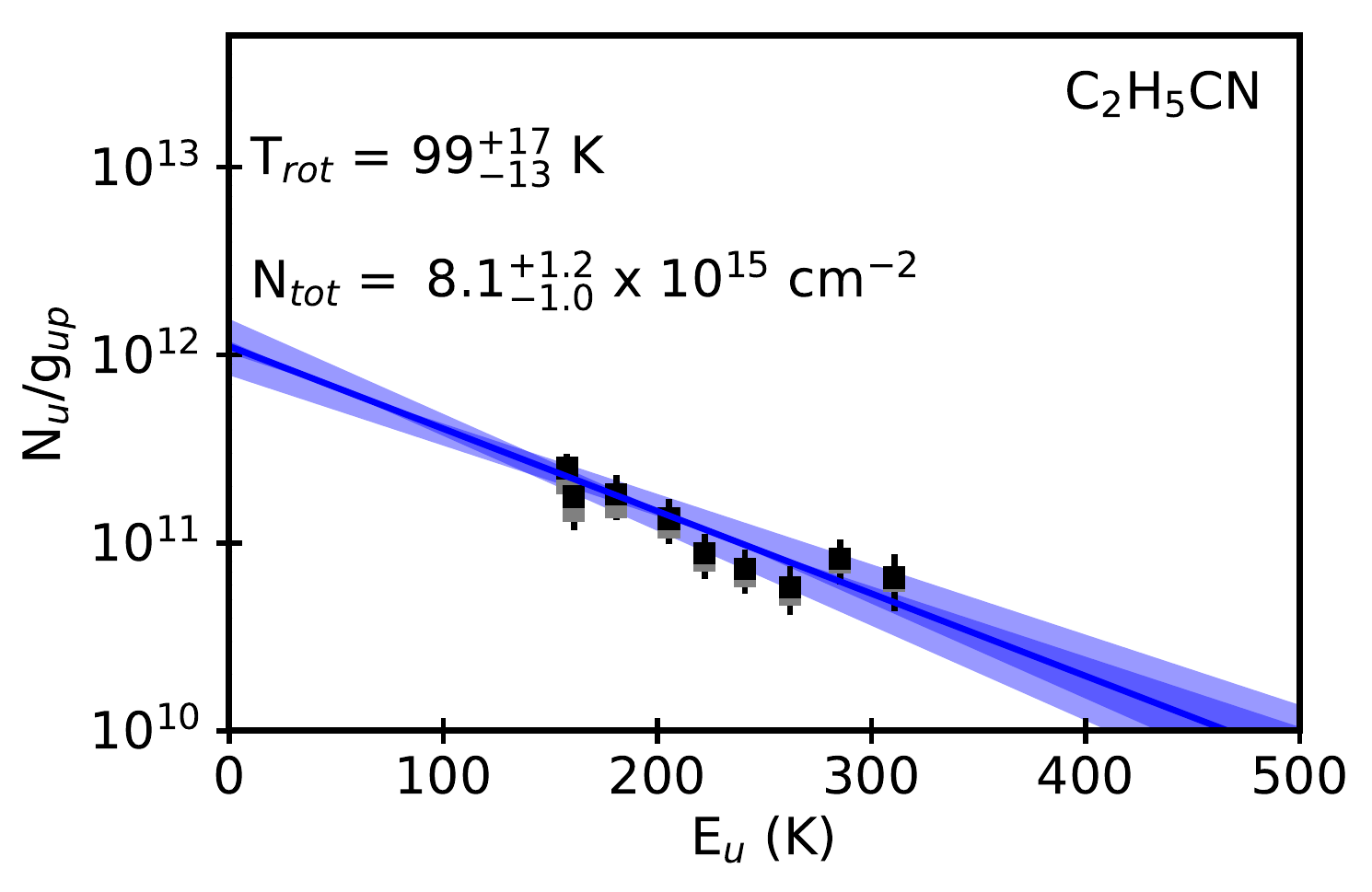}{0.45\textwidth}{}}
\vspace{-7.5mm}
\caption{As Fig. \ref{orot} but for the detected N-bearing COMs toward Ser-emb 11 W (black squares). 
%The squares correspond to the different emission lines observed toward the continuum peak in Ser-emb 11 W. 
%The plotted 1$\sigma$ error bars include a 10\% absolute calibration uncertainty,  
%to the integrated flux density errors provided by the Gaussian fit to account for calibration errors, 
%since lines observed with two different frequency settings were used. 
%The best exponential fit to the data according to Eq. \ref{rot} is represented by a solid line, and t
%The upper level populations are corrected according to Eq. \ref{nucorr}. 
%Values not corrected from the emission line optical depths ($C_{\tau}$) are shown as grey squares. 
%The MCMC fitting to the population diagrams is shown in blue, along with the 1$\sigma$ confidence region (shadowed zone). 
%The estimated rotational temperature and column density are indicated in every panel.
%A maximum source size of $\Omega_{source}$ = 0.17$\arcsec$ $\times$ 0.11$\arcsec$ was assumed (see the text). 
%The observed NH$_2$CHO, $^{13}$CH$_3$CN, and CH$_3$C$^{15}$N transitions did not span a wide enough range of E$_{up}$, and the exponential fit could not be properly performed. 
The average rotational temperature derived for CH$_2$DCN and CH$_3$CH$_2$CN was adopted in order to estimate the NH$_2$CHO, $^{13}$CH$_3$CN, and CH$_3$C$^{15}$N column densities. In these cases, the MCMC fitting is represented with dashed lines.}
\label{nrot}
\end{figure*}

\begin{deluxetable*}{cccc}
\caption{COM column densities and rotational temperatures in the hot corino of Ser-emb 11 W assuming a source size of $\Omega_{source}$ = 0.17$\arcsec$ $\times$ 0.11$\arcsec$. %Values are estimated from a population diagram analysis (see the text). %Small molecules and tentatively detected COMs are not included.
\label{abundances}}
\tablehead{
\colhead{Molecule} & \colhead{T$_{rot}$} & \colhead{$N_T$} & \colhead{$N_T$/$N_T$(CH$_3$OH)}\\
& \colhead{(K)} & \colhead{(m$^{-2}$)} & \colhead{(\%)}}
\startdata
CH$_3$OH & 171$^{+8}_{-8}$ & (3.7$^{+0.4}_{-0.3}$) $\times$ 10$^{18}$ & 100 \\ 
CH$_2$DOH & 188$^{+17}_{-15}$  & (4.4$^{+0.4}_{-0.4}$) $\times$ 10$^{17}$ & 12\\
\hline
C$_2$H$_5$OH & 171$^a$ & (1.13$^{+0.06}_{-0.06}$) $\times$ 10$^{17}$ & 3.1 \\
\hline
%CH$_3$OCH$_3$ & 281$^{+17}_{-15}$  & (8.2$^{+0.5}_{-0.6}$) $\times$ 10$^{17}$ & 22\\
CH$_3$OCH$_3$ & 171$^a$  & (5.4$^{+0.1}_{-0.1}$) $\times$ 10$^{17}$ & 15\\
\hline
CH$_3$OCHO & 110$^{+7}_{-5}$ & (7.8$^{+0.6}_{-0.8}$) $\times$ 10$^{17}$ & 21
\\%with vibrational corrections
\hline
CH$_3$COCH$_3$ & 171$^a$ & (7.1$^{+0.8}_{-0.8}$) $\times$ 10$^{16}$ & 1.9  \\ 
\hline
\hline
NH$_2$CHO & 171$^a$ & (5.0$^{+1.0}_{-1.0}$) $\times$ 10$^{15}$ & 0.14  \\
NH$_2$CHO& 115$^b$ & (3.9$^{+0.9}_{-0.8}$) $\times$ 10$^{15}$ & 0.11 \\
\hline
CH$_2$DCN & 130$^{+19}_{-15}$  & (2.8$^{+0.2}_{-0.1}$) $\times$ 10$^{15}$ & 0.076 \\
$^{13}$CH$_3$CN & 115$^b$ & (1.23$^{+0.08}_{-0.08}$) $\times$ 10$^{15}$ & 0.032\\
CH$_3$C$^{15}$N & 115$^b$ & (8.2$^{+0.5}_{-0.5}$) $\times$ 10$^{14}$ & 0.022\\
\hline
C$_2$H$_5$CN & 99$^{+17}_{-13}$ & (8.1$^{+1.2}_{-1.0}$) $\times$ 10$^{15}$ & 0.22\\%with vibrational corrections
\enddata
\tablecomments{%These column densities are calculated assuming a maximum source size of $\Omega_{source}$ = 0.17$\arcsec$ $\times$ 0.11$\arcsec$, as extracted from a 2D Gaussian fit of the CH$_3$OH 10$_{3,7}$ $-$ 11$_{2,9}$ E moment 0 map (after deconvolving the beam size), and should therefore be considered lower limits (see the text).\\
%
%The CH$_3$OCHO column density takes into account a correction factor of 1.47 to include higher torsional states in the partition function of this species reported in the JPL catalog. \\
%
$^a$The C$_2$H$_5$OH, CH$_3$COCH$_3$, and NH$_2$CHO column densities were estimated assuming the CH$_3$OH rotational temperature (Sect. \ref{sec:results_col}). 
$^b$The NH$_2$CHO, $^{13}$CH$_3$CN and CH$_3$C$^{15}$N column densities were estimated assuming the average rotational temperature derived for the complex nitriles. 
%
%Two column densities are reported for NH$_2$CHO, depending on which rotational temperature was assumed. 
%The reported rotational temperature and column densities are consistent with those estimated from a Gaussian fit of the observed spectrum using the MADCUBAIJ software, except for CH$_3$OCH$_3$ and the CH$_2$DOH rotational temperature (see Table \ref{abundances_MADCUBA}).
}
\end{deluxetable*}

Figures \ref{orot} and \ref{nrot} present the population diagrams of the detected O-bearing and N-bearing COMs, respectively. 
The upper level populations ($N_u$) are corrected according to Eq. \ref{nucorr} using the corresponding optical depth correction factor ($C_{\tau}$) and a filling factor ($\Omega_{beam}$/$\Omega_{source}$) of 14.3, according to the deconvolved size of the CH$_3$OH emitting region in Sect. \ref{sec:results_spat} (and assuming a uniform extent of both the beam and the source as a first approximation).  
%The more rigurous beam filling factor assuming Gaussian extents for both the beam and the source is 15.3. 
The resulting column densities are lower limits, since the emitting region could be smaller (Sect. \ref{sec:results_spat}).
%The exponential fit to the population diagrams according to Eq. \ref{rot} was performed with the function \texttt{curve\_fit} in Python, 
%
%Assuming the rotational temperature derived from the population diagram analysis, most of the observed CH$_3$OH lines had optical depths $\tau$ $>$ 0.3. 
%The corresponding CH$_3$OH line integrated intensities should thus be corrected by a factor of $C_{\tau}$ $>$ 1.2, which in turn would affect the estimated rotational temperature. 
%As a consequence, the excitation temperature presented in the top left panel of Fig. \ref{orot} was overestimated, while the calculated column density was underestimated. 
%Therefore, an iterative process was needed in order to derive the correct $T_{rot}$ and $N_{tot}$ values. 
%On the other hand, the vast majority of the observed lines corresponding to O- and N-bearing COMs other than CH$_3$OH had optical depths $\tau$ $<$ 0.2, leading to correction factors $<$ 1.1 that were considered negligible and not taken into account.  
We used the affine-invariant MCMC package \texttt{emcee} \citep{foreman13} with 10$^{10}$ cm$^{-2}$ $<$ $N_T$ $<$ 10$^{20}$ cm$^{-2}$ and 10 K $<$ $T_{rot}$ $<$ 500 K flat priors to 
estimate the COM column densities and excitation temperatures that best fit (i.e., the 50th percentile from the posteriors)
the observed population diagrams according to Eq. \ref{rot}. 
%, and estimate the COM excitation temperatures and column densities. 
%This package takes the optical depth of the observed emission lines into account during the fitting process. 
Additional information on the MCMC fitting can be found in \citet{ryan18} and \citet{jenny19}. 
%The results are included in Figures \ref{orot} and \ref{nrot}. 
%
%We note that most of the CH$_3$OH emission lines were optically thick. Therefore, the observed $N_u$/$g_{up}$ ratios (black squares in the top left panel of Fig. \ref{orot}) were lower than the expected values from the MCMC fitting, that used the corrected  $N_u$/$g_{up}$ ratios according to Eq. \ref{nucorr}. 

Table \ref{abundances} presents the estimated excitation temperatures and column densities for the detected O- and N-bearing COMs toward the continuum peak of Ser-emb 11 W. %We note that the estimated column densities should be considered lower limits since we have used the minimum filling factor. 
The observed CH$_3$CH$_2$OH, CH$_3$OCH$_3$, CH$_3$COCH$_3$, NH$_2$CHO, $^{13}$CH$_3$CN, and CH$_3$C$^{15}$N transitions did not span a wide enough range of upper level energies. %and the exponential fit to their population diagrams could not be properly performed. 
In these cases a fixed rotational temperature was adopted in order to estimate their column densities. In particular, the CH$_3$OH temperature was used for the O-bearing COMs CH$_3$CH$_2$OH, CH$_3$OCH$_3$, and CH$_3$COCH$_3$, and the average value of $T_{rot}$(CH$_2$DCN) and $T_{rot}$(CH$_3$CH$_2$CN) for the N-bearing COMs $^{13}$CH$_3$CN and CH$_3$C$^{15}$N. 
Since NH$_2$CHO is both O- and N-bearing, we estimated two column densities using the CH$_3$OH and the complex cyanide excitation temperatures. 
%In particular, 
%Eq. \ref{rot} was then applied to every individual %CH$_3$CH$_2$OH and CH$_3$COCH$_3$ 
%transition. 
%with a measured $N_u$/$g_{up}$ ratio 
%assuming $T_{rot}$ =  $T_{rot}($CH$_3$OH$)$. This allowed us to get an approximate $N_T$ value from every transition. 
%The resulting $N_T$ average value was finally considered as the species column densities. 
%
%Likewise, the average $T_{rot}$ derived for the cyanide molecules (CH$_2$DCN and CH$_3$CH$_2$CN) was used to estimate the $^{13}$CH$_3$CN and CH$_3$C$^{15}$N column densities, while two NH$_2$CHO column densities were estimated using both rotational temperatures, since this species could be treated as either an O-bearing or a N-bearing molecule. 
%%We note that the $N_u$/$g_{up}$ ratio of the NH$_2$CHO and $^{13}$CH$_3$CN transitions were corrected for their optical depths, since this correction was not negligible in those cases. 
We note that the partition functions of all species except CH$_3$OH and CH$_3$OCH$_3$ do not take into account higher vibrational states ($\nu_T$ $\ge$ 1) that could be populated at T $>$ 100 K. 
The CH$_3$OCHO and CH$_3$CH$_2$CN partition functions (and hence column densities) were corrected by a factor of 1.11 (corresponding to a rotational temperature of 110 K and 99 K, respectively, see Table \ref{abundances}) according to \citet{favre14} and \citet{heise81}. 
For the rest of species, the vibrational contributions to the partition function were not yet available, but we expect these corrections to be smaller compared to other sources of uncertainty. 
%This correction must be done after completing the population diagram analysis
%
The values reported in Table \ref{abundances} were cross-checked with those estimated with the SLIM tool within the MADCUBA package (see Appendix \ref{app-madcuba}).  
The excitation temperatures were within 35\%, and  
the column densities and column density ratios were within a factor of 2.  
%The resulting column density ratios were near identical.

In all cases where excitation temperatures could be constrained, they were above 100 K, confirming the detection of a hot corino toward Ser-emb 11 W. 
The O-bearing COMs (except for CH$_3$OCHO) presented excitation temperatures in the 170$-$188 K range, higher than the values estimated for the N-bearing COMs (99$-$130 K). 
This pattern is the opposite to that recently observed in the IRAS 16293-2422A hot corino \citep{vantHoff20}. 
%It is unclear whether the different ranges of estimated excitation temperatures are due to a chemical differentiation between O- and N- bearing species, or differences in the excitation conditions of both families. 
%This could indicate that the O- and N-bearing COMs are not co-spatial within the beam. 
A lower excitation temperature for the N-bearing COMs could indicate that the emission of these species is coming from a colder (or, rather, less warm) region of the inner envelope where COMs are observed. However, neither the emission of O-bearing COMs, nor N-bearing COMs was resolved within our beam, so a potential different spatial origin of both molecular families could not be evaluated.
Higher resolution observations would thus be needed in order to address these differences. 
The lower excitation temperature found for CH$_3$OCHO compared to the rest of O-bearing COMs, combined with the narrower FWHM of its emission lines, may indicate that this species is more efficiently destroyed through gas-phase chemical reactions at high temperatures than other O-bearing COMs, and would be therefore missing in the innermost regions of the hot corino. 

The estimated column densities for O-bearing COMs other than CH$_3$OH range between 7.1 $\times$ 10$^{16}$ cm $^{-2}$ and 7.8 $\times$ 10$^{17}$ cm $^{-2}$, while values between 8.2 $\times$ 10$^{14}$ cm $^{-2}$ and 8.1 $\times$ 10$^{15}$ cm $^{-2}$ were found for N-bearing COMs. 
With a CH$_3$OH column density of 3.7 $\times$ 10$^{18}$ cm $^{-2}$, this implies column density ratios with respcect to CH$_3$OH of 2$-$21\%, for O-bearing COMs, and 0.02$-$0.22\% for N-bearing COMs.

\subsection{Isotopic ratios in Ser-emb 11 W}\label{sec:isot}
The reported observations enabled us to estimate three isotopic ratios: CH$_2$DOH/CH$_3$OH, CH$_2$DCN/CH$_3$CN, and CH$_3$C$^{15}$N/CH$_3$CN.

The CH$_2$DOH/CH$_3$OH abundance ratio is around 12\% (Table \ref{abundances}), similar to what has been found in other Class 0/I sources \citep[see, e.g.,][and ref. therein]{bianchi19b}, 
%including IRAS 16293-2422B \citep[7.1\%,][]{jorgensen18}
and four orders of magnitude higher than the elemental D/H ratio  $\sim$ 10$^{-5}$ \citep{caselli12}. 
Previous works have associated high D/H ratios with a solid-state formation pathway at low temperatures \citep[see, e.g.,][]{vanGelder20}.    
%and subsequent desorption into the gas phase of the hot corino. 
%\citet{bianchi19b} evaluated the evolution of the deuteration fraction of COMs across the different stages of the star-formation process, with mixed results depending on the species. However, in our observations methanol is the only COM for which both the main isotopologue and at least one deuterated isotopologue are detected. 
%
%The following most abundant detected COMs are CH$_3$OCH$_3$ and CH$_3$OCHO, with roughly five times lower column densities than CH$_3$OH, while the CH$_3$CH$_2$OH and CH$_3$COCH$_3$ measured column densities are up to two orders of magnitude lower. 
%The N-bearing COMs, on the other hand, presented abundances on the order of 10$^{-3}$ with respect to CH$_3$OH. 
%These abundances are compared to those observed in other Class 0 and Class I hot corinos in Sect. \ref{sec:comp}. 

Assuming a $^{12}$C/$^{13}$C ratio for the local ISM of 81.3 \citep{botelho20}, we estimate a CH$_3$CN column density of 1.0 $\times$ 10$^{17}$ cm$^{-2}$ from the $^{13}$CH$_3$CN column density listed in Table \ref{abundances}. 
This would imply a CH$_2$DCN/CH$_3$CN abundance ratio of $\sim$3\%, which is lower than that found for CH$_3$OH, 
but similar to what was recently measured toward the Class 0 hot corino IRAS 16293-2422 \citep[][]{calcutt18}. 
%We note that if CH$_3$CN was enriched in $^{13}$C as suggested by the low $^{12}$C/$^{13}$C ratio measured for some O-bearing COMs in hot corinos \citep{jorgensen16,jorgensen18} and hot cores \citep{bogelund19,charles21}, then the estimated CH$_3$CN column density would be lower, leading to a higher CH$_2$DCN/CH$_3$CN abundance ratio. 
%However, the $^{12}$C/$^{13}$C ratio measured in IRAS 16293-2422 for CH$_3$CN is $\sim$71 \citep{calcutt18}, which would not explain the lower D/H ratio measured in Ser-emb 11 W from CH$_3$CN compared to CH$_3$OH. 

Similarly, the assumed $^{12}$C/$^{13}$C ratio of 81.3 would imply a CH$_3$C$^{15}$N/CH$_3$CN column density ratio of $\sim$0.8\%, corresponding to a $^{14}$N/$^{15}$N ratio of $\sim$123. 
This is $\sim$2 times higher than the CH$_3$C$^{15}$N/CH$_3$CN ratio measured in IRAS 16293-2422 \citep[][]{calcutt18}. 
We note that the CH$_3$CN column density in that source was derived from the $\nu_8$ = 1 state, as the main CH$_3$CN emission lines were optically thick. 
On the other hand, our $^{15}$N/$^{14}$N value in CH$_3$CN is on the same order as the $^{15}$N/$^{14}$N ratio measured for HCN in Class 0/I sources, Class II disks, and comets \citep[and references therein]{jenny20}.  
%toward Ser-emb 1, 8, and 17 \citep[0.4\%, 0.9\%, and 1.3\%, respectively]{jenny20}, 
%as well as in protostellar outflows \citep{benedettini21}, and protoplanetary disks \citep{guzman17}.  
%and up to a factor of $\sim$10 higher than those measured around high-mass protostars in HCN and HNC \citep{colzi18}. 

\subsection{Small molecules in Ser-emb 11}\label{sec:small}

\begin{figure*}
\centering
\includegraphics[width=\textwidth]{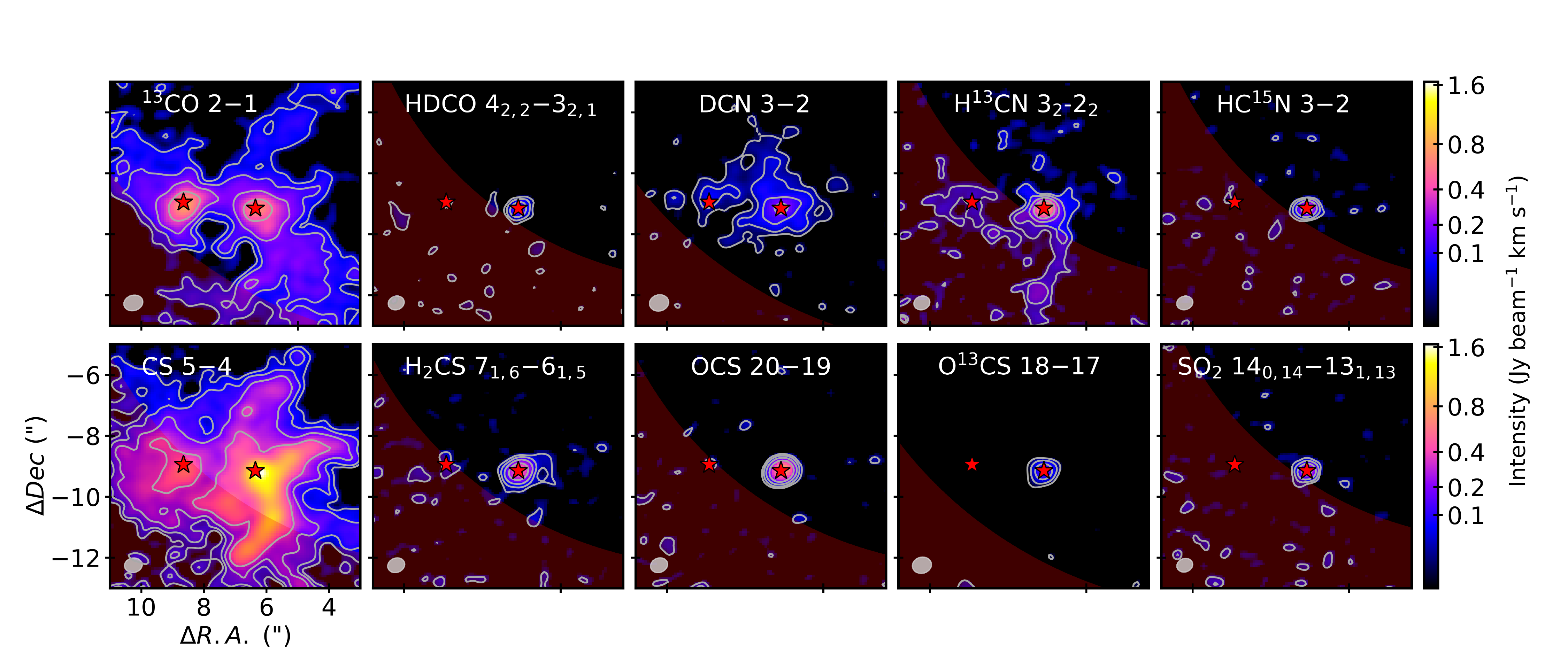}    
\caption{
Integrated moment 0 maps toward Ser-emb 11 E and Ser-emb 11 W with the 3, 6, 12, and 24$\sigma$ contours of the observed emission lines corresponding to the small species listed in Tables \ref{lines6} and \ref{lines_small} of Appendix \ref{app-lines}. 
%(contours for each map using the corresponding value of $\sigma$)
The position of the continuum emission peak is marked with a star symbol in each panel. 
Maps are clipped at 2$\sigma$. 
%(with $\sigma$ = 10 mJy for all panels, since all panels use the same color scale)
%The upper level energy of the transition is indicated in every panel. 
The size of the synthesized beam 
%($\sim$ 0.57$\arcsec$ x 0.47$\arcsec$) 
is shown on the lower left corner of every panel. 
The region of the field of view observed outside the half power beam width of the primary beam is colored in red.
}
    \label{fig:small}
\end{figure*}

Besides the complex organic species listed in Tables \ref{lines1}$-$\ref{lines6}, a number of small O-, N-, and S-bearing molecules (CO, H$_2$CO, DCN, SO$_2$, CS, H$_2$CS, OCS), along with some isotopologues ($^{13}$CO, C$^{18}$O, HDCO, H$^{13}$CN, HC$^{15}$N, and O$^{13}$CS)  were also observed toward the continuum peak of Ser-emb 11 W (and Ser-emb 11 E in some cases), in the two 1.875 GHz wide spectral windows presented in Fig. \ref{fig:spec}, and dedicated, narrower spectral windows not shown in Fig. \ref{fig:spec}. 
The observed transitions corresponding to these smaller species are listed in Table \ref{lines_small} of Appendix \ref{app-lines}. 

Figure \ref{fig:small} presents moment 0 maps of observed emission lines corresponding to the small species listed in Table \ref{lines_small}, except for CO, C$^{18}$O and H$_2$CO, shown in Fig. \ref{overview}. 
Half of the molecules were detected toward both Ser-emb 11 W and Ser-emb 11 E, although the emission was brighter toward Ser-emb 11 W (except for C$^{18}$O and $^{13}$CO), while
the rest were only observed toward Ser-emb 11 W.  
In particular, 
CO, $^{13}$CO and C$^{18}$O were detected toward both sources. %but the CO 2 $-$ 1 emission was fainter toward Ser-emb 11 E (Fig. \ref{overview}). 
So were the H$_2$CO (3$_{0,3}$ $-$ 2$_{0,2}$), shown in the fourth panel in Fig. \ref{overview}, and (3$_{2,2}$ $-$ 2$_{2,1}$) emission lines, that presented a lower intensity toward Ser-emb 11 E, while HDCO was not detected at all toward this component. 
DCN (and tentatively H$^{13}$CN) was observed toward both protostars (although with a higher intensity toward Ser-emb 11 W), with the other isotopologue (HC$^{15}$N) not detected toward Ser-emb 11 E. 
Among the S-bearing species, only CS was observed toward both sources, but again the emission was fainter toward Ser-emb 11 E.

\subsection{Possible rotation signature in Ser-emb 11 W}\label{sec:results_rot}

\begin{figure}
\centering
\includegraphics[width=8cm]{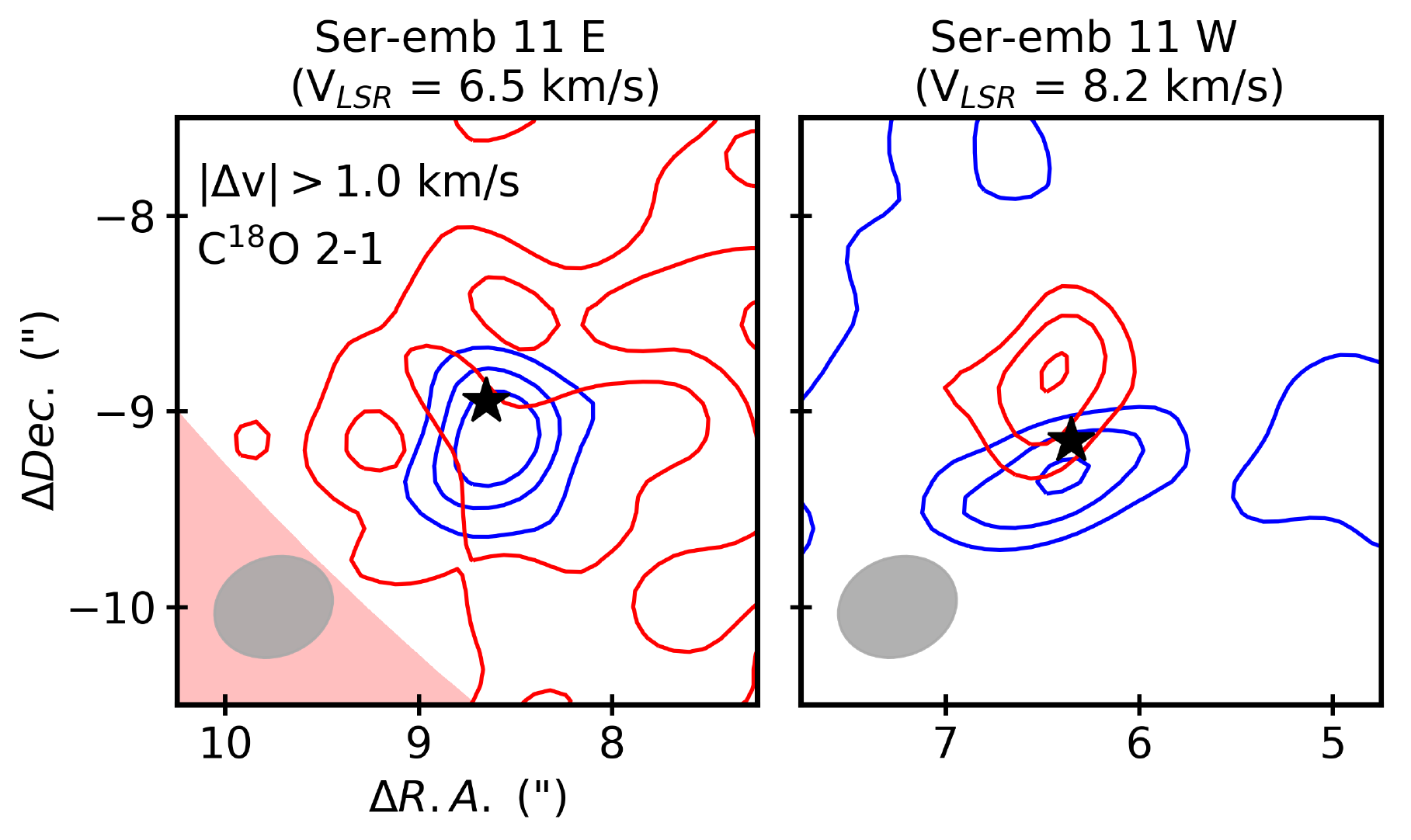}
\caption{6, 9, and 12$\sigma$ contours ($\sigma$ $\sim$ 4.5 mJy) %(half of $\sigma$ of the whole mom0 map). 
%and 6, 9, 12, 15, 18, 24, and 30$\sigma$ contours (with $\sigma$ $\sim$ 9 mJy) 
of the high velocity ($\Delta$v $>$ 1 km/s) integrated blue- and red-shifted C$^{18}$O 2$-1$ 
%(top panels) and $^{13}$CO (bottom panels) 
emission toward Ser-emb 11 E (left panel) and Ser-emb 11 W (right panel). 
%In Ser-emb 11 E (left panel), the high velocity, blue-shifted moment 0 map corresponds to the emission integrated over the channels with v = 4.2 $-$ 5.5 km/s, while the red-shifted moment 0 map integrates the emission in the v = 7.5 $-$ 8.2 km/s channels. 
%In Ser-emb 11 W (right panel), the high velocity, blue-shifted moment 0 map corresponds to the emission integrated over the channels with v = 6.3 $-$ 7.2 km/s, while the red-shifted moment 0 map integrates the emission in the v = 9.2 $-$ 9.7 km/s channels. 
The position of the continuum emission peaks is marked with a star symbol in each panel. 
%The dashed green lines represent the 6$\sigma$ contours corresponding to the dust continuum emission around the Ser-emb 11 E and Ser-emb 11 W central protostars. 
The size of the synthesized beam 
%($\sim$ 0.57$\arcsec$ x 0.47$\arcsec$) 
is shown on the lower left corner of each panel. 
The region of the field of view observed outside the half power beam width of the primary beam is colored in red.}
\label{fig:c18o_cont}
\end{figure}

%Among the small species detected in both Ser-emb 11 E and Ser-emb 11 W, only t
The C$^{18}$O J = 2$-$1 emission was 
%evenly distributed toward the two protostars (see third panel of Fig. \ref{overview}). This transition were thus 
selected to study the gas kinematics around both central objects. 
Fig. \ref{fig:c18o_cont} shows the high velocity ($\Delta$V $>$ 1 km/s) integrated redshifted and blueshifted C$^{18}$O J = 2$-$1 emission with respect to the systemic velocity of the components 
($V_{LSR}$). For the purpose of this Section, we adopted as the $V_{LSR}$ the velocity position of the C$^{18}$O 2$-$1 intensity peak within a 2$\arcsec$ region around the continuum peak of each component ($\sim$ 6.5 km/s and $\sim$ 8.2 km/s for Ser-emb 11 E and Ser-emb 11 W, respectively)
We note that both components are part of the same cloud, and the differences in $V_{LSR}$ (as defined above) may be due to a combination of coarse spectral resolution, self-absorption, and/or the presence of infalling material or outflows along the line of sight.  
%\footnote{The emission coming from the small molecules detected toward Ser-emb 11 E and Ser-emb 11 W presented the intensity peak at a different velocity toward each source. 
%In the case of the C$^{18}$O 2 $-$ 1 emission detected in a 2$\arcsec$ region around the Ser-emb 11 E and Ser-emb 11 W continuum peaks, the systemic velocity (i.e., the velocity of the line emission peak) was $V_{LSR}$ $\sim$ 6.5 km/s and $V_{LSR}$ $\sim$ 8.2 km/s, respectively. 
%On the other hand, we measured a $V_{LSR}$ of $\sim$ 7.0 km/s and $\sim$ 9.7 km/s for Ser-emb 11 E and Ser-emb 11 W, respectively, with the observed $^{13}$CO 2 $-$ 1 emission.
%%IT DOES NOT MATTER WHAT VLSR I CHOOSE FOR 13CO, THE MOM 0 MAPS ARE ALMOST IDENTICAL
%
%We note that, unlike in the integrated intensity maps shown in Figures \ref{overview} and \ref{overview_com}, we have not represented the 3$\sigma$ contours, in order to focus on the more intense emission located closer to the Ser-emb 11 E and W central protostars. 

While in Ser-emb 11 E the high-velocity, redshifted C$^{18}$O emission is located in the same region as the blueshifted emission (left panel of Fig. \ref{fig:c18o_cont}),  
in Ser-emb 11 W the emission is spatially distributed from redshifted to blueshifted velocities in the N-S direction of a compact region around the central protostar (top right panel of Fig. \ref{fig:c18o_cont}). 
A velocity gradient across a protostar can originate from outflows, infall, and/or rotation of the molecules around the central object. 
We note that even when rotation is observed, it does not necessarily imply the presence of a rotationally supported disk, since it could also trace the rotating motion of the infalling envelope.
%Usually, if the observed velocity gradient is orthogonal to an outflow driven by the central object, it is likely that the emission is originated in a  disk-like structure.   
%
In the case of Ser-emb 11 W, the presence of a protostellar disk was suggested in \citet{enoch11} due to the substantial $\sim$230 GHz continuum flux detected at intermediate \textit{uv} distances (30$-$100 k$\lambda$). 
The velocity gradient observed for the C$^{18}$O emission in the right panel of Fig. \ref{fig:c18o_cont} 
%would be consistent with the rotation of this species around the Ser-emb 11 W protostar.   
could be due to this putative disk. 
However, our current observations do not allow us to exclude other possibilities. 
For example, 
it could be related to  
the v-shaped outflow seen in the CO J = 2$-$1 emission (Fig. \ref{fig:co_cont} in Appendix \ref{app-co-cont}).  Unfortunately, the $^{12}$CO and $^{13}$CO emission is too complex to deduce the precise outflow direction.  
%In addition, this compact velocity gradient is not clearly observed in the case of the $^{13}$CO 2 $-$ 1 transition, where the red- and blue-shifted emission is more extended around the central object (Fig. \ref{fig:c18o_cont}, bottom right panel). 
%On the other hand, the red-shifted  is completely obscured toward Ser-emb 11 E (Fig. \ref{fig:c18o_cont}, bottom left panel)
%Therefore, 
Higher spatial resolution observations are needed to confirm the presence of a protostellar disk in Ser-emb 11 W.

\begin{figure*}[ht!]
\centering
\includegraphics[width=15cm]{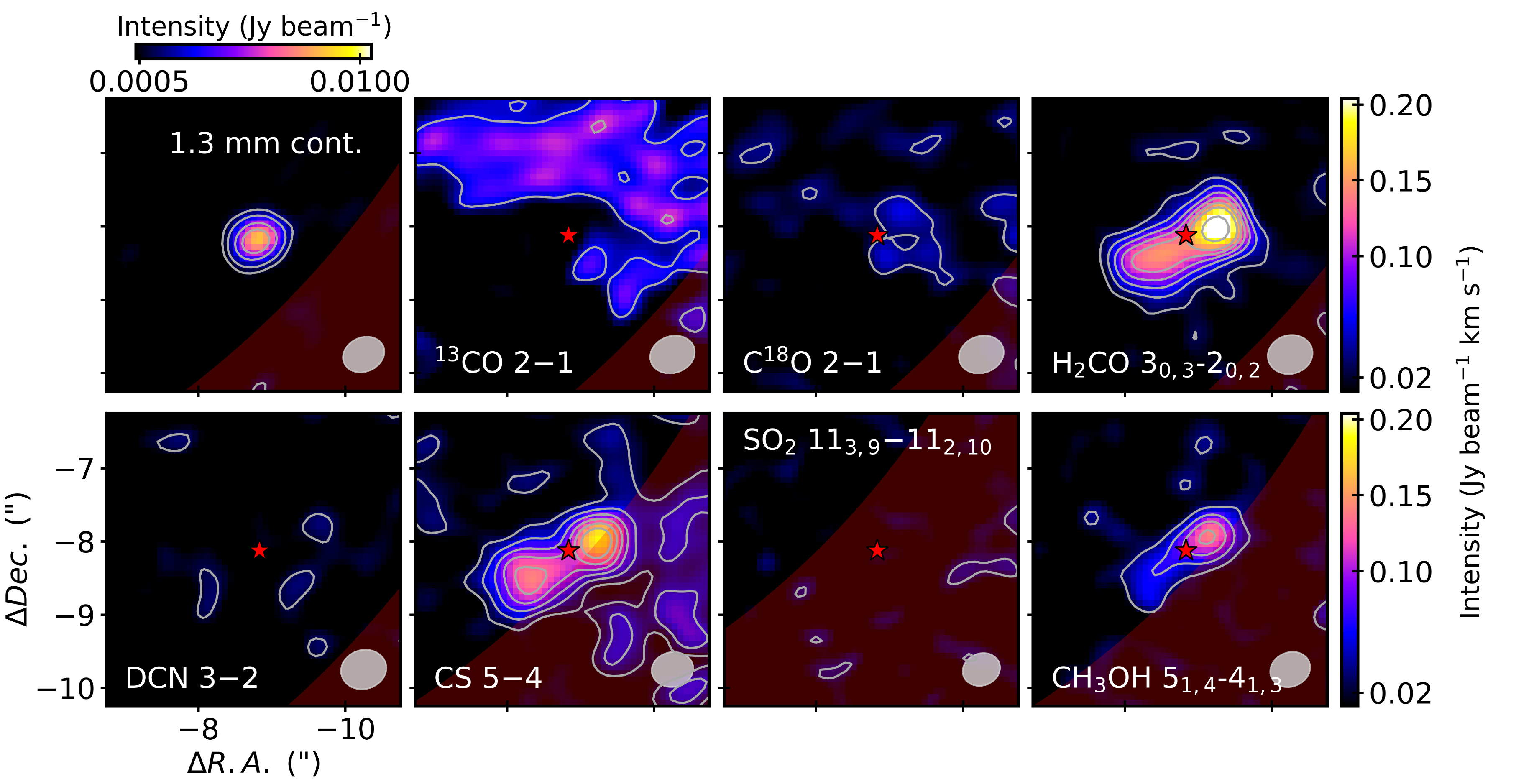}
\caption{
%The top left panel shows the continuum map centered at 232 GHz corresponding to the ID 68 YSO, 
232 GHz continuum, $^{13}$CO, C$^{18}$O, H$_2$CO, DCN, CS, SO$_2$, and CH$_3$OH integrated emission maps toward Ser-emb ALMA 1, 
along with the 3, 6, 12, and 18$\sigma$ contours in white.  %($\sigma$ = 0.35 mJy beam$^{-1}$). 
%The 6$\sigma$ contour around the continuum peak is presented as green dashed lines in every panel in order to indicate the approximate position of the central protostar. 
The position of the continuum emission peak is marked with a star symbol in each panel. 
%The rest of the panels show the integrated moment 0 maps along with the 3, 6, 9, 12, 15, 18, and 24$\sigma$ contours for the (from left to right and from top to bottom)  $^{13}$CO 2 $-$ 1 ($\sigma$ = 18 mJy beam$^{-1}$), C$^{18}$O 2 $-$ 1 ($\sigma$ = 11 mJy beam$^{-1}$), H$_2$CO 3$_{0,3}$ $-$ 2$_{0,2}$ ($\sigma$ = 8.5 mJy beam$^{-1}$), DCN 3 $-$ 2 ($\sigma$ = 8 mJy beam$^{-1}$), CS 5 $-$ 4 ($\sigma$ = 8.5 mJy), SO$_2$ 11$_{3,9}$ $-$ 11$_{2,10}$ ($\sigma$ = 11 mJy beam$^{-1}$), and CH$_3$OH 5$_{1,4}$ $-$ 4$_{1,3}$ ($\sigma$ = 11 mJy beam$^{-1}$) transitions, respectively. 
Maps are clipped at 1$\sigma$. 
The size of the synthesized beam 
%($\sim$ 0.57$\arcsec$ $\times$ 0.47$\arcsec$) 
is shown on the lower right corner of each panel. 
The region of the field of view observed outside the half power beam width of the primary beam is colored in red.}
\label{overview68}
\end{figure*}

\begin{figure*}[ht!]
\centering
\includegraphics[width=11cm]{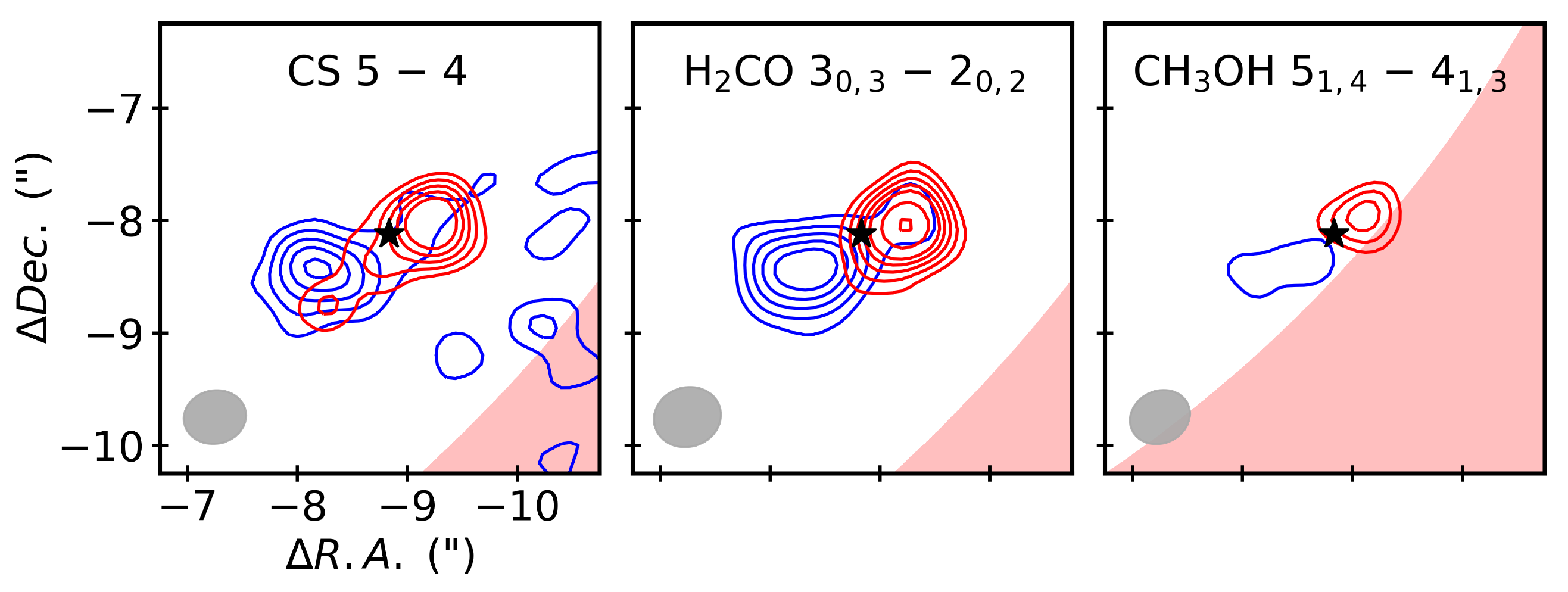}
\caption{6, 9, 12, 15, and 18$\sigma$ contours ($\sigma$ $\sim$ 6 mJy) of the integrated blue- and red-shifted  CS (5$-$4), H$_2$CO (3$_{0,3}$ $-$ 2$_{0,2}$), and CH$_3$OH (5$_{1,4}$ $-$ 4$_{1,3}$) emission toward the additional millimeter source  
($|$$\Delta$v$|$ $\sim$ 0$-$2.5 km/s,     
$V_{LSR}$ is assumed to be 8.3 km/s, Sect. \ref{sec:results_id68}). 
%The dashed green line represents the 6$\sigma$ contour corresponding to the dust continuum emission around the  central protostar. 
The position of the continuum emission peak is marked with a star symbol in each panel. 
The synthesized beam %($\sim$ 0.57$\arcsec$ x 0.47$\arcsec$)
is shown on the lower left corner of every panel. 
The region of the field of view observed outside the half power beam width of the primary beam is colored in red.}
\label{fig:id68_cont}
\end{figure*}

\subsection{Organic chemistry toward Ser-emb ALMA 1}\label{sec:results_id68}

%Among the 235 YSOs present in the Serpens Molecular Cloud,  \citet{harvey07} highlighted 15 YSOs that displayed the coldest spectral energy distributions, including ID 68. This source had not been previously observed in millimeter/submillimeter wavelengths.
%
Fig. \ref{overview68} presents an overview of the additional millimeter source detected in Fig. \ref{fig:fov}, as traced by the millimeter continuum emission, as well as the line emission corresponding to 
%different species, similarly to the overview of the Ser-emb 11 binary source presented in Fig. \ref{overview}. 
$^{13}$CO, C$^{18}$O, CS, SO$_2$, DCN, H$_2$CO, and CH$_3$OH. 
The millimeter continuum emission peak is located at $R.A._{\rm J2000}$ = 18h 29m 05.61s, $Dec._{\rm J2000}$ = +00$^{\circ}$ 30$^{\prime}$ 34.9$^{\prime\prime}$. 
%These coordinates are slightly different from those reported for the ID 68 infrared source in \citet{harvey07} ($RA_{\rm J2000}$ = 18h 29min 06.75s, $DEC_{\rm J2000}$ = +00$^{\circ}$ 30$^{\prime}$ 34.3$^{\prime\prime}$).  
This source does not have an IR counterpart in any of the four IRAC bands or the 24 $\mu$m and 70 $\mu$m MIPS bands observed as part of the Spitzer Legacy project "From Molecular Cores to Planet-forming Disks" \citep["c2d",][]{evans03}.  
%toward five star-forming clouds including the Serpens Molecular Cloud. 
Therefore, it was not included in the catalog of 235 Serpens YSOs identified in \citet{harvey07}, from which 34 sources (Ser-emb 1$-$34) were further classified as embedded protostars associated with Bolocam 1.1 mm cores in \citet{enoch09}. 
We have named this previously not catalogued millimeter source Ser-emb ALMA 1. 
%This source could be at a very early stage of the star-formation process. 

Interestingly, no emission coming from any CO isotopologue was detected toward this new millimeter source. %(the faint emission observed in the vicinity of ID 68 is likely related to the outflow driven by the Ser-emb 17 central protostar, or to any other large-scale structure partially filtered out in our observations). 
Simple N-bearing molecules were not detected either. 
Among the small molecules detected in Ser-emb 11 and listed in Table \ref{lines_small}, only CS (but no other S-bearing species) was detected toward Ser-emb ALMA 1, along with the COM precursor H$_2$CO. 
In addition, the CH$_3$OH (5$_{1,4}$ $-$ 4$_{1,3}$) emission line (E$_{up}$ = 49.66 K) was also detected, but 
%either other CH$_3$OH transitions with higher $E_{up}$, nor any other complex organic species 
no other methanol or COM lines 
were detected. 
%Therefore, no hot corino chemistry was detected toward this very young source, but the detection of organic chemistry  could be worth studying with follow-up observations. 
Therefore, Ser-emb ALMA 1 does probably not host a hot corino, even though it presents CH$_3$OH emission.

In the case of the CS (5$-$4), H$_2$CO (3$_{0,3}$ $-$ 2$_{0,2}$), and CH$_3$OH (5$_{1,4}$ $-$ 4$_{1,3}$) transitions shown in Fig. \ref{overview68}, the emission is resolved around the continuum emission peak,  
%The emission coming from the three detected species is offset with respect to the assumed position of the protostar (i.e., the continuum emission peak), since 
and shows 
%two different line emission peaks can be observed 
an elongated distribution 
across the north-west (NW) to south-east (SE) direction with respect to the continuum peak. 
Fig. \ref{fig:id68_cont} shows the integrated emission maps of the detected lines decomposed into the redshifted and blueshifted channels with respect to the systematic velocity of the source ($V_{LSR}$ = 8.3 km/s, according to the intensity peak of the H$_2$CO (3$_{0,3}$ $-$ 2$_{0,2}$) emission detected at the continuum peak of this source). 
%I USED THE CONTINUUM PEAK AND NOT A 2" REGION AROUND IT AS FOR THE C18O EMISSION IN SER-EMB 11 BECAUSE I DID NOT HAVE SELF-ABSORPTION ISSUES (AT LEAST NOT AS IMPORTANT) IN THIS CASE. 
In all cases, the redshifted emission is predominantly detected toward the NW 
%position with respect to the central protostar, 
of the continuum peak, 
while the blueshifted emission is located to the SE. %of the central object. 
%As explained in Sect. \ref{sec:results_rot}, t
This could indicate the presence of an outflow, infall, and/or rotation of the molecules around the central object. 
%Since this is a very cold source, possibly at a very early stage of the star-formation process, it is likely that the emission could be originated in a small outflow in the NW-SE direction. 
%Alternatively, the emission could indicate the presence of a disk around an older, obscured source. 

\section{Discussion}\label{sec:discussion}

%\subsection{The hot corino in Ser-emb 11 W}

The detection of complex organic chemistry (Sect. \ref{sec:results_hotcorino}) in a compact region around the western component of the Ser-emb 11 binary system (Sect. \ref{sec:results_spat}), with excitation temperatures above 100 K (Sect. \ref{sec:results_col}) indicates that this source hosts a hot corino. If we adopt the classification established in \citet{enoch09} (see Sect. \ref{sec:intro}), Ser-emb 11 W would represent only the fourth hot corino detected toward a Class I source to date.

It is unclear why the number of currently detected Class 0 hot corino sources is much higher than the number of Class I hot corinos. 
%
%Thus far, only a few studies had focused on the chemistry of Class I protostars, often targeting the outer envelopes, or some specific molecules \citep[][and references therein]{bianchi20}. 
%Therefore, the low number of currently detected Class I hot corinos may not be significant. 
%
\citet{yang21} recently investigated the relation between the protostellar physical properties and the occurrence of COMs in the PEACHES survey sample, 
and found no clear correlation between the detection of COMs and the bolometric temperature. 
%They classified the sources in groups according to the number of detected COMs, and found that the highest bolometric temperature measured across the sources of each group was lower in those groups where a higher number of COMs had been detected. %was lower than the maximum T$_{bol}$ found among of the sources with no COM detection. Even though this may suggest that more evolved sources (with higher T$_{bol}$) may have less COMs in the gas-phase, the survey did not show any obvious differences for the Class 0 and Class I sources regarding COM detection.
In addition, they did not find any obvious trend between the number of detected COMs and the bolometric luminosity of the sources either, even though lower luminosities should correspond to a smaller size for the regions where the temperature would be above the ice desorption temperature. %This should result in fainter COM emission, reducing their detectability.
The explanation behind the smaller number of hot corinos currently detected toward Class I sources could be instead related to the physical structure (density and temperature profile) of the envelope of the Class I sources on the disk formation scales \citep{villarmois19}. 
In a different survey toward a sample of 12 Class I sources in the Ophiuchus star-forming region, no CH$_3$OH emission toward the continuum peak of any of the sources was detected. 
%the result of a lower ice sublimation rate around Class I protostars, due to the combination of lower accretion rate of material into the inner region of the envelope, and lower luminosity of the central protostar compared to Class 0 sources, that would lead to fainter, harder to detect COM emission.
However, when a constant density profile corresponding to the flattening of the inner envelope was assumed (as opposed to a centrally-peaked power law density profile typical of Class 0 sources), the resulting upper limits were comparable to the detected abundances toward Class 0 hot corinos \citep{villarmois19}.
%Es decir, si en lugar de asumir la misma estructura fisica que en el caso de las Class 0 sources, se asumen otras disitintas, la region en la que se reunirian las condiciones necesarias para que el CH3OH estuviera en la fase gaseosa tendria un tamano menor, por lo que la emission seria mas dificil de detectar, y la no deteccion no seria tan significativa, correspondiendose con upper limits altos comparables con las abundancias detectadas en Class 0 hot corinos. 
%LESS LUMINOUS IN CLASS I
In any case, more observations and higher statistics are needed to evaluate whether the occurrence of hot corinos is actually lower toward Class I sources.
%
%The observed COMs include five O-bearing and three N-bearing species. %with N-bearing COMs other than NH$_2$CHO being detected in a Class I hot corino for the first time. 

%CH$_3$OH is the most abundant COM, with a column density on the order of $\sim$10$^{18}$ cm$^{-2}$.
%
%Even though \citet{martin19} suggested a possible connection between the source bolometric luminosity %\citep[that can be used as a proxy of the warm, inner envelope size,][]{dunham14}, 
%and the measured CH$_3$OH column density in hot corinos, the value measured toward Ser-emb 11 W does not follow this trend, since it is one order of magnitude higher than that found in the Ser-emb 1 Class 0 and Ser-emb 17 Class I hot corinos  \citep[respectively]{martin19,jenny19}, for a similar bolometric 
%luminosity\footnote{The detectability of COMs in YSOs, on the other hand, has been proven to be independent of the bolometric luminosity of the source, as well as the bolometric temperature or the gas mass column density \citep{belloche20,vanGelder20,yang21}; although \citet{belloche20} suggested that the internal luminosity is the source parameter that most affects the COM chemical composition in Class 0/I sources.}. 

\subsection{Chemical differentiation in binary sources}

%Ser-emb 11 W is also the fifth hot corino detected toward a binary source, along with the four binary hot corinos listed in Table \ref{tab:hc}. %(without considering B1-bN and B1-bS as a binary system).
%(more binary hot corinos have been detected in the PEACHES sample).
%
In the Ser-emb 11 binary source, hot corino chemistry has been  detected only toward the western component, while the eastern component presented no signs of COM emission (Sect. \ref{sec:results_hotcorino}). 
Two of the four binary sources with observed hot corino chemistry in Table \ref{tab:hc} also presented a similar chemical differentiation, with COMs initially 
detected in only one of the two components \citep[IRAS 4A2, and SVS13-A VLA4A, see][respectively]{lopezsepulcre16,bianchi19}.
In L1551-IRS5, on the other hand, the hot corino emission is brighter toward the N component, but a second hot corino associated with the S component cannot be excluded \citep{bianchi20}. 
Even the binary sources with hot corino chemistry detected toward both components present some chemical differences. 
In IRAS 16293-2422,  both the A and B components harbor a hot corino. However, significant differences in the abundance of some COMs relative to CH$_3$OH between the two components have been reported \citep{manigand20}. %\footnote{The IRAS 16293-2422 binary source is presented as an example of a binary source with no chemical differentiation in \citet{belloche20}, although \citet{desimone20} highlights that the COM emission is brighter toward the A component.}, and maybe 
This apparent chemical differentiation in binary systems has also been found in sources from the CALYPSO \citep{belloche20} and PEACHES \citep{yang21} surveys. 
%On this regard, \citet{belloche20} suggested that all YSOs could eventually go through a phase showing COM emission. 

The origin of these differences between binary components is not clear, and could be different for different sources. 
As mentioned in the introduction, the two components of a chemically differentiated binary system could be at different evolutionary stages or present different physical conditions.
In the case of Ser-emb 11, the lower intensity of the continuum emission detected toward Ser-emb 11 E may indicate that the temperature of this source is lower. Assuming that the COMs detected toward Ser-emb 11 W were sublimated from the ice mantles upon passive heating of the inner envelope (see also Sect. \ref{sec:results_spat}), this could mean that the temperature toward Ser-emb 11 E is not high enough to induce thermal desorption of the ice mantles in the inner regions of its envelope. 

Another possibility is that the dust opacity could be different toward the two components of the binary system, hindering the COM detection in one of them.
In the case of the IRAS 4A binary source, 
recent observations of complex organics (CH$_3$OH, $^{13}$CH$_3$OH, CH$_2$DOH, and CH$_3$CHO) seen in absorption at millimeter wavelengths \citep{sahu19}, as well as CH$_3$OH emission observed at centimeter wavelengths \citep{desimone20} suggest that IRAS 4A1 could also harbor a hot corino similar to that observed toward IRAS 4A2, but obscured by the dust continuum opacity. 
In 
Section %\ref{sec:results_overview} and
\ref{sec:small} 
%Figures \ref{overview} and \ref{fig:small} 
we show that the emission of some of the small species observed in Ser-emb 11 is fainter toward Ser-emb 11 E. %, where no COMs are detected. 
This could be also due to a higher dust continuum opacity in Ser-emb 11 E that could be blocking the COM millimeter emission. 
However, the lower intensity of the continuum emission detected toward this component compared to Ser-emb 11 W makes this possibility less likely than in the case of IRAS 4A (although it could just mean that the (maybe optically thick) dust in Ser-emb 11 E is at a lower temperature). 
Follow-up observations at centimeter wavelengths would be needed to check whether the potential Ser-emb 11 E COM millimeter emission is actually blocked by dust.

\subsection{Organic composition in Class 0 and Class I hot corinos}\label{sec:comp}

The variation in the chemical composition across different hot corino sources is currently unclear. 
In the literature, there is evidence supporting both
differences 
in the COM relative abundances  
of more than one order of magnitude 
from one hot corino to another \citep[e.g.,][]{jenny19,belloche20}, 
%\citet{jenny19} reported differences of up to two orders of magnitude in the COM abundance ratios with respect to CH$_3$OH measured toward different Class 0 hot corinos, but similar COM abundance ratios between the Ser-emb 17 Class I and the Ser-emb 1 and 8 Class 0 hot corinos. although the uncertainties in the CH$_3$OH emission optical depths could have affected the COM relative abundances reported for Ser-emb 1, Ser-emb 8, and Ser-emb 17. 
%\citet{belloche20} classified the 9 CALYPSO sources with COM emission (not necessarily hot corinos) and two additional Class 0 sources in 3 groups according to the relative abundances of the detected COMs with respect to CH$_3$OH, and claimed that the COM composition was not homogeneous. 
and relatively similar COM abundance ratios 
%(differences of only a factor of a few) 
between different hot corinos \citep[e.g.,][]{vanGelder20,yang21}.  
%On the other hand, \citet{vanGelder20} %(p.9) 
%and \citet{yang21} %(p.20 and Fig. 17 in p.27) 
%have recently claimed that the COM abundance ratios were similar toward the different sources in their corresponding samples\footnote{We note that the sample in the PEACHES survey included both Class 0 and Class I sources.}. %(even though the estimated absolute column densities span up to 3 orders of magnitude). 
%In particular, \citet{yang21} claimed that the PEACHES survey showed similar abundance ratios between COMs, and that the COM abunance ratios were similar to the ones measured in the CALYPSO source (they even used a single value to represent each COM relative abundance in the PEACHES survey). 
%
%%Even though a slight correlation was observed between the measured ratios of CH$_3$OCHO and CH$_3$OCH$_3$ with respect to CH$_3$OH with the bolometric luminosity and temperature  of the sources in the PEACHES sample, \citet{yang21} suggested that these protostellar properties have little impact on the COM abundances. 
%Larger differences were observed, though, between the abundance ratios measured in the PEACHES sample and those reported for other Class 0 hot corinos \citep{yang21}, in line with \citet{jenny19}. 
%These differences were attributed to a possible underestimation of the CH$_3$OH column density due to the dust optical depth in the PEACHES sample. 
%
One potential issue when comparing reported abundances across different sources is that the COM relative abundances with respect to CH$_3$OH may be affected by large uncertainties in the estimated CH$_3$OH column densities, due to line optical depth effects that are treated differently across different works. 
\citep[see, e.g., the order of magnitude difference in the Ser-emb 8 COM relative abundances reported in][]{jenny19,vanGelder20}. 
%For example, the CH$_3$OCH$_3$ and CH$_3$OCHO abundances toward Ser-emb 8 reported in \citet{jenny19} and \citet{vanGelder20} differ in one order of magnitude. 
%The discrepancies between both works are probably due to different approaches to the optical depth corrections during the estimation of the CH$_3$OH column density.
%This suggests that either the optical depth correction for CH$_3$OH lines in \citet{jenny19}, or the assumed $^{16}$O/$^{18}$O ratio in \citet{vanGelder20} (or both) are not accurate. 
%
%We note that the study of correlations between the column densities of different species are not completely useful, since a correlation does not necessarily imply a chemical link \citep[see, e.g.,][]{quenard18,belloche20}.
%Furthermore, evidence of differences in the COM abundance ratios are also found when COMs detected in physical structures other than hot corinos are included. 
%For example, the sources of the CALYPSO sample 
%were classified in three groups according to their COM relative abundances \citep{belloche20} %p.19. 
%However, the authors did not find any correlations between the chemical composition of the different groups and the possible origin of the COMs (either a hot corino, an outflow, or an accretion shock/disk atmosphere), the cloud environment of the sources, or their evolutionary status.  

\begin{figure*}[ht!]
     %\centering
     \includegraphics[width=0.95\textwidth]{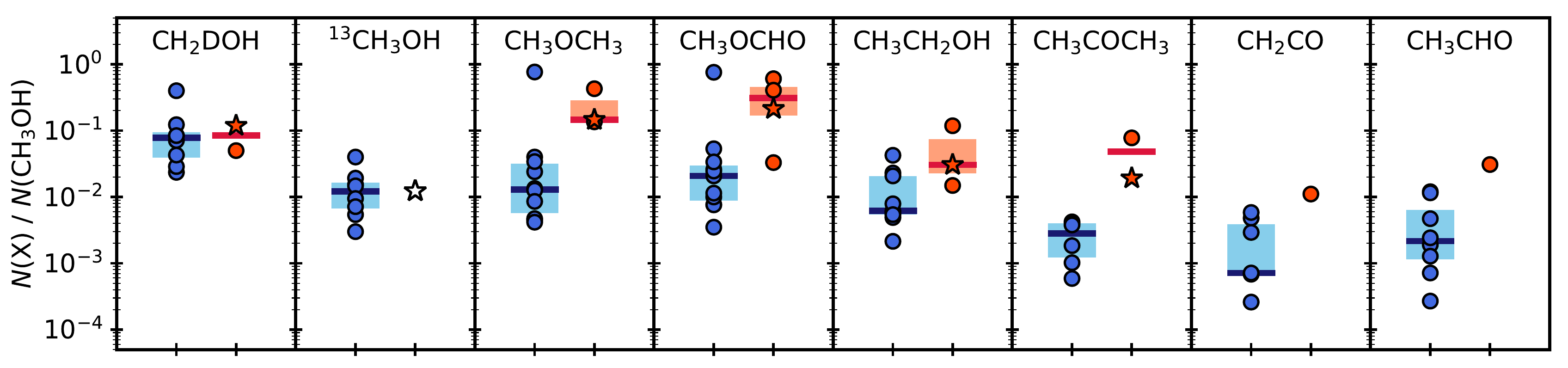}
     \includegraphics[width=0.84\textwidth]{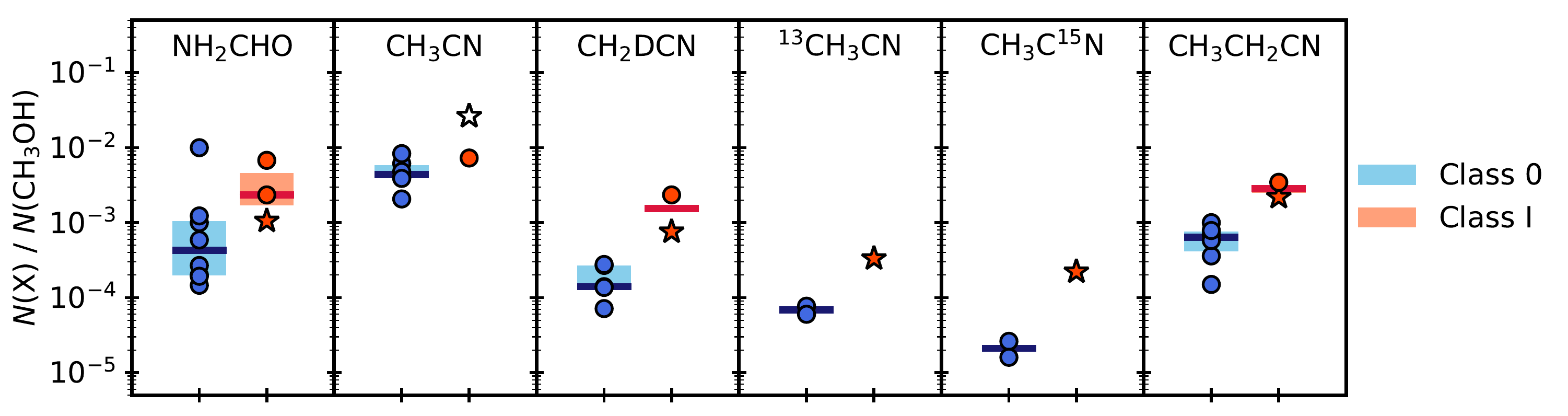}
     \caption{COM relative abundances with respect to CH$_3$OH for a series of O-bearing (top panels) and N-bearing (bottom panels) species, measured toward Class 0 (blue markers) and Class I (red markers) hot corinos. 
     The values are extracted from the literature (Appendix \ref{app-hot_corinos}), except for the Class I Ser-emb 11 W hot corino, reported in Table \ref{abundances} (red stars). 
     The $^{13}$CH$_3$OH abundance in Ser-emb 11 W (white marker) corresponds to a $^{12}$C/$^{13}$C ratio of 81.3 for the local ISM \citep{botelho20}. 
     The CH$_3$CN abundance in Ser-emb 11 W (white marker) has been estimated from the $^{13}$CH$_3$CN abundance assuming the above $^{12}$C/$^{13}$C ratio. 
     For those species with detections in more than one Class 0 and/or Class I hot corino, solid lines indicate the median abundances found in the sample.  When the species is detected in more than two Class 0 and/or Class I hot corinos, color bars represent the range between the lower and upper abundance quartiles 
     (i.e., half of the hot corino sources have abundances in that range).  %and considered as the representative abundance range for comparison purposes.  
     }
     \label{fig:comp_ME}
 \end{figure*}

\begin{figure*}[ht!]
     %\centering
     \includegraphics[width=0.95\textwidth]{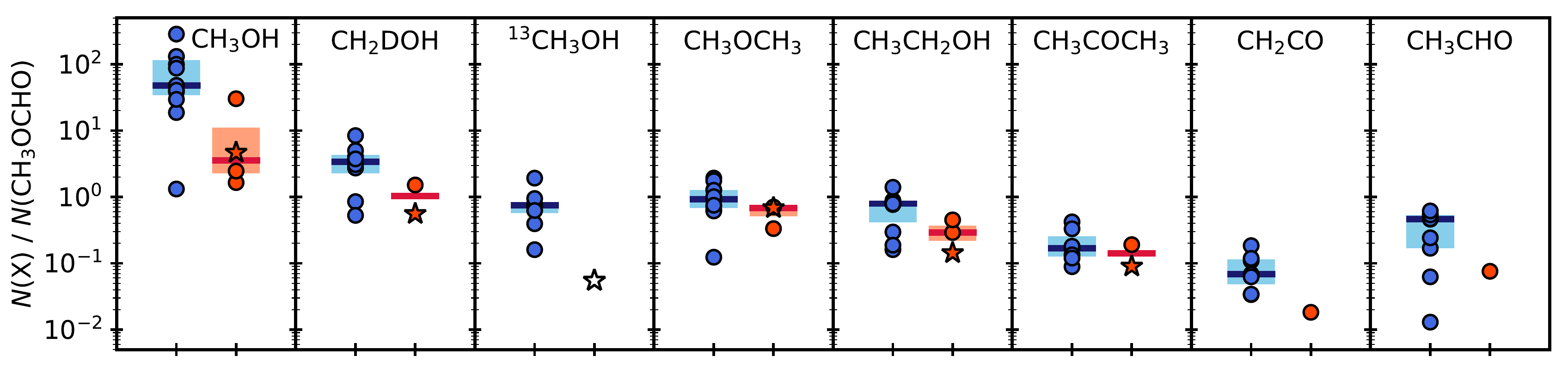}
     \includegraphics[width=0.84\textwidth]{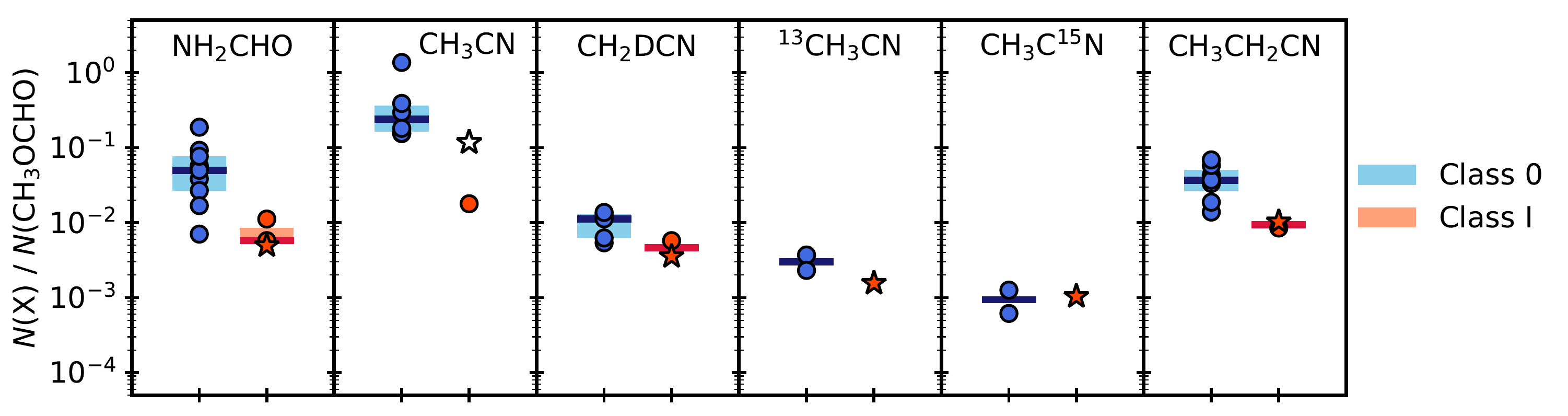}
     \caption{Same as Fig. \ref{fig:comp_ME} but with COM relative abundances calculated with respect to CH$_3$OCHO. 
     }
     \label{fig:comp_MF}
 \end{figure*}

Previous studies focusing on the COM composition of Class I hot corinos compared to Class 0 hot corinos found no substantial differences between the two groups, but the sample of Class I hot corinos was small  \citep[SVS13-A, Ser-emb 17, and L1551 IRS5][respectively]{bianchi19b,jenny19,bianchi20}.
%\citet{bianchi19b} reported that the COM abundance ratios with respect to CH$_3$OCHO of the SVS13-A Class I source appeared to be similar to those measured for Class 0 hot corinos. 
%Similarly, the COM abundance ratios with respect to CH$_3$OH in the Ser-emb 17 Class I hot corino were similar to those measured toward the Ser-emb 1 and Ser-emb 8 Class 0 hot corinos \citep{jenny19},  
%while the chemical composition of the L1551 IRS5 Class I hot corino was also similar to that observed in Class 0 sources \citep{bianchi20}.  
%
According to the classification in \citet{enoch09}, Ser-emb 11 W would represent the fourth hot corino detected toward a Class I source to date, although its low bolometric temperature \citep[77 K,][]{enoch09} may indicate that this source is rather in a transitional stage between the Class 0 and Class I stages (Sect. \ref{sec:intro}). In any case,
in order to compare the chemical composition across Class 0 and Class I hot corinos,  
Fig. \ref{fig:comp_ME} presents the COM relative abundances with respect to CH$_3$OH
%\footnote{COM abundance ratios of O-bearing species with respect to CH$_3$OH are usually thought of as conversion efficiencies, due to the parental role assigned in the literature to CH$_3$OH in the COM formation process \citep[see, e.g.,][]{jenny19}.} 
%Since neither the column density of CH$_3$OH, nor that of any other isotopologue was reported in \citet{imai16}, the B335 Class 0 hot corino was not included in Fig. \ref{fig:comp_ME}. 
%As reported in \citet{jenny19}, the relative abundances with respect to CH$_3$OH of most of the selected COMs span up to two orders of magnitude in both Class 0 and Class I hot corinos. 
of commonly detected O- and N-bearing COMs, as reported in the literature for the hot corinos in Table \ref{tab:hc}
\citep[except for IRAS 4B, since only single-dish observations have been reported toward this source,][]{bottinelli07}, along with those presented in this work for Ser-emb 11 W. 
In Fig. \ref{fig:comp_ME} (and also Fig. \ref{fig:comp_MF}), we have considered Ser-emb 11 W as a Class I source, but we note that the subsequent analysis would not be affected if this source was considered instead a transitional Class 0/I source.
The reported COM column densities in the different Class 0 and Class I hot corinos are listed in 
Tables \ref{abundances_0} and \ref{abundances_1} (respectively) of Appendix \ref{app-hot_corinos}. 

The COM relative abundances estimated in this work for the Ser-emb 11 W hot corino (red stars in Fig. \ref{fig:comp_ME}) are similar to those found toward the three Class I hot corinos previously reported in the literature (red circles in Fig. \ref{fig:comp_ME}), with typical differences around a factor of a few. 
A larger scatter (around 1$-$2 orders of magnitude) is observed for the COM relative abundances in Class 0 hot corinos (blue circles in Fig. \ref{fig:comp_ME}), as previously suggested in the literature \citep[e.g.,][]{jenny19,belloche20}. 

In order to compare the COM relative abundances in Class I hot corinos with those measured toward Class 0 hot corinos, Fig. \ref{fig:comp_ME} also shows the median relative abundances, along with the lower and upper abundance quartiles, found for the O- and N-bearing COMs toward the two groups. 
The range between the lower and upper abundance quartiles are comparable in Class 0 and Class I hot corinos for CH$_2$DOH, C$_2$H$_5$OH, and CH$_3$CN; 
while the rest of studied O- and N-bearing COMs  
%On the other hand, CH$_3$OCH$_3$, CH$_3$OCHO, CH$_3$COCH$_3$, CH$_3$CHO, CH$_2$DCN, $^{13}$CH$_3$CN, CH$_3$C$^{15}$N, and C$_2$H$_5$CN 
present higher abundances with respect to CH$_3$OH toward Class I hot corinos. 
We note that CH$_2$CO has only been detected in the Ser-emb 17 Class I hot corino, CH$_3$CHO only in the SVS-13A Class I hot corino, and $^{13}$CH$_3$CN, and CH$_3$C$^{15}$N only in the Class I (or Class 0/I) Ser-emb 11 hot corino. %Their reported abundances could thus be not  representative of the chemical composition in Class I hot corinos.
These putative differences thus need to be confirmed. 

\citet{bianchi19b,bianchi20} stated that a more reliable comparison of COM relative abundances across different sources can be obtained using species with reliably optically thin emission as references. 
Fig. \ref{fig:comp_MF} presents the COM relative abundances with respect to CH$_3$OCHO (though we note that a few CH$_3$OCHO emission lines were optically thick toward our source). 
%This allows us to avoid the aforementioned optical depth problems that could affect the estimation of the CH$_3$OH column densities, and therefore, of the corresponding relative abundances. 
%\citep[see, e.g.,][]{jenny19,bianchi19b,bianchi20}. 
%While COM abundance ratios of O-bearing species with respect to CH$_3$OH are usually thought of as conversion efficiencies, due to the parental role assigned in the literature to CH$_3$OH in COM formation process \citep[see, e.g.,][]{jenny19}, the presented COM abundance ratios relative to CH$_3$OCHO could be instead considered as product branching ratios.  
In this case, the ranges between the lower and upper abundance quartiles of all COMs other than CH$_3$OH are within a factor of a few in Class 0 and Class I hot corinos, 
%\citet{bianchi19} also observed a tighter correlation in the COM relative abundances of Class 0 and Class I hot corinos when calculated with respect to CH$_3$OCHO (see middle panels of Fig. 3 in that work). 
%Actually, \citet{bianchi20} mentioned the contradictory results found in \citet{bianchi19} when comparing COM relative abundances with respect to CH$_3$OH in Class 0 and Class I sources. 
except for NH$_2$CHO and CH$_3$CN. 
\citet{nazari21} have recently reported a large scatter in the NH$_2$CHO abundances measured across various sources (including low- and high-mass YSOs). 
These differences were attributed to the multiple formation pathways proposed in the literature for this species, whose abundance could depend on the local source conditions to a larger extent compared to other COMs. 
We also note that CH$_3$CN has only been detected toward only one Class I hot corino, and the abundance in Ser-emb 11 W estimated from the observed $^{13}$CH$_3$CN column density is consistent with that measured toward Class 0 hot corinos. 
Recent works have shown that a tight correlation in the N-bearing COM relative abundances across different sources is found when another N-bearing species such as HNCO or CH$_3$CN is used as reference \citep{jorgensen20,nazari21}. 
However, we have not carried out such comparison in this work, since neither HNCO nor CH$_3$CN column densities have been measured toward Ser-emb 11 W, 
and they have not been reported either for more than half of the sources in our sample. 

Based on these comparisons, we do not find any evidence of a significant evolution in the complex organic chemistry between Class 0 and Class I hot corinos. 
%\citet{jenny19} had already suggested that this chemical inheritance could span from the prestellar cores to the cometary composition observed in the Solar System, although the chemical diversity hinted at the hot corino stage prevented us from drawing any final conclusions. 
%From a more general point of view, this could suggest that the complex organic composition is roughly universal between different star-forming regions, although local source properties could also play a role in particular cases \citep{vanGelder20}.  
This suggests that the complex organic composition around low-mass protostars is likely shaped by the ice chemistry that takes place in the parental molecular cloud and the chemical pathways that are activated during the warm-up of the ice mantles prior to their desorption \citep{oberg20}.

\section{Conclusions}\label{sec:conclusions}

\begin{enumerate}
    \item Five O-bearing COMs (CH$_3$OH, CH$_3$CH$_2$OH, CH$_3$OCH$_3$, CH$_3$OCHO, and CH$_3$COCH$_3$) and three  N-bearing COMs (NH$_2$CHO, CH$_2$DCN, and CH$_3$CH$_2$CN), as well as three isotopologues (CH$_2$DOH, $^{13}$CH$_3$CN, and CH$_3$C$^{15}$N) have been detected toward the W component of the Ser-emb 11 binary YSO, a Class I source near the Class 0/I cutoff. 
    %This is the first time that N-bearing COMs other than NH$_2$CHO are detected in a Class I hot corino. 
    
    \item No complex organic emission was observed toward the Ser-emb 11 E component. 
    This chemical differentiation observed in the Ser-emb 11 binary system is in line with similar observations toward other binary sources. We speculate that the COM emission toward Ser-emb 11 E could be blocked by optically thick dust, as recently observed in the IRAS 4A binary source. 
    
    \item The emission of the complex organic species was unresolved in our observations, implying a compact emission region around the central protostar in Ser-emb 11 W (0.17$\arcsec$ x 0.11$\arcsec$ size according to the deconvolved CH$_3$OH emission). 
    In addition, a population diagram analysis of five of the observed COMs revealed excitation temperatures above 100 K, i.e., the desorption temperature of water ice. 
    This implies the presence of a hot corino in Ser-emb 11 W. 
    
    %\item Comparison of the relative COM column densities with respect to CH$_3$OH in Class 0 and Class I hot corinos shows mixed results, with some species presenting similar abundances in both groups of sources, and other molecules being more abundant in Class I hot corinos. 
    \item The COM relative abundances with respect to CH$_3$OH in the Ser-emb 11 W Class I hot corino are consistent with those found toward the other three Class I hot corinos previously reported in the literature. 
    
    \item The estimated COM relative abundances are similar in the sample of observed Class 0 and Class I hot corinos when they are calculated with respect to a species with optically thin emission lines such as CH$_3$OCHO. This suggests some level of continuity in the underlying chemistry during the star-formation process. 
    
    \item The measured CH$_2$DOH/CH$_3$OH column density ratio was $\sim$11\%, four orders of magnitude higher than the elemental D/H ratio. Assuming a $^{12}$C/$^{13}$C ratio of 81.3, we also estimated CH$_2$DCN/CH$_3$CN and CH$_3$C$^{15}$N/CH$_3$CN column density ratios of $\sim$3\% and $\sim$0.8\%, respectively (corresponding to a $^{14}$N/$^{15}$N of $\sim$123).
    
    \item The high-velocity C$^{18}$O J = 2$-$1 emission was spatially distributed from redshifted to blueshifted velocities in the N-S direction of a compact region around the central protostar in Ser-emb 11 W. This is compatible with the presence of a protostellar disk in this source, as previously proposed in the literature. However, our current observations do not allow us to exclude other possibilities.

    \item Complex organic chemistry (in particular, H$_2$CO and CH$_3$OH emission) was also detected toward Ser-emb ALMA 1, a nearby millimeter source that had not been previously catalogued. %YSO ID 68, possibly originating in a small outflow in the NW-SE direction. 
    
\end{enumerate}

\acknowledgments

This paper makes use of the following ALMA data: ADS/JAO.ALMA\#2015.1.00964.S. 
ALMA is a partnership of ESO (representing its member states), NSF (USA) and NINS (Japan), together with NRC (Canada), MOST and ASIAA (Taiwan), and KASI (Republic of Korea), in cooperation with the Republic of Chile. The Joint ALMA Observatory is operated by ESO, AUI/NRAO and NAOJ.
This work was supported by an award from the Simons Foundation (SCOL \# 321183, KO). 
J.B.B. acknowledges funding from the National Science Foundation Graduate Research Fellowship under Grant DGE1144152. 
C.J.L. acknowledges funding from the National Science Foundation Graduate Research Fellowship under Grant DGE1745303.
The group of JKJ is supported by the European Research Council (ERC) under the European Union's Horizon 2020 research and innovation programme through ERC Consolidator Grant ''S4F'' (grant agreement No~646908). Research at Centre for Star and Planet Formation is funded by the Danish National Research Foundation.

\facility{ALMA}
\software{CASA (v4.5.3, v4.70, 5.4.1, \& v5.4.2; \citep{mcmullin07}, MADCUBA \citep{rivilla16a,rivilla16b}, \texttt{NumPy} \citep{vanderWalt11}, \texttt{SciPy} \citep{virtanen20}}

%\bibliography{sample63}{}
%\bibliographystyle{aasjournal}



\appendix

\section{Observed spectral windows}\label{app-spw}

Table \ref{tab:spw} lists the spectral windows observed with the two Band 6 frequency settings described in Sect. \ref{sec:obs}. 

\begin{deluxetable}{ccc}\label{tab:spw}
\tabletypesize{\small}
\tablecaption{Observed spectral windows}
\tablehead{
\colhead{Frequency range} & \colhead{Spectral resolution} & \colhead{Beam size} \\
\colhead{(GHz)} & \colhead{(km s$^{-1}$)} & \colhead{$\arcsec$ $\times$ $\arcsec$}}
%\colnumbers
\startdata
217.209 $-$ 217.268 & 0.195 & 0.62 $\times$ 0.53\\ 
218.193 $-$ 218.251 & 0.194 & 0.62 $\times$ 0.53 \\
218.446 $-$ 218.505 & 0.194 & 0.60 $\times$ 0.52 \\
218.703 $-$ 218.762 & 0.193 & 0.60 $\times$ 0.52 \\ 
219.502 $-$ 219.619 & 0.193 & 0.62 $\times$ 0.50 \\
220.340 $-$ 220.457 & 0.192 & 0.62 $\times$ 0.50 \\
230.479 $-$ 230.597 & 0.183 & 0.59 $\times$ 0.47 \\
231.262 $-$ 231.380 & 0.183 & 0.58 $\times$ 0.47 \\
231.481 $-$ 233.356 & 1.250 & 0.58 $\times$ 0.47 \\
242.978 $-$ 244.853 & 1.191 & 0.56 $\times$ 0.47 \\
244.164 $-$ 244.281 & 0.173 & 0.57 $\times$ 0.47 \\
244.877 $-$ 244.994 & 0.173 & 0.57 $\times$ 0.47 \\
258.098 $-$ 258.216 & 0.164 & 0.53 $\times$ 0.44 \\
258.953 $-$ 259.070 & 0.163 & 0.52 $\times$ 0.45 \\
260.459 $-$ 260.577 & 0.162 & 0.52 $\times$ 0.44 \\
262.150 $-$ 262.267 & 0.161 & 0.52 $\times$ 0.44 \\%
\enddata
\end{deluxetable}

\section{Identified molecular transitions toward Ser-emb 11 W}\label{app-lines}

Tables \ref{lines1}$-$\ref{lines6} list the identified COM transitions in Fig. \ref{fig:spec}. 
In particular, Tables \ref{lines1}$-$\ref{lines5} present the O-bearing COM detected transitions, while Table \ref{lines6} lists those corresponding to N-bearing COMs. %Table \ref{lines6} presents the three observed transitions in Fig. \ref{fig:spec} corresponding to S-bearing species. 
In addition, Table \ref{lines_small} presents all the transitions corresponding to small (4 atoms or less) O-, N-, and S-bearing molecules observed in Fig. \ref{fig:spec} and in dedicated, narrow spectral windows not shown in Fig. \ref{fig:spec}.  
The line parameters were extracted from the JPL catalog for CH$_3$OH, CH$_2$DOH, and CH$_3$OCHO, and from the CDMS catalog for CH$_3$CH$_2$OH, CH$_3$OCH$_3$, CH$_3$COCH$_3$, and the five N-bearing COMs. 
%In particular, the values for CH$_3$OH, CH$_2$DOH, CH$_3$OCH$_3$, and CH$_3$OCHO used in Sect. \ref{sec:rot} are extracted from \citet{xu08}, \citet{pearson12}, \citet{endres09}, and \citet{ilyushin09}, respectively.
Tentatively detected lines (observed with SNR $<$ 3) corresponding to confirmed species are marked with a * in Tables \ref{lines1}$-$\ref{lines6}, and they were not considered in the discussion. Species with only one line observed at SNR $>$ 3 were considered tentative detections, and are also marked with a *.

As explained in Sect. \ref{sec:results_col}, the spectral features observed in Fig. \ref{fig:spec} were fitted with a Gaussian using the Levenberg-Marquardt minimization implementation of \texttt{scipy.optimize.curve\_fit} in Python (except those observed with a SNR $<$ 3). 
The position, width, integrated intensities, and uncertainties estimated from the Gaussian fits of the lines are also presented. 

Transitions marked with an $^a$ in Tables \ref{lines1}$-$\ref{lines6} appeared partially blended in Fig. \ref{fig:spec} with a transition from a different detected species. As the partially blended emission lines were resolved, different Line IDs were assigned to every transition marked with an $^a$. 
A multiple Gaussian fitting was performed in these cases in order to get the position, linewidth, and integrated intensities of the individual components. 
For some blended emission lines, as well as isolated, weak lines with SNR close to 3, the linewidth of the different components was fixed as the average linewidth of the two strongest, resolved transitions of the corresponding species. This ensured that emission from neighboring lines and/or noise was not taken into account for the Gaussian fit of every component. 
In those particular cases, the adopted FWHM is indicated with no uncertainties in Tables \ref{lines1}$-$\ref{lines6}. 

In other cases two or more transitions from different species appeared completely blended, and only one emission line was observed in Fig. \ref{fig:spec}. These transitions are indicated with a $^b$ in Tables \ref{lines1}$-$\ref{lines6}, and were assigned the same Line ID. The multiple Gaussian fitting was not performed for these lines, as the contribution from the different components could not be disentangled. 

When the completely blended transitions corresponded to the same species, a single Gaussian fit was performed in the cases where the blended transitions had the same upper level energy (E$_{up}$) and Einstein coefficient (A$_{ul}$). The integrated intensities of the different components were calculated from the total integrated intensity according to their upper level degeneracies (g$_{up}$). 

Finally, the identified transitions marked with a $^c$ were partially blended with other unidentified transitions. In most cases, a multiple Gaussian fitting allowed the estimation of the position, linewidth and integrated intensity of the identified transition. A fixed linewidth was adopted when necessary, as explained above.

We note that the uncertainties in the position and FWHM of the lines only take into account the fitting errors. The real uncertainties are expected to be larger due to the spectral resolution of the observations (Table \ref{tab:spw}).

\begin{deluxetable*}{cccccccccc} 
%\begin{ThreePartTable}
%\begin{TableNotes}
%\begin{list}{}
%\item $^a$Blended with an identified transition from a different species. $^b$Likely blended with other unidentified (U) transition.
%For blended emission lines, a Gaussian deconvolution was performed in order to estimate the integrated intensity of the different components. 
%Values in italics indicate that contribution from neighboring components could not be ruled out.
%Blank spaces indicate that either the different blended transitions were completely overlapped and the Gaussian deconvolution was not possible, or that the signal-to-noise ratio of the line was lower than 3 (only for confirmed species). 
%Completely overlapped in a --> same line number
%Completely overlapped in b --> +U in the plot. 
%\item A * in a confirmed species indicates that that particular emission line is barely detected above the noise level. The integrated intensity is usually not reported in those cases.
%\end{list}
%\end{TableNotes}
%\begin{longtable*}[c]{cccccccccc} 
\tablecaption{Identified O-bearing COM transitions toward the continuum peak of Ser-emb 11 W. 
The list includes 3 species and 1 isotopologue (CH$_3$OCH$_3$ and CH$_3$OCHO detected transitions are presented in Tables \ref{lines2}$-$\ref{lines5}). 
Tentative detections are indicated with a $^{*}$. 
Integrated intensity maps of boldfaced transitions are presented in Fig. \ref{overview_com}.
\label{lines1}}
\tablehead{\colhead{Line ID} &
\colhead{Molecule} & \colhead{Transition} & \colhead{Frequency} & \colhead{log$_{10}$(A$_{ul}$)} & \colhead{$g_{up}$} & \colhead{$E_{up}$} & \colhead{$V_{LSR}$} & \colhead{FWHM} & \colhead{Integrated Intensity}\\
%\hline
%\hline
%Line & Molecule & Transition & Frequency & log(A$_{ul}$) & $g_{up}$ & $E_{up}$ & $V_{LSR}$ & FWHM & Integrated Intensity\\
 &  & & (GHz) & (s$^{-1}$) &  & (K) & \multicolumn{2}{c}{(km s$^{-1}$)}  & (mJy beam$^{-1}$ km s$^{-1}$)}
 %%the transitions are spectrally integrated (km s-1), but not spatilally (/beam)
%\hline
%\endfirsthead
%\hline
%\hline
%Line & Molecule & Transition & Frequency & log(A$_{ul}$) & $g_{up}$ & $E_{up}$ & $V_{LSR}$ & FWHM & Integrated Intensity\\
% &  & & (GHz) & (s$^{-1}$) &  & (K) & \multicolumn{2}{c}{(km s$^{-1}$)}  & (mJy beam$^{-1}$ km s$^{-1}$)\\
%\hline
%\endhead
%\hline
%\endfoot
%\hline
%\insertTableNotes
%\endlastfoot
\startdata
%55.5 & SiC$_4^b$* & 76 $-$ 75 & 233.031 & -2.54 & 153 & 430.66 & 7.00 $\pm$ 0.12 & 2.33 $\pm$ 0.37 & 57.4 $\pm$ 11.6\\
%\hline
1 &t-HCOOH* & 10$_{1,9}$ $-$ 9$_{1,8}$ & 231.506 & -3.88 & 21 & 64.47 & 9.87 $\pm$ 0.42 & 4.00 $\pm$ 0.76 & 80.1 $\pm$ 18.0 \\
\hline
123 &CH$_2$CO* & 12$_{1,11}$ $-$ 11$_{1,10}$ & 244.714 & -3.79 & 75 & 89.42 & 11.11 $\pm$ 0.20 & 3.47 $\pm$ 0.49 & 80.4 $\pm$ 14.7\\
\hline
35 &CH$_3$OH$^a$ & 10$_{2,8}$ $-$ 9$_{3,7}$ A & 232.418 & -4.73 & 21 & 165.40 & 8.18 $\pm$ 0.06 & 4.46 $\pm$ 0.16 & 522.1 $\pm$ 22.6 \\
44 &CH$_3$OH & 18$_{3,16}$ $-$ 17$_{4,13}$ A & 232.783 & -4.66 & 37 & 446.53 & 8.18 $\pm$ 0.05& 4.14 $\pm$ 0.13 & 393.2 $\pm$ 16.2\\
\textbf{50} &\textbf{CH$_3$OH} & \textbf{10$_{3,7}$ $-$ 11$_{2,9}$ E} & 232.946 & -4.67 & 21 & \textbf{190.37} & 9.00 $\pm$ 0.05 & 4.22 $\pm$ 0.13 & 487.7 $\pm$ 19.4\\
89 &CH$_3$OH & 18$_{6,13}$ $-$ 19$_{5,14}$ A & 243.397 & -4.70 & 37 & 590.28 & 8.43 $\pm$ 0.07 & 4.52 $\pm$ 0.19 & (386.6 $\pm$ 21.3)/2\\
%The two overlapping emission lines presented the same transition parameters in terms of Einstein coefficient, degeneracy of the upper level, and upper level energy. Therefore, their corresponding integrated intensities were expected to be the same, i.e., half of the calculated integrated intensity for the sum of both lines. Tau = 0.32 so this is valid. 
%Since the small difference in the line frequencies (261KHz) is lower than the channel spacing of the observations (488.281KHz, not to mention the spectral resolution that is 2 times that) there is no way the two transitions could be resolved. Therefore, no need to to a two component fitting. 
89 &CH$_3$OH & 18$_{6,12}$ $-$ 19$_{5,15}$ A & 243.398 & -4.70 & 37 & 590.28 & 8.43 $\pm$ 0.07 & 4.52 $\pm$ 0.19 & (386.6 $\pm$ 21.3)/2\\
90 &CH$_3$OH & 23$_{3,20}$ $-$ 23$_{2,21}$ A & 243.413 & -4.11 & 47 & 690.08 & 9.31 $\pm$ 0.10 & 3.88 $\pm$ 0.25 & 276.3 $\pm$ 23.1\\
97 &CH$_3$OH & 5$_{1,4}$ $-$ 4$_{1,3}$ A & 243.916 & -4.22 & 11 & 49.66 & 9.00 $\pm$ 0.04 & 4.36 $\pm$ 0.14 & 754.8 $\pm$ 30.1\\ 
111 &CH$_3$OH$^a$ & 22$_{3,19}$ $-$ 22$_{2,20}$ A & 244.330 & -4.11 & 45 & 636.75 & 8.02 $\pm$ 0.12 & 4.13 $\pm$ 0.30 & 391.4 $\pm$ 32.7\\
112 &CH$_3$OH $\nu$=1& 9$_{-1,9}$ $-$ 8$_{-0,8}$ E & 244.338 & -4.39 & 19 & 395.64 & 8.76 $\pm$ 0.07 & 4.01 $\pm$ 0.18 & 399.5 $\pm$ 22.8\\
\hline
10 &CH$_2$DOH$^a$ & 10$_{1,9}$ $-$ 9$_{2,8}$ e$_0$ & %231.840 & -4.78 & 21 & 123.70 & 7.95 $\pm$ 0.19 & 4.92 $\pm$ 0.49 & 128.7 $\pm$ 16.5\\
231.840 & -4.78 & 21 & 123.70 & 7.88 $\pm$ 0.13 & 3.54 & 99.1 $\pm$ 7.0\\
\textbf{19} &\textbf{CH$_2$DOH} & \textbf{9$_{2,7}$ $-$ 9$_{1,8}$ e$_0$} & 231.969 & -4.06 & 19 & \textbf{113.11} & 9.08 $\pm$ 0.08 & 4.08 $\pm$ 0.20 & 279.5 $\pm$ 17.6\\
60 &CH$_2$DOH & 14$_{2,12}$ $-$ 14$_{1,13}$ o$_1$ & 233.142 & -4.52 & 29 & 260.58 & 9.47 $\pm$ 0.10 & 3.03 $\pm$ 0.26 & 105.0 $\pm$ 11.7 \\ 
66 &CH$_2$DOH$^{a,c}$ & 17$_{2,15}$ $-$ 17$_{1,16}$ e$_0$ & 233.210 & -4.05 & 35 & 347.06 & \nodata & \nodata & \nodata\\ 
%66 &CH$_2$DOH$^{a,b}$ & 17$_{2,15}$ $-$ 17$_{1,16}$ e$_0$ & 233.210 & -4.05 & 35 & 347.06 & 9.3 & 3.6 & \textit{130.4 $\pm$ 18.1}\\ 
%%The fitted line FWHM was a factor of 3 higher than the estimated FWHM of the rest of observed CH$_2$DOH lines. We thus assumed that the line was blended to other unidentified transition(s), in addition to the unresolved CH$_3$OCHO multiplet detected at a slightly higher frequency (Line 67 in Table 2). 
%%As a first approximation, we fixed the position (V$_{LSR}=9.28 km s$^{-1}$) and width (FWHM = 3.5 km s$^{-1}$) of the fitting Gaussian to the average value of the two unblended CH$_2$DOH lines (Lines 19 and 60) in order to estimate the integrated intensity of the CH$_2$DOH transition.  
%%In any case, this line was not included in the rotational diagram analysis. 
85 &CH$_2$DOH$^b$ & 5$_{2,3}$ $-$ 5$_{1,4}$ e$_0$& 243.226 & -4.18 & 11 & 48.40 & \nodata & \nodata & \nodata\\
%%The emission line completely overlapped with a transition from a different species, and the two components could not be deconvolved. 
% & CH$_2$DOH & 14$_{2,12}$ $-$ 13$_{3,11}$ e$_0$& 243.465 & -4.71 & 29 & 243.02 & \\ 
%The transition was observed with a signal-to-noise ratio $<$ 3. 
98 &CH$_2$DOH & 16$_{1,16}$ $-$ 15$_{2,14}$ e$_1$& 243.960 & -4.96 & 33 & 304.06 & 8.17 $\pm$ 0.19 & 2.75 $\pm$ 0.46 & 47.3 $\pm$ 10.3\\
%The estimated integrated intensity was higher than expected from the transition parameters, compared to the integrated intensity calculated for other observed CH$_2$DOH lines.  
%The line is perhaps blended to other unidentified transition(s), but since the position and FWHM of the line is consistent with other transitions, we have treated this line as it corresponded only to CH2DOH. 
%The fitted integrated instensity may thus be considered an upper limit. 
114 &CH$_2$DOH$^{b}$ & 11$_{2,10}$ $-$ 11$_{1,11}$ e$_1$& 244.490 & -4.89 & 23 & 167.89 & \nodata & \nodata & \nodata\\
%%The emission line completely overlapped with a transition from a different species, and the two components could not be deconvolved. 
120 &CH$_2$DOH$^{b}$ & 7$_{1,6}$ $-$ 6$_{2,4}$ o$_1$& 244.588 & -4.73 & 15 & 83.33 &  \nodata & \nodata & \nodata\\
%%The emission line completely overlapped with a transition from a different species, and the two components could not be deconvolved. 
%125 &CH$_2$DOH$^{c}$ & 18$_{3,16}$ $-$ 17$_{4,13}$ e$_0$& 244.825 & -4.59 & 37 & 403.20 & 8.33 $\pm$ 0.28  & 2.44 $\pm$ 0.72 & 37.3 $\pm$ 18.9\\ 
125 & CH$_2$DOH$^{c}$ & 18$_{3,16}$ $-$ 17$_{4,13}$ e$_0$& 244.825 & -4.59 & 37 & 403.20 & 8.02 $\pm$ 0.28  & 3.54 & 54.7 $\pm$ 6.7\\
126 &CH$_2$DOH$^a$ & 4$_{2,2}$ $-$ 4$_{1,3}$ e$_0$ & 244.841 & -4.32 & 9 & 37.59 & 8.59 $\pm$ 0.07 & 3.00 $\pm$ 0.17 & 147.9 $\pm$ 11.1\\
\hline
%\textbf{2} &\textbf{t-CH$_3$CH$_2$OH} & \textbf{21$_{5,17}$ $-$ 21$_{4,18}$} & 231.559 & -4.08 & 43 & \textbf{225.89} & 9.35 $\pm$ 0.36 & 4.08 $\pm$ 0.90 & 75.3 $\pm$ 18.9 \\
%\textbf{2} &\textbf{t-CH$_3$CH$_2$OH$^{c}$} & \textbf{21$_{5,16}$ $-$ 21$_{4,17}$} & 231.561 & -4.08 & 41 & \textbf{208.15} & 8.64 $\pm$ 0.12 & 1.80 $\pm$ 0.31 & 51.6 $\pm$ 11.0\\
\textbf{2} &\textbf{t-CH$_3$CH$_2$OH} & \textbf{21$_{5,17}$ $-$ 21$_{4,18}$} & 231.559 & -4.08 & 43 & \textbf{225.89} & 9.98 $\pm$ 0.35 & 3.55 & 55.3 $\pm$ 8.9 \\
\textbf{2} &\textbf{t-CH$_3$CH$_2$OH$^{c}$} & \textbf{21$_{5,16}$ $-$ 21$_{4,17}$} & 231.561 & -4.08 & 41 & \textbf{208.15} & 8.70 $\pm$ 0.24 & 3.55 & 85.2 $\pm$ 9.1\\
3 &g-CH$_3$CH$_2$OH & 14$_{1,14}$ $-$ 13$_{1,13}$ & 231.669 & -3.96 & 29 & 141.86 & 9.01 $\pm$ 0.15 & 2.46 $\pm$ 0.37 & 49.8 $\pm$ 9.9 \\
5 &t-CH$_3$CH$_2$OH$^{b}$ & 19$_{5,15}$ $-$ 19$_{4,16}$ & 231.738 & -4.13 & 39 & 191.32 & \nodata & \nodata & \nodata \\
%%The emission line completely overlapped with a transition from a different species, and the two components could not be deconvolved. 
7 &t-CH$_3$CH$_2$OH & 22$_{5,18}$ $-$ 22$_{4,19}$ & 231.790 & -4.07 & 45 & 244.47 & 8.84 $\pm$ 0.11 & 2.00 $\pm$ 0.26 & 36.2 $\pm$ 6.2  \\
22 &t-CH$_3$CH$_2$OH & 18$_{5,14}$ $-$ 18$_{4,15}$ & 232.035 & -4.09 & 37 & 175.24 & 9.77 $\pm$ 0.22 & 3.28 $\pm$ 0.55 & 63.5 $\pm$ 13.8\\
23 &t-CH$_3$CH$_2$OH$^{b,c}$ & 15$_{5,10}$ $-$ 15$_{4,11}$ & 232.075 & -4.15 & 31 & 132.29 & \nodata & \nodata & \nodata \\
32 &t-CH$_3$CH$_2$OH$^c$ & 23$_{5,19}$ $-$ 23$_{4,20}$ & 232.318 & -4.07 & 47 & 263.91 & 8.07 $\pm$ 0.34 & 3.50 $\pm$ 0.89 & 51.8 $\pm$ 16.2\\
%%The FWHM of the modeled Gaussian and the shape of the observed line suggests that the line is blended with an unidentified transition. 
34 &t-CH$_3$CH$_2$OH & 17$_{5,13}$ $-$ 17$_{4,14}$ & 232.405 & -4.10 & 35 & 160.07 & 9.23 $\pm$ 0.19 & 2.53 $\pm$ 0.48 & 38.2 $\pm$ 9.4\\
37 &g-CH$_3$CH$_2$OH & 14$_{0,14}$ $-$ 13$_{0,13}$ & 232.491 & -3.95 & 29 & 141.75 & 8.61 $\pm$ 0.21 & 4.07 $\pm$ 0.52 & 67.9 $\pm$ 11.2\\
39 &g-CH$_3$CH$_2$OH$^b$ & 14$_{1,14}$ $-$ 13$_{1,13}$ & 232.597 & -3.95 & 29 & 146.54 & \nodata & \nodata & \nodata \\
%39 &g-CH$_3$CH$_2$OH$^a$ & 14$_{1,14}$ $-$ 13$_{1,13}$ & 232.597 & -3.95 & 29 & 146.54 & 9.04 $\pm$ 0.11 & 3.49 $\pm$ 0.26 & 117.6 $\pm$ 11.6 \\
46 &t-CH$_3$CH$_2$OH & 16$_{5,12}$ $-$ 16$_{4,13}$ & 232.809 & -4.10 &33 & 145.73 & 8.98 $\pm$ 0.44 & 3.50 $\pm$ 1.10 & 36.3 $\pm$ 14.9\\
49 &t-CH$_3$CH$_2$OH & 14$_{5,9}$ $-$ 14$_{4,10}$ & 232.928 & -4.12 & 29 & 119.63 & 8.66 $\pm$ 0.30 & 3.55 & 56.5 $\pm$ 8.5\\ 
59 & g-CH$_3$CH$_2$OH & 20$_{4,17}$ $-$ 20$_{3,17}$ & 233.096 & -4.83 & 41 & 256.10 &  8.04 $\pm$ 0.24 & 3.09 $\pm$ 0.70 & 49.0 $\pm$ 14.1 \\
81 &g-CH$_3$CH$_2$OH & 14$_{4,11}$ $-$ 13$_{4,10}$ & 243.120 & -3.91 & 29 & 163.70 & 8.38 $\pm$ 0.18 & 2.63 $\pm$ 0.44 & 40.7 $\pm$ 9.0 \\
83 &g-CH$_3$CH$_2$OH$^c$ & 14$_{4,10}$ $-$ 13$_{4,9}$ & 243.207 & -3.94 & 29 & 168.57 & 9.03 $\pm$ 0.18 & 2.88 $\pm$ 0.45 & 59.2 $\pm$ 11.3\\ 
88 &g-CH$_3$CH$_2$OH & 14$_{4,10}$ $-$ 13$_{4,9}$ & 243.350 & -3.91 & 29 & 163.73 & 9.23 $\pm$ 0.24 & 3.76 $\pm$ 0.60 & 73.9 $\pm$ 15.5 \\
93 &t-CH$_3$CH$_2$OH* & 8$_{2,6}$ $-$ 7$_{1,7}$ & 243.557 & -4.23 & 17 & 35.57 & \nodata & \nodata & \nodata \\
%%The emission line is weak and blended with a transition from a different species, and the two components could not be deconvolved. 
122 &g-CH$_3$CH$_2$OH & 14$_{1,13}$ $-$ 13$_{1,12}$ & 244.634 & -3.84 & 29 & 151.72 & 8.98 $\pm$ 0.17 & 3.33 $\pm$ 0.42 & 70.4 $\pm$ 11.7\\
\hline
86 & CH$_3$COCH$_3$ & 19$_{6,13}$ $-$ 18$_{7,12}$  & 243.286 & -3.36 & 624 & 133.29 & 8.15 $\pm$ 0.23 & 3.71 $\pm$ 0.57 & (61.2 $\pm$ 12.1)/2\\
86 & CH$_3$COCH$_3$ & 19$_{7,13}$ $-$ 18$_{6,12}$  & 243.286 & -3.36 & 624 & 133.29 & 8.15 $\pm$ 0.23 & 3.71 $\pm$ 0.57 & (61.2 $\pm$ 12.1)/2\\
%The two overlapping emission lines presented the same transition parameters in terms of Einstein coefficient, degeneracy of the upper level, and upper level energy. Therefore, their corresponding integrated intensities were expected to be the same, i.e., half of the calculated integrated intensity for the sum of both lines. 
109 & CH$_3$COCH$_3$ & 20$_{5,15}$ $-$ 19$_{6,14}$  & 244.286 & -3.30 & 656 & 139.46 & 9.07 $\pm$ 0.25 & 3.30 $\pm$ 0.60 & (55.6 $\pm$ 13.3)/2\\
109 & CH$_3$COCH$_3$ & 20$_{6,15}$ $-$ 19$_{5,14}$  & 244.286 & -3.30 & 656 & 139.46 & 9.07 $\pm$ 0.25 & 3.30 $\pm$ 0.60 & (55.6 $\pm$ 13.3)/2\\
%The two overlapping emission lines presented the same transition parameters in terms of Einstein coefficient, degeneracy of the upper level, and upper level energy. Therefore, their corresponding integrated intensities were expected to be the same, i.e., half of the calculated integrated intensity for the sum of both lines. 
\enddata
\begin{list}{}
\item \item $^a$Partially blended with a transition from a different species. $^b$Completely blended with a transition from a different species. $^c$Partially blended with an unidentified (U) transition.

Gaussian fits were not performed for completely blended lines or lines observed with SNR $<$ 3. 
A FWHM with no uncertainties indicates that a fixed linewidth was adopted for the fit (Appendix \ref{app-lines}).

%For blended emission lines, a Gaussian deconvolution was performed in order to estimate the integrated intensity of the different components. 
%Values in italics indicate that contribution from neighboring lines could not be ruled out. 

%Blank spaces indicate that either the different blended transitions were completely overlapped and the Gaussian deconvolution was not possible, or that the signal-to-noise ratio of the line was lower than 3 (only for confirmed species). 
%Completely overlapped in a --> same line number
%Completely overlapped in b --> +U in the plot. 

%\item A * in a confirmed species indicates that the particular emission line was not detected with a signal-to-noise ratio above 3. A Gaussian fit was not performed in those cases.
\end{list}
\end{deluxetable*}

\begin{deluxetable*}{cccccccccc} 
\tablecaption{Identified CH$_3$OCH$_3$ and CH$_3$OCHO transitions toward the continuum peak of Ser-emb 11 W. 
Tentative detections are indicated with a $^{*}$. 
Integrated intensity maps of boldfaced transitions are presented in Fig. \ref{overview_com}.} %Tentative assignments are indicated with a $^{*}$. 
\label{lines2}
\tablehead{\colhead{Line ID} &
\colhead{Molecule} & \colhead{Transition} & \colhead{Frequency} & \colhead{log$_{10}$(A$_{ul}$)} & \colhead{$g_{up}$} & \colhead{$E_{up}$} & \colhead{$V_{LSR}$} & \colhead{FWHM} & \colhead{Integrated Intensity}\\% & \colhead{Mom. 0 rms$^{(a)}$}\\
%%the transitions are spectrally integrated (km s-1), but not spatilally (/beam)
\colhead{} & \colhead{} & \colhead{} & \colhead{(GHz)} & \colhead{(s$^{-1}$)} & \colhead{} & \colhead{(K)} 
& \multicolumn{2}{c}{(km s$^{-1}$)}  & \colhead{(mJy beam$^{-1}$ km s$^{-1}$)}}
\startdata
%20 &CH$_3$OCH$_3^a$ & 13$_{0,13}$ $-$ 12$_{1,12}$ (0) & 231.988 & -4.04 & 270 & 80.92 & 8.49 $\pm$ 0.06 & \textit{3.49 $\pm$ 0.16} & \textit{(425.8 $\pm$ 24.7)} x 270/972\\
%20 &CH$_3$OCH$_3^a$ & 13$_{0,13}$ $-$ 12$_{1,12}$ (1) & 231.988 & -4.04 & 432 & 80.92 &8.49 $\pm$ 0.06 & \textit{3.49 $\pm$ 0.16}  & \textit{(425.8 $\pm$ 24.7)} x 432/972\\
%20 &CH$_3$OCH$_3^a$ & 13$_{0,13}$ $-$ 12$_{1,12}$ (5) & 231.988 & -4.04 & 162 & 80.92 & 8.49 $\pm$ 0.06 & \textit{3.49 $\pm$ 0.16}  & \textit{(425.8 $\pm$ 24.7)} x 162/972\\
%20 &CH$_3$OCH$_3^a$ & 13$_{0,13}$ $-$ 12$_{1,12}$ (3) & 231.988 & -4.04 & 108 & 80.92 & 8.49 $\pm$ 0.06 & \textit{3.49 $\pm$ 0.16}  & \textit{(425.8 $\pm$ 24.7)} x 108/972\\
%%I did not deconvolute the C2H5OH line in the blue wing of this line. Its contribution may be negligible. FInally, I am not considering this line. 20 &CH$_3$OCH$_3^a$ & 13$_{0,13}$ $-$ 12$_{1,12}$ (0) & 231.988 & -4.04 & 270 & 80.92 & 8.49 $\pm$ 0.06 & \textit{3.49 $\pm$ 0.16} & \textit{(425.8 $\pm$ 24.7)} x 270/972\\
20 &CH$_3$OCH$_3^b$ & 13$_{0,13}$ $-$ 12$_{1,12}$ (0) & 231.988 & -4.04 & 270 & 80.92 & \nodata & \nodata & \nodata\\
20 &CH$_3$OCH$_3^b$ & 13$_{0,13}$ $-$ 12$_{1,12}$ (1) & 231.988 & -4.04 & 432 & 80.92 & \nodata & \nodata & \nodata\\
20 &CH$_3$OCH$_3^b$ & 13$_{0,13}$ $-$ 12$_{1,12}$ (5) & 231.988 & -4.04 & 162 & 80.92 & \nodata & \nodata & \nodata\\
20 &CH$_3$OCH$_3^b$ & 13$_{0,13}$ $-$ 12$_{1,12}$ (3) & 231.988 & -4.04 & 108 & 80.92 & \nodata & \nodata & \nodata\\
62 &CH$_3$OCH$_3^b$ & 22$_{2,21}$ $-$ 21$_{1,22}$ (5) & 233.189 & -4.56 & 270 & 232.92 & \nodata & \nodata & \nodata\\
62 &CH$_3$OCH$_3^b$ & 22$_{2,21}$ $-$ 21$_{1,22}$ (3) & 233.189 & -4.56 & 180 & 232.92 &\nodata & \nodata & \nodata\\
63 &CH$_3$OCH$_3$ & 22$_{2,21}$ $-$ 21$_{1,22}$ (1)& 233.193 & -4.56 &720 & 232.92 &7.80 $\pm$ 0.07 & 2.08 $\pm$ 0.18 & 88.9 $\pm$ 10.1\\
%64 &CH$_3$OCH$_3^a$ & 22$_{2,21}$ $-$ 21$_{1,22}$ (0)& 233.198 & -4.56 & 450 & 232.92 & 9.04 $\pm$ 0.15 & 3.33 $\pm$ 0.43 & 72.0 $\pm$ 12.2\\
64 &CH$_3$OCH$_3^a$ & 22$_{2,21}$ $-$ 21$_{1,22}$ (0)& 233.198 & -4.56 & 450 & 232.92 & 9.15 $\pm$ 0.13 & 2.07 & 50.2 $\pm$ 6.8\\
\textbf{95} &\textbf{CH$_3$OCH$_3$}& \textbf{23$_{5,18}$ $-$ 23$_{4,19}$ (5)} & 243.739 & -4.10 & 282 & \textbf{286.98} & 8.49 $\pm$ 0.07 & 4.01 $\pm$ 0.16 & (313.0 $\pm$ 16.4) x 282/1692\\
\textbf{95} &\textbf{CH$_3$OCH$_3$} & \textbf{23$_{5,18}$ $-$ 23$_{4,19}$ (3)} & 243.739 & -4.10 & 188 & \textbf{286.98} & 8.49 $\pm$ 0.07 & 4.01 $\pm$ 0.16 & (313.0 $\pm$ 16.4) x 188/1692\\
\textbf{95} &\textbf{CH$_3$OCH$_3$}& \textbf{23$_{5,18}$ $-$ 23$_{4,19}$ (1)}& 243.740 & -4.10 & 752 & \textbf{286.98} & 8.49 $\pm$ 0.07 & 4.01 $\pm$ 0.16 & (313.0 $\pm$ 16.4) x 752/1692\\
\textbf{95} &\textbf{CH$_3$OCH$_3$}& \textbf{23$_{5,18}$ $-$ 23$_{4,19}$ (0)} & 243.741 & -4.10 & 470 & \textbf{286.98} & 8.49 $\pm$ 0.07 & 4.01 $\pm$ 0.16 & (313.0 $\pm$ 16.4) x 470/1692\\
115 &CH$_3$OCH$_3$ & 23$_{2,22}$ $-$ 23$_{1,23}$ (5) & 244.503 & -4.31 & 94 & 253.35 & 8.13 $\pm$ 0.09 & 2.91 $\pm$ 0.23 & (81.8 $\pm$ 8.4) x 94/282\\
115 &CH$_3$OCH$_3$ & 23$_{2,22}$ $-$ 23$_{1,23}$ (3) & 244.503 & -4.31 & 188 & 253.35 & 8.13 $\pm$ 0.09 & 2.91 $\pm$ 0.23 & (81.8 $\pm$ 8.4) x 188/282\\
116 &CH$_3$OCH$_3$ & 23$_{2,22}$ $-$ 23$_{1,23}$ (1) & 244.508 & -4.31 & 752 & 253.35 & 8.23 $\pm$ 0.07 & 2.05 $\pm$ 0.16 & 74.3 $\pm$ 7.5 \\
117 &CH$_3$OCH$_3$ & 23$_{2,22}$ $-$ 23$_{1,23}$ (0) & 244.513 & -4.31 & 282 & 253.35 & 8.68 $\pm$ 0.18 & 2.71 $\pm$ 0.45 & 41.4 $\pm$ 8.9\\
\hline
38 & CH$_3$OCHO & 19$_{9,11}$ $-$ 19$_{8,12}$ E & 232.579 & -4.82 & 78 & 165.98 & 8.04 $\pm$ 0.17 & 3.31 $\pm$ 0.42 & 62.7 $\pm$ 10.5 \\
39 & CH$_3$OCHO$^{b}$ & 19$_{9,10}$ $-$ 19$_{8,11}$ E & 232.597 & -4.82 & 78 & 165.98 & \nodata & \nodata & \nodata\\
40 & CH$_3$OCHO* & 19$_{9,10}$ $-$ 19$_{8,11}$ A & 232.617 & -4.82 & 78 & 165.97 & \nodata & \nodata & \nodata\\ %%too noisy
41 & CH$_3$OCHO & 19$_{9,11}$ $-$ 19$_{8,12}$ A & 232.625 & -4.82 & 78 & 165.97 & 8.71 $\pm$ 0.11 & 2.33 $\pm$ 0.28 & 61.7 $\pm$ 9.8\\
61 & CH$_3$OCHO & 18$_{9,9}$ $-$ 18$_{8,10}$ A & 233.166 & -4.83 & 74 & 154.72 &\nodata & \nodata & \nodata \\
61 & CH$_3$OCHO & 19$_{18,1}$ $-$ 18$_{18,0}$ A & 233.167 & -4.71 & 78 & 326.91 & \nodata & \nodata & \nodata \\ %fit for the three lines whose contribution cannot be disentangled
61 & CH$_3$OCHO & 19$_{18,2}$ $-$ 18$_{18,1}$ A & 233.167 & -4.71 & 78 & 326.91 &\nodata & \nodata & \nodata \\
62 & CH$_3$OCHO$^{b}$ & 19$_{18,2}$ $-$ 18$_{18,1}$ E & 233.189 & -4.71 & 78 & 326.90 & \nodata & \nodata & \nodata \\
65 & CH$_3$OCHO$^{a}$ & 19$_{17,2}$ $-$ 18$_{17,1}$ A & 233.200 & -4.42 & 78 & 303.72 & 7.76 $\pm$ 0.10  & 2.06 $\pm$ 0.29 & (52.3 $\pm$ 8.9)/2\\
65 & CH$_3$OCHO$^{a}$ & 19$_{17,3}$ $-$ 18$_{17,2}$ A & 233.200 & -4.42 &78 & 303.72 & 7.76 $\pm$ 0.10  & 2.06 $\pm$ 0.29 & (52.3 $\pm$ 8.9)/2\\
67 & CH$_3$OCHO$^{a}$ & 19$_{17,2}$ $-$ 18$_{17,1}$ E & 233.213 & -4.42 & 78 & 303.71 & \nodata  & \nodata & \nodata \\
67 & CH$_3$OCHO$^{a}$ & 19$_{4,16}$ $-$ 18$_{4,15}$ E & 233.213 & -3.74 & 78 & 123.26 & \nodata  & \nodata & \nodata\\
%Eran dificiles de deconvolucionar y encima no podia separarlas entre los dos porque Eup y Aul eran distintas
68 & CH$_3$OCHO & 19$_{17,3}$ $-$ 18$_{17,2}$ E & 233.222 & -4.42 & 78 & 303.70 & 7.83 $\pm$ 0.16 & 2.68 $\pm$ 0.43 & 51.6 $\pm$ 10.6\\
\textbf{69} & \textbf{CH$_3$OCHO} & 19$_{4,16}$ $-$ 18$_{4,15}$ A & 233.227 & -3.74 & 78 & \textbf{123.25} & 8.66 $\pm$ 0.04 & 2.94 $\pm$ 0.10 & 225.3 $\pm$ 9.6\\
70 & CH$_3$OCHO & 19$_{16,3}$ $-$ 18$_{16,2}$ A & 233.246 & -4.25 & 78& 281.85 & 7.14 $\pm$ 0.06 & 2.06 $\pm$ 0.16 & (80.6 $\pm$ 8.1)/2\\
70 & CH$_3$OCHO & 19$_{16,4}$ $-$ 18$_{16,3}$ A & 233.246 & -4.25 & 78& 281.85 & 7.14 $\pm$ 0.06 & 2.06 $\pm$ 0.16 & (80.6 $\pm$ 8.1)/2\\
71 & CH$_3$OCHO$^{c}$ & 19$_{16,3}$ $-$ 18$_{16,2}$ E & 233.256 & -4.25 & 78 & 281.84 & 8.09 $\pm$ 0.24 & 2.78 & 55.2 $\pm$ 7.9\\
72 & CH$_3$OCHO$^{c}$ & 19$_{16,4}$ $-$ 18$_{16,3}$ E & 233.269 & -4.25 & 78 & 281.83 & 8.33 $\pm$ 0.19 & 2.56 $\pm$ 0.40 & 56.5 $\pm$ 10.9\\ 
73 & CH$_3$OCHO & 19$_{15,4}$ $-$ 18$_{15,3}$ A & 233.310 & -4.14 & 78 & 261.30 & 8.15 $\pm$ 0.07 & 2.53 $\pm$ 0.16 & (112.6 $\pm$ 9.6)/2 \\
73 & CH$_3$OCHO & 19$_{15,5}$ $-$ 18$_{15,4}$ A & 233.310 & -4.14 &78 & 261.30 & 8.15 $\pm$ 0.07 & 2.53 $\pm$ 0.16 & (112.6 $\pm$ 9.6)/2 \\
74 & CH$_3$OCHO & 19$_{15,4}$ $-$ 18$_{15,3}$ E & 233.316 & -4.14 & 78 & 261.29 & 8.46 $\pm$ 0.08 & 2.09 $\pm$ 0.18 & 66.7 $\pm$ 7.7\\
75 & CH$_3$OCHO & 19$_{15,5}$ $-$ 18$_{15,4}$ E & 233.331 & -4.14 & 78 & 261.28 & 7.65 $\pm$ 0.25 & 2.78 & 54.0 $\pm$8.6\\
114 & CH$_3$OCHO$^{b}$ & 32$_{10,22}$ $-$ 32$_{9,23}$ A & 244.490 & -4.66 & 130 & 379.67 & \nodata & \nodata & \nodata\\
119 & CH$_3$OCHO & 20$_{4,17}$ $-$ 19$_{4,16}$ E & 244.580 & -3.68 & 82 & 134.99 & 8.20 $\pm$ 0.04 & 2.62 $\pm$ 0.11 & 217.1 $\pm$ 12.3\\
121 & CH$_3$OCHO$^a$ & 20$_{4,17}$ $-$ 19$_{4,16}$ A & 244.594 & -3.68 & 82 & 134.98 & 8.48 $\pm$ 0.04 & 2.67 $\pm$ 0.11 & 213.7 $\pm$ 11.0 \\
\enddata
\begin{list}{}
\item $^a$Partially blended with a transition from a different species. $^b$Completely blended with a transition from a different species. $^c$Partially blended with an unidentified (U) transition.

Gaussian fits were not performed for completely blended lines or lines observed with SNR $<$ 3. 
A FWHM with no uncertainties indicates that a fixed linewidth was adopted for the fit (Appendix \ref{app-lines}).

%For blended emission lines, a Gaussian deconvolution was performed in order to estimate the integrated intensity of the different components. 
%Values in italics indicate that contribution from neighboring lines could not be ruled out. 

%Blank spaces indicate that either the different blended transitions were completely overlapped and the Gaussian deconvolution was not possible, or that the signal-to-noise ratio of the line was lower than 3 (only for confirmed species). 
%Completely overlapped in a --> same line number
%Completely overlapped in b --> +U in the plot. 

%\item A * in a confirmed species indicates that the particular emission line was not detected with a signal-to-noise ratio above 3. A Gaussian fit was not performed in those cases.
\end{list}
\end{deluxetable*}

\begin{deluxetable*}{cccccccccc} 
\tablecaption{Identified CH$_3$OCHO $\nu$=1 transitions toward the continuum peak of Ser-emb 11 W in the 231.262 $-$ 233.380 GHz range. }%Tentative assignments are indicated with a $^{*}$. 
\label{lines4}
\tablehead{\colhead{Line ID} &
\colhead{Molecule} & \colhead{Transition} & \colhead{Frequency} & \colhead{log$_{10}$(A$_{ul}$)} & \colhead{$g_{up}$} & \colhead{$E_{up}$} & \colhead{$V_{LSR}$} & \colhead{FWHM} & \colhead{Integrated Intensity}\\% & \colhead{Mom. 0 rms$^{(a)}$}\\
%%the transitions are spectrally integrated (km s-1), but not spatilally (/beam)
\colhead{} &\colhead{} & \colhead{} & \colhead{(GHz)} & \colhead{(s$^{-1}$)} & \colhead{} & \colhead{(K)} 
& \multicolumn{2}{c}{(km s$^{-1}$)} & \colhead{(mJy beam$^{-1}$ km s$^{-1}$)}}%& \multicolumn{2}{c}{(mJy beam$^{-1}$ km s$^{-1}$)}}
\startdata
\hline
4 &CH$_3$OCHO $\nu$=1 & 18$_{4,14}$ $-$ 17$_{4,13}$ E & 231.724 & -3.75 & 74 & 300.80 & 8.27 $\pm$ 0.05 & 2.45 $\pm$ 0.13 & 134.8 $\pm$ 9.3 \\
5 &CH$_3$OCHO$^{b}$ $\nu$=1 & 19$_{14,6}$ $-$ 18$_{14,5}$ E & 231.735 & -4.06 & 78 & 429.95 & \nodata & \nodata & \nodata\\
6 &CH$_3$OCHO$^c$ $\nu$=1 & 19$_{10,9}$ $-$ 18$_{10,8}$ E & 231.749 & -3.87 & 78 & 366.05 & 8.09 $\pm$ 0.16 & 2.78 & 48.6 $\pm$ 7.2\\
8 & CH$_3$OCHO $\nu$=1 & 19$_{16,3}$ $-$ 18$_{16,2}$ A & 231.801 & -4.26 &78 & 469.99 & \nodata & \nodata & \nodata\\
8 & CH$_3$OCHO $\nu$=1 & 19$_{16,4}$ $-$ 18$_{16,3}$ A & 231.801 & -4.26 & 78 & 469.99 & \nodata & \nodata & \nodata\\
8 & CH$_3$OCHO $\nu$=1 & 19$_{17,2}$ $-$ 18$_{17,1}$ A & 231.802 & -4.24 & 78 & 492.08 & \nodata & \nodata & \nodata\\
8 & CH$_3$OCHO $\nu$=1 & 19$_{17,3}$ $-$ 18$_{17,2}$ A & 231.802 & -4.24 & 78 & 492.08 & \nodata & \nodata & \nodata\\
8 & CH$_3$OCHO $\nu$=1 & 19$_{18,1}$ $-$ 18$_{18,0}$ A & 231.803 & -4.71 & 78 & 515.51  & \nodata & \nodata & \nodata\\
8 & CH$_3$OCHO $\nu$=1 & 19$_{18,2}$ $-$ 18$_{18,1}$ A& 231.803& -4.71 & 78 & 515.51 &\nodata & \nodata & \nodata\\
8 & CH$_3$OCHO $\nu$=1 & 24$_{9,15}$ $-$ 24$_{8,16}$ E & 231.804 & -4.78 & 98 & 418.01 &\nodata & \nodata & \nodata\\
8 & CH$_3$OCHO $\nu$=1 & 19$_{15,4}$ $-$ 18$_{15,3}$ A & 231.804 & -4.15 & 78 & 449.25 & \nodata & \nodata & \nodata\\
8 & CH$_3$OCHO $\nu$=1 & 19$_{15,5}$ $-$ 18$_{15,4}$ A & 231.804 & -4.15 & 78 & 449.25 & \nodata & \nodata & \nodata\\
%9 &CH$_3$OCHO $\nu$=1 & 19$_{14,5}$ $-$ 18$_{14,4}$ A & 231.817 & -4.06 & 78 & 429.85 & 8.80 $\pm$ 0.22 & 4.29 $\pm$ 0.60 & (84.3 $\pm$ 14.8)/2\\
%9 &CH$_3$OCHO $\nu$=1 & 19$_{14,6}$ $-$ 18$_{14,5}$ A & 231.817 & -4.06 & 78 & 429.85 & 8.80 $\pm$ 0.22 & 4.29 $\pm$ 0.60 & (84.3 $\pm$ 14.8)/2\\
9 &CH$_3$OCHO $\nu$=1 & 19$_{14,5}$ $-$ 18$_{14,4}$ A & 231.817 & -4.06 & 78 & 429.85 & 8.62 $\pm$ 0.17 & 2.78 & (60.2 $\pm$ 7.0)/2\\
9 &CH$_3$OCHO $\nu$=1 & 19$_{14,6}$ $-$ 18$_{14,5}$ A & 231.817 & -4.06 & 78 & 429.85 & 8.62 $\pm$ 0.17 & 2.78 & (60.2 $\pm$ 7.0)/2\\
11 &CH$_3$OCHO $\nu$=1 & 19$_{13,6}$ $-$ 18$_{13,5}$ A & 231.847 & -4.00 &78  & 411.79 & 8.59 $\pm$ 0.09 & 2.68 $\pm$ 0.23 & (87.1 $\pm$ 9.6)/2\\
11 &CH$_3$OCHO $\nu$=1 & 19$_{13,7}$ $-$ 18$_{13,6}$ A & 231.847 & -4.00 & 78 & 411.79 &8.59 $\pm$ 0.09 & 2.68 $\pm$ 0.23 & (87.1 $\pm$ 9.6)/2\\
14 &CH$_3$OCHO $\nu$=1 & 19$_{4,16}$ $-$ 18$_{4,15}$ E & 231.896 & -3.75 & 78 & 309.75 & 8.24 $\pm$ 0.08 & 2.46 $\pm$ 0.18 & 118.8 $\pm$ 10.8\\
15 &CH$_3$OCHO$^b$ $\nu$=1 & 19$_{12,7}$ $-$ 18$_{12,6}$ A & 231.904 & -3.95 & 78 & 395.09 & \nodata & \nodata & \nodata \\
15 &CH$_3$OCHO$^b$ $\nu$=1 & 19$_{12,8}$ $-$ 18$_{12,7}$ A & 231.904 & -3.95 & 78 & 395.09 &\nodata & \nodata & \nodata \\
21 &CH$_3$OCHO$^c$ $\nu$=1 & 19$_{11,8}$ $-$ 18$_{11,7}$ A & 232.003 & -3.90 & 78 & 397.73 & 9.03 $\pm$ 0.11 & 2.99 $\pm$ 0.28 & (103.4 $\pm$ 12.4)/2\\
21 &CH$_3$OCHO$^c$ $\nu$=1 & 19$_{12,8}$ $-$ 18$_{12,7}$ A & 232.003 & -3.90 & 78 & 397.73 & 9.03 $\pm$ 0.11 & 2.99 $\pm$ 0.28 & (103.4 $\pm$ 12.4)/2\\
%24 &CH$_3$OCHO$^{*}$ $\nu$=1 & 19$_{12,8}$ $-$ 18$_{12,7}$ E & 232.129 & -3.95 & & 394.99 & & \\
%25 &CH$_3$OCHO$^a$ $\nu$=1 & 19$_{9,10}$ $-$ 18$_{9,9}$ E & 232.160 & -3.83 & 78 & 353.32 & 7.88 $\pm$ 0.19 & 4.03 $\pm$ 0.48 & 95.8 $\pm$ 14.6\\
25 &CH$_3$OCHO$^a$ $\nu$=1 & 19$_{9,10}$ $-$ 18$_{9,9}$ E & 232.160 & -3.83 & 78 & 353.32 & 7.98 $\pm$ 0.14 & 2.78 & 74.9 $\pm$ 6.7\\
26 &CH$_3$OCHO$^b$ $\nu$=1 & 19$_{10,9}$ $-$ 18$_{10,8}$ A & 232.164 & -3.86 & 78 & 365.73 & \nodata & \nodata & \nodata\\
26 &CH$_3$OCHO$^b$ $\nu$=1 & 19$_{10,10}$ $-$ 18$_{10,9}$ A & 232.164 & -3.86 & 78 & 365.73 & \nodata & \nodata & \nodata\\
33 &CH$_3$OCHO $\nu$=1 & 19$_{11,9}$ $-$ 18$_{11,8}$ E & 232.378 & -3.90 & 78 & 379.55 & 9.38 $\pm$ 0.14 & 2.11 $\pm$ 0.35 & 35.2 $\pm$ 7.6\\
36 &CH$_3$OCHO$^a$ $\nu$=1 & 19$_{9,10}$ $-$ 18$_{9,9}$ A & 232.423 & -3.83 & 78 & 353.09 & 8.39 $\pm$ 0.11 & 2.70 $\pm$ 0.27 & (122.9 $\pm$ 16.0)/2\\
36 &CH$_3$OCHO$^a$ $\nu$=1 & 19$_{9,11}$ $-$ 18$_{9,10}$ A & 232.423 & -3.83 & 78 & 353.09 &  8.39 $\pm$ 0.11 & 2.70 $\pm$ 0.27 & (122.9 $\pm$ 16.0)/2\\
42 &CH$_3$OCHO $\nu$=1 & 19$_{10,10}$ $-$ 18$_{10,9}$ E & 232.684 & -3.86 & 78 & 365.47 & 8.69 $\pm$ 0.10 & 2.18 $\pm$ 0.24 & 66.1 $\pm$ 9.6\\
43 &CH$_3$OCHO $\nu$=1 & 19$_{8,11}$ $-$ 18$_{8,10}$ E & 232.739 & -3.80 & 78 & 341.98 & 9.04 $\pm$ 0.14 & 2.19 $\pm$ 0.34 & 49.4 $\pm$ 10.1\\
%47 &CH$_3$OCHO $\nu$=1 & 19$_{8,12}$ $-$ 18$_{8,11}$ A & 232.836 & -3.80 & 78 & 341.83 & 8.30 $\pm$ 0.07 & 1.88 $\pm$ 0.18 & 54.6 $\pm$ 6.6\\
47 &CH$_3$OCHO $\nu$=1 & 19$_{8,12}$ $-$ 18$_{8,11}$ A & 232.836 & -3.80 & 78 & 341.83 & 8.29 $\pm$ 0.13 & 2.78 & 64.8 $\pm$ 5.5\\
48 &CH$_3$OCHO $\nu$=1 & 19$_{8,11}$ $-$ 18$_{8,10}$ A & 232.840 & -3.80 & 78 & 341.83 & 8.99 $\pm$ 0.08 & 2.53 $\pm$ 0.20 & 73.6 $\pm$ 7.7\\
57 &CH$_3$OCHO$^c$ $\nu$=1 & 19$_{9,11}$ $-$ 18$_{9,10}$ E & 233.080 & -3.83 & 78 & 352.77 & \nodata & \nodata & \nodata\\
\enddata
\begin{list}{}
\item $^a$Partially blended with a transition from a different species. $^b$Completely blended with a transition from a different species. $^c$Partially blended with an unidentified (U) transition.

Gaussian fits were not performed for completely blended lines or lines observed with SNR $<$ 3. 
A FWHM with no uncertainties indicates that a fixed linewidth was adopted for the fit (Appendix \ref{app-lines}).

\end{list}
\end{deluxetable*}

\begin{deluxetable*}{cccccccccc} 
\tablecaption{Identified CH$_3$OCHO $\nu$=1 transitions toward the continuum peak of Ser-emb 11 W in the 242.978 $-$ 244.853 GHz range. 
Tentative detections are indicated with a $^{*}$. }%Tentative assignments are indicated with a $^{*}$. 
\label{lines5}
\tablehead{\colhead{Line} &
\colhead{Molecule} & \colhead{Transition} & \colhead{Frequency} & \colhead{log(A$_{ul}$)} & \colhead{$g_{up}$} & \colhead{$E_{up}$} & \colhead{$V_{LSR}$} & \colhead{FWHM} & \colhead{Integrated Intensity}\\% & \colhead{Mom. 0 rms$^{(a)}$}\\
%%the transitions are spectrally integrated (km s-1), but not spatilally (/beam)
\colhead{} &\colhead{} & \colhead{} & \colhead{(GHz)} & \colhead{(s$^{-1}$)} & \colhead{} & \colhead{(K)} 
& \multicolumn{2}{c}{(km s$^{-1}$)} & \colhead{(mJy beam$^{-1}$ km s$^{-1}$)}}%& \multicolumn{2}{c}{(mJy beam$^{-1}$ km s$^{-1}$)}}
\startdata
%79 &CH$_3$OCHO$^{a}$ $\nu$=1 & 20$_{16,4}$ $-$ 19$_{16,3}$ E & 243.057 & -4.11 & 82 & 482.38 & 8.58 $\pm$ 0.18 & 1.90 $\pm$ 0.43 & 29.8 $\pm$ 8.9\\
79 &CH$_3$OCHO$^{a}$ $\nu$=1 & 20$_{16,4}$ $-$ 19$_{16,3}$ E & 243.057 & -4.11 & 82 & 482.38 & 8.51 $\pm$ 0.29 & 2.78 & 33.6 $\pm$ 6.4\\
85 &CH$_3$OCHO$^b$ $\nu$=1 & 20$_{4,17}$ $-$ 19$_{4,16}$ E & 243.223 & -3.68 & 82 & 321.42 & \nodata & \nodata & \nodata\\
87 &CH$_3$OCHO* $\nu$=1 & 20$_{13,7}$ $-$ 19$_{13,6}$ E & 243.325 & -3.90 & 82 & 424.03 & \nodata & \nodata & \nodata\\%%too noisy
91 &CH$_3$OCHO$^c$ $\nu$=1 & 20$_{12,8}$ $-$ 19$_{12,7}$ E & 243.511 & -3.85 &82  & 407.25 & 7.87 $\pm$ 0.20 & 2.78 & 64.7 $\pm$ 7.1\\
96 &CH$_3$OCHO$^b$ $\nu$=1 & 20$_{11,9}$ $-$ 19$_{11,8}$ E & 243.766 & -3.82 &82 & 391.83 & 7.60 $\pm$ 0.12 & 3.88 $\pm$ 0.29 & 131.4 $\pm$ 12.9\\
99 &CH$_3$OCHO* $\nu$=1 & 20$_{14,7}$ $-$ 19$_{14,6}$ E & 244.000 & -3.95 & 82 & 441.66 & \nodata & \nodata & \nodata\\
100 &CH$_3$OCHO$^b$ $\nu$=1 & 20$_{15,5}$ $-$ 19$_{15,4}$ E & 244.048 & -4.02 & 82 & 460.96 & \nodata & \nodata & \nodata\\
100 &CH$_3$OCHO$^b$ $\nu$=1 & 20$_{15,6}$ $-$ 19$_{15,5}$ E & 244.048 & -4.02 & 82 & 460.96 & \nodata & \nodata & \nodata \\
101 &CH$_3$OCHO $\nu$=1 & 19$_{4,15}$ $-$ 18$_{4,14}$ A & 244.067 & -3.68 & 78 & 312.75 & 8.92 $\pm$ 0.06 & 2.14 $\pm$ 0.15 & 97.7 $\pm$ 9.1 \\
%102 &CH$_3$OCHO $\nu$=1 & 20$_{14,6}$ $-$ 19$_{14,5}$ A & 244.074 & -3.95 & 82 & 441.56 & 8.67 $\pm$ 0.11 & 1.65 $\pm$ 0.27 & (39.8 $\pm$ 8.4)/2\\
%102 &CH$_3$OCHO $\nu$=1 & 20$_{14,7}$ $-$ 19$_{14,6}$ A & 244.074 & -3.95 & 82 & 441.56 &8.67 $\pm$ 0.11 & 1.65 $\pm$ 0.27 & (39.8 $\pm$ 8.4)/2\\
102 &CH$_3$OCHO $\nu$=1 & 20$_{14,6}$ $-$ 19$_{14,5}$ A & 244.074 & -3.95 & 82 & 441.56 & 8.79 $\pm$ 0.21 & 2.78 & (49.5 $\pm$ 7.1)/2\\
102 &CH$_3$OCHO $\nu$=1 & 20$_{14,7}$ $-$ 19$_{14,6}$ A & 244.074 & -3.95 & 82 & 441.56 & 8.79 $\pm$ 0.21 & 2.78 & (49.5 $\pm$ 7.1)/2\\
103 &CH$_3$OCHO $\nu$=1 & 20$_{10,10}$ $-$ 19$_{10,9}$ E & 244.112 & -3.78 & 82 & 377.76 & 8.19 $\pm$ 0.16 & 3.92 $\pm$ 0.38 & 97.6 $\pm$ 12.5 \\ %% maybe too wide?
104 &CH$_3$OCHO $\nu$=1 & 20$_{13,7}$ $-$ 19$_{13,6}$ A & 244.120 & -3.90 & 82 & 423.51 & 8.55 $\pm$ 0.13 & 3.68 $\pm$ 0.32 & (92.1 $\pm$ 10.5)/2\\ %% maybe too wide?
104 &CH$_3$OCHO $\nu$=1 & 20$_{13,8}$ $-$ 19$_{13,7}$ A & 244.120 & -3.90 & 82 & 423.51 & 8.55 $\pm$ 0.13 & 3.68 $\pm$ 0.32 & (92.1 $\pm$ 10.5)/2\\ %% maybe too wide?
106 &CH$_3$OCHO $\nu$=1 & 20$_{12,8}$ $-$ 19$_{12,7}$ A & 244.198 & -3.85 & 82 & 406.81 & 8.27 $\pm$ 0.16 & 2.78 & (81.9 $\pm$ 8.5)/2\\
106 &CH$_3$OCHO $\nu$=1 & 20$_{12,9}$ $-$ 19$_{12,8}$ A & 244.198 & -3.85 & 82 & 406.81 & 8.27 $\pm$ 0.16 & 2.78 & (81.9 $\pm$ 8.5)/2 \\
107 &CH$_3$OCHO* $\nu$=1 & 20$_{13,8}$ $-$ 19$_{13,7}$ E & 244.207 & -3.89 & 82 & 423.51 & \nodata & \nodata & \nodata\\
%110 &CH$_3$OCHO$^a$ $\nu$=1 & 20$_{11,9}$ $-$ 19$_{11,8}$ A & 244.326 & -3.81 & 82 & 391.46 & 7.86 $\pm$ 0.27 & 3.72 $\pm$ 0.64 & (150.2 $\pm$ 29.9)/2\\
%110 &CH$_3$OCHO$^a$ $\nu$=1 & 20$_{11,10}$ $-$ 19$_{11,9}$ A & 244.326 & -3.81 & 82 & 391.46 & 7.86 $\pm$ 0.27 & 3.72 $\pm$ 0.64 & (150.2 $\pm$ 29.9)/2 \\
110 &CH$_3$OCHO$^a$ $\nu$=1 & 20$_{11,9}$ $-$ 19$_{11,8}$ A & 244.326 & -3.81 & 82 & 391.46 & 8.11 $\pm$ 0.19 & 2.78 & (112.5 $\pm$ 12.9)/2\\
110 &CH$_3$OCHO$^a$ $\nu$=1 & 20$_{11,10}$ $-$ 19$_{11,9}$ A & 244.326 & -3.81 & 82 & 391.46 & 8.11 $\pm$ 0.19 & 2.78 & (112.5 $\pm$ 12.9)/2\\
%113 &CH$_3$OCHO $\nu$=1 & 20$_{12,9}$ $-$ 19$_{12,8}$ E & 244.446 & -3.85 & 82 & 406.72 & 9.19 $\pm$ 0.14 & 1.63 $\pm$ 0.33& 29.4 $\pm$ 7.9\\
113 &CH$_3$OCHO* $\nu$=1 & 20$_{12,9}$ $-$ 19$_{12,8}$ E & 244.446 & -3.85 & 82 & 406.72 & \nodata & \nodata & \nodata\\
118 &CH$_3$OCHO $\nu$=1 & 20$_{10,10}$ $-$ 19$_{10,9}$ A & 244.529 & -3.78 & 82 & 377.46 & 9.08 $\pm$ 0.09 & 1.87 $\pm$ 0.21 & (62.0 $\pm$ 9.1)/2\\
118 &CH$_3$OCHO $\nu$=1 & 20$_{10,11}$ $-$ 19$_{10,10}$ A & 244.529 & -3.78 & 82 & 377.46 & 9.08 $\pm$ 0.09 & 1.87 $\pm$ 0.21 & (62.0 $\pm$ 9.1)/2\\
120 &CH$_3$OCHO$^{b}$ $\nu$=1 & 20$_{9,11}$ $-$ 19$_{9,10}$ E & 244.588 & -3.75 & 82 & 365.06 & \nodata & \nodata & \nodata\\
%124 &CH$_3$OCHO $\nu$=1 & 20$_{11,10}$ $-$ 19$_{11,9}$ E & 244.730 & -3.81 & 82 & 391.29 & 8.73 $\pm$ 0.17 & 4.24 $\pm$ 0.43 & 96.0 $\pm$ 12.5\\
124 &CH$_3$OCHO $\nu$=1 & 20$_{11,10}$ $-$ 19$_{11,9}$ E & 244.730 & -3.81 & 82 & 391.29 & 8.57 $\pm$ 0.13 & 2.78 & 72.1 $\pm$ 6.2\\
127 &CH$_3$OCHO$^a$ $\nu$=1 & 20$_{9,11}$ $-$ 19$_{9,10}$ A & 244.845 & -3.75 & 82 & 364.84 & 7.90 $\pm$ 0.09  & 2.30 $\pm$ 0.25 & (79.8 $\pm$ 11.0)/2\\
127 &CH$_3$OCHO$^a$ $\nu$=1 & 20$_{9,12}$ $-$ 19$_{9,11}$ A & 244.845 & -3.75 & 82 & 364.84 & 7.90 $\pm$ 0.09  & 2.30 $\pm$ 0.25 & (79.8 $\pm$ 11.0)/2\\
\hline
\enddata
\begin{list}{}
\item $^a$Partially blended with a transition from a different species. $^b$Completely blended with a transition from a different species. $^c$Partially blended with an unidentified (U) transition.

Gaussian fits were not performed for completely blended lines or lines observed with SNR $<$ 3. 
A FWHM with no uncertainties indicates that a fixed linewidth was adopted for the fit (Appendix \ref{app-lines}).

\end{list}
\end{deluxetable*}
%\end{longtable*}
%\end{ThreePartTable}

\begin{deluxetable*}{cccccccccc} 
\tablecaption{Identified N-bearing COM transitions toward the continuum peak of Ser-emb 11 W. %The list includes 3 species and 2 additional isotopologues. 
%1 tentative assignment is indicated with a $^{*}$. 
Tentative detections are indicated with a $^{*}$
Integrated intensity maps of boldfaced transitions are presented in Fig. \ref{overview_com}.\label{lines6}}
\tablehead{\colhead{Line ID} &
\colhead{Molecule} & \colhead{Transition} & \colhead{Frequency} & \colhead{log$_{10}$(A$_{ul}$)} & \colhead{$g_{up}$} & \colhead{$E_{up}$} & \colhead{$V_{LSR}$} & \colhead{FWHM} & \colhead{Integrated Intensity}\\% & \colhead{Mom. 0 rms$^{(a)}$}\\
%%the transitions are spectrally integrated (km s-1), but not spatilally (/beam)
\colhead{} & \colhead{} & \colhead{} & \colhead{(GHz)} & \colhead{(s$^{-1}$)} & \colhead{} & \colhead{(K)} 
& \multicolumn{2}{c}{(km s$^{-1}$)} & \colhead{(mJy beam$^{-1}$ km s$^{-1}$)}}%& \multicolumn{2}{c}{(mJy beam$^{-1}$ km s$^{-1}$)}}
\startdata
13 & HNCO* & 28$_{1,28,27}$ $-$ 29$_{0,29,28}$ & 231.873 & -4.18 & 55 & 469.87 & 9.18 $\pm$ 0.20 & 4.86 $\pm$ 0.49 & (141.6 $\pm$ 18.5) x 55/171\\
13 & HNCO* & 28$_{1,28,29}$ $-$ 29$_{0,29,30}$ & 231.873 & -4.18 & 59 & 469.87 & 9.18 $\pm$ 0.20 & 4.86 $\pm$ 0.49 & (141.6 $\pm$ 18.5) x 59/171\\
13 & HNCO* & 28$_{1,28,28}$ $-$ 29$_{0,29,29}$ & 231.873 & -4.18 & 57 & 469.87 & 9.18 $\pm$ 0.20 & 4.86 $\pm$ 0.49 & (141.6 $\pm$ 18.5) x 57/171 \\ 
\hline
31 & NH$_2$CHO$^c$ & 11$_{2,10}$ $-$ 10$_{2,9}$ & 232.274 & -3.05 & 23 & 78.93 & 10.28 $\pm$ 0.36 & 3.50 $\pm$ 0.82 & 89.9 $\pm$ 23.6\\
\textbf{92} & \textbf{NH$_2$CHO}$^c$ & \textbf{12$_{1,12}$ $-$ 11$_{1,11}$} & 243.521 & -2.98 & 25 & \textbf{79.17} & 10.30 $\pm$ 0.69 & 5.19 $\pm$ 1.14 & 178.4 $\pm$ 43.1\\
\hline
76 & CH$_2$DCN & 14$_{5,9}$ $-$ 13$_{5,8}$ & 243.013 & -2.97 & 29 & 222.27 & 8.27 $\pm$ 0.15 & 3.27 $\pm$ 0.35 & (80.4 $\pm$ 11.4)/2 \\
76 & CH$_2$DCN & 14$_{5,10}$ $-$ 13$_{5,9}$ & 243.013 & -2.97 & 29 & 222.27 & 8.27 $\pm$ 0.15 & 3.27 $\pm$ 0.35 & (80.4 $\pm$ 11.4)/2\\
%The two overlapping emission lines presented the same transition parameters in terms of Einstein coefficient, degeneracy of the upper level, and upper level energy. Therefore, their corresponding integrated intensities were expected to be the same, i.e., half of the calculated integrated intensity for the sum of both lines. 
77 & CH$_2$DCN & 14$_{0,14}$ $-$ 13$_{0,13}$ & 243.041 & -2.91 & 29 & 87.50 & 8.27 $\pm$ 0.12 & 3.63 $\pm$ 0.30 & 123.9 $\pm$ 13.4\\
78 & CH$_2$DCN & 14$_{4,10}$ $-$ 13$_{4,9}$ & 243.051 & -2.94 & 29 & 173.77 & 7.78 $\pm$ 0.10 & 3.46 $\pm$ 0.25 & (142.2 $\pm$ 13.4)/2\\
78 & CH$_2$DCN & 14$_{4,11}$ $-$ 13$_{4,10}$ & 243.051 & -2.94 & 29 & 173.77 &7.78 $\pm$ 0.10 & 3.46 $\pm$ 0.25 & (142.2 $\pm$ 13.4)/2\\
\textbf{80} & \textbf{CH$_2$DCN} & \textbf{14$_{3,12}$ $-$ 13$_{3,11}$} & 243.083 & -2.93 & 29 & \textbf{136.04} & \nodata & \nodata & \nodata\\
\textbf{80} & \textbf{CH$_2$DCN}& \textbf{14$_{2,13}$ $-$ 13$_{2,12}$} & 243.083 & -2.92 & 29 & \textbf{109.08} & \nodata & \nodata & \nodata\\
\textbf{80} & \textbf{CH$_2$DCN}& \textbf{14$_{3,11}$ $-$ 13$_{3,10}$} & 243.083 & -2.93 & 29 & \textbf{136.04} & \nodata & \nodata & \nodata\\
82 & CH$_2$DCN & 14$_{2,12}$ $-$ 13$_{2,11}$ & 243.152 & -2.92 & 29 & 109.09 & 8.59 $\pm$ 0.10 & 2.52 $\pm$ 0.23 & 94.7 $\pm$ 11.5\\
105 & CH$_2$DCN & 14$_{1,13}$ $-$ 13$_{1,12}$ & 244.143 & -2.90 & 29 & 93.28 & 8.65 $\pm$ 0.12 & 3.36 $\pm$ 0.29 & 125.2 $\pm$ 14.1\\
\hline
23 & $^{13}$CH$_3$CN$^{b,c}$ & 13$_6$ $-$ 12$_6$ & 232.077 & -3.07 & 108 & 355.51 & \nodata & \nodata & \nodata\\
%Includes F=13-13, 13-12, 12-13, 14-13, 12-11 and 12-12 in SPLATALOGE. Intensity from QN=0 in SPLATALOGE that has same JPL intensity as JPL catalog
24 & $^{13}$CH$_3$CN*$^{c}$ & 13$_5$ $-$ 12$_5$ & 232.125 & -3.04 & 54 & 256.88 & \nodata & \nodata & \nodata\\ 
%Includes F=13-13, 13-12, 12-11, 14-13, 12-13 and 12-12 in SPLATALOGE. Intensity from QN=0 in SPLATALOGE that has same JPL intensity as JPL catalog
26 & $^{13}$CH$_3$CN$^{b}$ & 13$_4$ $-$ 12$_4$ & 232.164 & -3.01 & 54 & 192.51 &\nodata & \nodata & \nodata\\
%Includes F=13-13, 13-12, 12-11, 14-13, 12-13 and 12-12 in SPLATALOGE. Intensity from QN=0 in SPLATALOGE that has same JPL intensity as JPL catalog
\textbf{27} & \textbf{$^{13}$CH$_3$CN} & 13$_3$ $-$ 12$_3$ & 232.195 & -2.99 & 108 & \textbf{142.43} & 8.50 $\pm$ 0.11 & 2.94 $\pm$ 0.27 & 92.5 $\pm$ 11.1 \\ 
%Includes F=13-13, 13-12, 12-11, 14-13, 12-13 and 12-12 in SPLATALOGE. Intensity from QN=0 in SPLATALOGE that has same JPL intensity as JPL catalog
28 & $^{13}$CH$_3$CN & 13$_2$ $-$ 12$_2$ & 232.216 & -2.98 & 54 & 106.65 & 7.35 $\pm$ 0.11 & 2.74 $\pm$ 0.28 & 79.0 $\pm$ 10.5\\ 
%Includes F=13-13, 12-11, 13-12, 14-13, 12-13 and 12-12 in SPLATALOGE. Intensity from QN=0 in SPLATALOGE that has same JPL intensity as JPL catalog
29 & $^{13}$CH$_3$CN & 13$_1$ $-$ 12$_1$ & 232.230 & -2.97 & 54 & 85.18 & 8.53 $\pm$ 0.10 & 3.51 $\pm$ 0.25 & 93.6 $\pm$ 8.6 \\ 
%Includes F=13-13, 13-12, 12-11, 14-13, 12-13 and 12-12 in SPLATALOGE. Intensity from QN=0 in SPLATALOGE that has same JPL intensity as JPL catalog
30 & $^{13}$CH$_3$CN & 13$_0$ $-$ 12$_0$ & 232.234 & -2.97 & 54 & 78.02 & 8.41 $\pm$ 0.09 & 2.73 $\pm$ 0.21 & 82.5 $\pm$ 8.2 \\ 
%Includes F=13-13, 12-11, 13-12, 14-13, 12-13 and 12-12 in SPLATALOGE. Intensity from QN=0 in SPLATALOGE that has same JPL intensity as JPL catalog
\hline
15 & CH$_3$C$^{15}$N$^b$ & 13$_3$ $-$ 12$_3$ & 231.902 & -2.99 & 108 & 142.34 & \nodata & \nodata & \nodata\\
16 & CH$_3$C$^{15}$N & 13$_2$ $-$ 12$_2$ & 231.924 & -2.98 & 54& 106.55 & 7.82 $\pm$ 0.16 & 3.08 $\pm$ 0.40 & 50.8 $\pm$ 8.7\\
\textbf{17} & \textbf{CH$_3$C$^{15}$N} & \textbf{13$_1$ $-$ 12$_1$} & 231.937 & -2.97 & 54& \textbf{85.08} & 6.65 $\pm$ 0.12 & 3.08 & 76.6 $\pm$ 5.0 \\
18 & CH$_3$C$^{15}$N & 13$_0$ $-$ 12$_0$ & 231.941 & -2.97 & 54& 77.93 & 7.76 $\pm$ 0.18 & 3.08 & 50.1 $\pm$ 5.0 \\
\hline
\textbf{12} & \textbf{CH$_3$CH$_2$CN} & \textbf{27$_{1,27}$ $-$ 26$_{1,26}$} & 231.854 & -2.98 & 55 & \textbf{157.73} & 8.16 $\pm$ 0.10 & 3.46 $\pm$ 0.26 & 94.1 $\pm$ 9.0\\ 
20 & CH$_3$CH$_2$CN$^b$ & 27$_{0,27}$ $-$ 26$_{0,26}$ & 231.990 & -2.98 & 55 & 157.71 & \nodata & \nodata & \nodata \\
45 & CH$_3$CH$_2$CN & 26$_{3,24}$ $-$ 25$_{3,23}$ & 232.790 & -2.98 & 53 & 161.02 & 8.78 $\pm$ 0.24 & 3.45 $\pm$ 0.64 & 64.7 $\pm$ 15.3\\
51 & CH$_3$CH$_2$CN & 26$_{10,16}$ $-$ 25$_{10,15}$ & 232.962 & -3.04 & 53 & 261.96 & 7.91 $\pm$ 0.13 & 2.12 $\pm$ 0.31 & (40.5 $\pm$ 7.8)/2\\
51 & CH$_3$CH$_2$CN & 26$_{10,17}$ $-$ 25$_{10,16}$ & 232.962 & -3.04 & 53 & 261.96 & 7.91 $\pm$ 0.13 & 2.12 $\pm$ 0.31 & (40.5 $\pm$ 7.8)/2\\
52 & CH$_3$CH$_2$CN & 26$_{9,17}$ $-$ 25$_{9,16}$ & 232.968 & -3.03 & 53 & 240.89 & 9.19 $\pm$ 0.14 & 2.72 $\pm$ 0.34 & (51.9 $\pm$ 8.5)/2\\
52 & CH$_3$CH$_2$CN & 26$_{9,18}$ $-$ 25$_{9,17}$ & 232.968 & -3.03 & 53 & 240.89 & 9.19 $\pm$ 0.14 & 2.72 $\pm$ 0.34 & (51.9 $\pm$ 8.5)/2 \\
53 & CH$_3$CH$_2$CN & 26$_{11,15}$ $-$ 25$_{11,14}$ & 232.976 & -3.06 & 53 & 285.23 & 8.92 $\pm$ 0.19 & 3.59 $\pm$ 0.47 & (57.5 $\pm$ 9.7)/2\\ 
53 & CH$_3$CH$_2$CN & 26$_{11,16}$ $-$ 25$_{11,15}$ & 232.976 & -3.06 & 53 & 285.23 & 8.92 $\pm$ 0.19 & 3.59 $\pm$ 0.47 & (57.5 $\pm$ 9.7)/2 \\
54 & CH$_3$CH$_2$CN & 26$_{8,18}$ $-$ 25$_{8,17}$ & 232.999 & -3.01 & 53 & 222.04  &9.68 $\pm$ 0.16 & 3.28 $\pm$ 0.43 & (65.4 $\pm$ 11.2)/2 \\
54 & CH$_3$CH$_2$CN & 26$_{8,19}$ $-$ 25$_{8,18}$ & 232.999 & -3.01 & 53 & 222.04 & 9.68 $\pm$  0.16 & 3.28 $\pm$ 0.43 & (65.4 $\pm$ 11.2)/2 \\
55 & CH$_3$CH$_2$CN & 26$_{12,14}$ $-$ 25$_{12,13}$ & 233.003 & -3.07 & 53 & 310.69 & 8.26 $\pm$ 0.23 & 3.18 $\pm$ 0.57 & (44.8 $\pm$ 10.5)/2\\
55 & CH$_3$CH$_2$CN & 26$_{12,15}$ $-$ 25$_{12,14}$ & 233.003 & -3.07 &53  & 310.69 & 8.26 $\pm$ 0.23 & 3.18 $\pm$ 0.57 & (44.8 $\pm$ 10.5)/2 \\
56 & CH$_3$CH$_2$CN & 26$_{7,19}$ $-$ 25$_{7,18}$ & 233.069 & -3.00 & 53 & 205.40 & 8.47 $\pm$ 0.19 & 3.61 $\pm$ 0.47 & (101.4 $\pm$ 17.2)/2\\
56 & CH$_3$CH$_2$CN & 26$_{7,20}$ $-$ 25$_{7,19}$ & 233.069 & -3.00 & 53 & 205.40 & 8.47 $\pm$ 0.19 & 3.61 $\pm$ 0.47 & (101.4 $\pm$ 17.2)/2 \\
58 & CH$_3$CH$_2$CN$^c$ & 26$_{14,12}$ $-$ 25$_{14,11}$ & 233.089 & -3.12 &53  & 368.17 & \nodata & \nodata & \nodata \\
58 & CH$_3$CH$_2$CN$^c$ & 26$_{14,13}$ $-$ 25$_{14,12}$ & 233.089 & -3.12 & 53 & 368.17 &\nodata & \nodata & \nodata\\
94 & CH$_3$CH$_2$CN & 27$_{4,23}$ $-$ 26$_{4,22}$ & 243.643 & -2.92 & 55 & 180.79 & 7.58 $\pm$ 0.18 & 3.34 $\pm$ 0.43 & 80.5 $\pm$ 13.7\\
\enddata
\begin{list}{}
\item $^a$Partially blended with a transition from a different species. $^b$Completely blended with a transition from a different species. $^c$Partially blended with an unidentified (U) transition.

Gaussian fits were not performed for completely blended lines or lines observed with SNR $<$ 3. 
A FWHM with no uncertainties indicates that a fixed linewidth was adopted for the fit (Appendix \ref{app-lines}).

\end{list}
\end{deluxetable*}

\begin{deluxetable*}{cccccccccc} 
\tablecaption{Identified transitions corresponding to small molecules (4 atoms or less) detected %in narrow, dedicated spectral windows 
toward the continuum peak of Ser-emb 11 W. %The list includes 4 distinct species and 5 additional isotopologues, in addition to SO$_2$, and O$^{13}$CS (these S-bearing species are already reported in Table \ref{lines7}).
} 
\label{lines_small}
\tablehead{\colhead{Line ID} &
\colhead{Molecule} & \colhead{Transition} & \colhead{Frequency} & \colhead{log$_{10}$(A$_{ul}$)} & \colhead{$g_{up}$} & \colhead{$E_{up}$} & \colhead{$V_{LSR}$} & \colhead{FWHM} & \colhead{Integrated Intensity}\\
%\\% & \colhead{Mom. 0 rms$^{(a)}$}\\
%%the transitions are spectrally integrated (km s-1), but not spatilally (/beam)
\colhead{} & \colhead{} & \colhead{} & \colhead{(GHz)} & \colhead{(s$^{-1}$)} & \colhead{} & \colhead{(K)} 
& \multicolumn{2}{c}{(km s$^{-1}$)}  & \colhead{(mJy beam$^{-1}$ km s$^{-1}$)}
}
\startdata
 %& SiO & 6 $-$ 5 & 260.518 & -3.04 & 13 & 43.74 \\
 & CO$^a$ & 2 $-$ 1 & 230.538 & -6.16 & 5 & 16.60 & $\sim$12.6 & $\sim$17.7 & $\gtrsim$11360\\
 %CO line had structure
 & $^{13}$CO$^a$ & 2 $-$ 1 & 220.399 & -6.22 & 5 & 15.87 & $\sim$8.1 & $\sim$5.2 & $\gtrsim$1380\\
 & C$^{18}$O$^{a}$ & 2 $-$ 1 & 219.560 & -6.22 & 5 & 15.81 & $\sim$7.9 & $\sim$2.0 & $\gtrsim$380\\
 \hline
 & H$_2$CO$^a$ & 3$_{0,3}$ $-$ 2$_{0,2}$ & 218.222 & -3.55 & 7 & 20.96  & $\sim$8.5 & $\sim$3.5 & $\gtrsim$920 \\ 
 & H$_2$CO & 3$_{2,2}$ $-$ 2$_{2,1}$ & 218.476 & -3.80 & 7 & 68.09 & 9.33 $\pm$ 0.04 & 3.53 $\pm$ 0.10 & 506.8 $\pm$ 18.7\\
%& H$_2$CO & 3$_{2,1}$ $-$ 2$_{2,0}$ & 218.760 & -3.8024 & 7 & 68.09 & \nodata$^{(b)}$ & \nodata$^{(b)}$ \\
 & HDCO & 4$_{2,2}$ $-$ 3$_{2,1}$ & 259.035 & -3.43 & 9 & 62.86 & 8.25 $\pm$ 0.11 & 3.91 $\pm$ 0.31 & 230.2 $\pm$ 23.0 \\
\hline
\hline
& DCN & 3 $-$ 2 & 217.238 & -3.49 & 33 & 20.85 & 8.44 $\pm$ 0.09 & 4.22 $\pm$ 0.21 & 291.9 $\pm$ 19.1\\
 & H$^{13}$CN$^{a}$ & 3$_2$ $-$ 2$_2$ & 259.012 & -3.34 & 33 & 24.86 & $\sim$8.7 & $\sim$6.1 & $\gtrsim$1000\\
 & HC$^{15}$N & 3 $-$ 2 & 258.157 & -3.09 & 7 & 24.78 & 8.52 $\pm$ 0.09 & 4.09 $\pm$ 0.24 & 409.3 $\pm$ 30.0\\
\hline
\hline
108 &SO$_2$ & 14$_{0,14}$ $-$ 13$_{1,13}$ & 244.254 & -3.79 & 29 & 93.90 & 8.77 $\pm$ 0.09 & 6.47 $\pm$ 0.23 & 554.6 $\pm$ 25.9 \\
 & SO$_2$ & 11$_{3,9}$ $-$ 11$_{2,10}$ & 262.257 & -3.85 &  23 & 82.80 & 8.79 $\pm$ 0.14 & 4.28 $\pm$ 0.41 & 276.4 $\pm$ 32.8\\
 \hline
& CS$^a$ & 5 $-$ 4 & 244.936 & -3.52 & 11 & 35.27 & $\sim$9.3 & $\sim$4.0 & $\gtrsim$1740\\
\hline
100 & H$_2$CS$^b$ & 7$_{1,6}$ $-$ 6$_{1,5}$ & 244.048 & -3.68 & 45 & 60.04 & 8.13 $\pm$ 0.05 & 3.22 $\pm$ 0.11 & \textit{459.9 $\pm$ 20.9}\\ 
\hline
 84 &OCS$^b$ & 20 $-$ 19 & 243.218 & -4.38 & 41 & 122.57 & 8.55 $\pm$ 0.03 & 2.98 $\pm$ 0.09 & 692.1 $\pm$ 25.9\\
 & O$^{13}$CS & 18 $-$ 17 & 218.199 & -4.52 & 37 & 99.49 & 8.34 $\pm$ 0.05 & 2.25 $\pm$ 0.11 & 178.5 $\pm$ 11.4\\
\enddata
\begin{list}{}
\item $^a$The self-absorption of the emission line prevents us from properly calculating its position, width or integrated intensity. Only an approximate lower limit is indicated in this case, calculated from the Gaussian fit of the emission line without taking into account the channels affected by the self-absorption. 
%Other deviations from a Gaussian profile are probably due to the higher spectral resolution of these observed lines. 
%
%$^b$Possibly blended with a CH$_3$OCHO $\nu$=1 transition not listed in Tables \ref{lines4} and \ref{lines5}. $^c$Possibly blended with a CH$_3$OCHO transition not listed in Table \ref{lines2}. 
%However, a higher column density than that derived from the rotational diagram analysis or the MADCUBAIJ spectral simulation would be needed to explain the observed emission intensity... 
$^b$Blended with a transition from a different species. 
%\item $^b$Possibly blended with other unidentified (U) transitions. 

For blended emission lines, a Gaussian deconvolution was performed in order to estimate the integrated intensity of the different components. 
Values in italics indicate that contribution from neighboring lines could not be ruled out.

\end{list}
\end{deluxetable*}

\section{Population diagram analysis}\label{app-rot}

According to \citet{goldsmith99}, if local thermodynamical equilibrium (LTE) is assumed in the observed region, the integrated intensity ($\int{S_{\nu} dv}$) of a transition is directly proportional to the population of the corresponding upper level ($N_u$), provided that the emission line is optically thin:

\begin{equation}
\int{S_{\nu} dv} = A_{ul}hc \times N_u \times \frac{\Omega}{4\pi},
%\label{nueq}
\end{equation}

where $A_{ul}$ is the Einstein coefficient of the transition (indicated in Tables \ref{lines1} $-$ \ref{lines_small} for every detected transition in our observations), $h$ is the Planck constant, $c$ the speed of light, and $\Omega$ is the solid angle subtended by the emitting region. 

The population of the upper level can thus be calculated for a particular transition from the observed integrated intensity according to Eq. \ref{nueq}:

\begin{equation}
N_{u,obs} = \frac{4\pi\int{S_{\nu,obs} dv}}{A_{ul}\Omega h c}. 
\label{nueq}
\end{equation}

The synthesized beam solid angle of the observations is used as $\Omega$ when deriving beam-averaged column densities.  
If the emitting region is smaller than the beam size (i.e., the emission is not spatially resolved), the observed integrated flux density would be assumed to be emitted from a region larger than the actual emitting region, 
and the resulting beam-averaged column density would be 
underestimated compared to the actual column density in the source. 
This so-called beam dilution effect can be corrected by applying the filling factor $\Omega_{beam}$/$\Omega_{source}$ to Eq. \ref{nueq} (assuming that both the beam and source extent are uniform). 

%The population diagram analysis described in \citet{goldsmith99} and summarized in Appendix \ref{app-rot} assumes optically thin lines. 
A correction factor $C_{\tau}$ should also be applied to the observed integrated intensities in Eq. \ref{nueq} to account for the optical depths of the emission lines \citep[see also][]{taquet15,ryan18,jenny19}: 

\begin{equation}\label{eq_corr}
    C_{\tau} = \frac{\tau}{1 - e^{-\tau}}
\end{equation}

where $\tau$ is the optical depth of any emission line,  determined by: 

\begin{equation}\label{eq_tau}
    \tau = \frac{c^3 A_{ul} N_u}{8\pi \nu^3 \Delta\nu}(e^{h\nu/k T_{rot}} - 1),
\end{equation}

where %$c$ is the speed of light, 
%$A_{ul}$ is the Einstein coefficient of the transition (indicated in Tables \ref{lines1} $-$ \ref{lines_small} for every detected transition in our observations), $N_u$ is the population of the upper level of the transition (calculated from the integrated line intensity with Eq. \ref{nueq}), 
$\nu$ is the line frequency, 
$\Delta\nu$ is the line FWHM (also indicated in Tables \ref{lines1} $-$ \ref{lines_small}), 
%$h$ is the Planck constant, 
$k$ is the Boltzmann constant, and 
$T_{rot}$ is the rotational temperature of the species. 

Altogether, the corrected upper level population can be calculated as:

\begin{equation}
    N_u = N_{u,obs} C_\tau \frac{\Omega_{beam}}{\Omega_{source}}\label{nucorr}
\end{equation}

At the same time, the upper level population $N_u$ corresponding to any transition is related to the total column density $N_T$ of the species and its excitation temperature $T_{rot}$ according to the Boltzmann equation:

\begin{equation}
{N_u} = N_T \frac{1}{e^{E_{up}/T_{rot}}}\frac{g_{up}}{Q(T_{rot})},
%\label{rot}
\end{equation}
%Equations

where $g_{up}$ and $E_{up}$ are the upper level degeneracy and upper level energy of the transition, respectively, in K (also indicated in Tables \ref{lines1} $-$ \ref{lines_small} for the detected transitions), and $Q(T_{rot})$ is the partition function of the species. The partition function of every species was extracted from the same catalog as the rest of line parameters, except for CH$_3$COCH$_3$, for which the partition function was extracted from the JPL catalog, since it was not available in the CDMS catalog. 

When several transitions spanning a wide range of upper level energies are observed for a particular species, the relation between the measured $N_u/g_{up}$ ratio and $E_{up}$ of every transition can be described with an exponential function:

\begin{equation}
\frac{N_u}{g_{up}} = \frac{N_T}{Q(T_{rot})} e^{-E_{up}/T_{rot}}. 
\label{rot}
\end{equation}
%Equations

%where the species column density $N_T$ and excitation temperature $T_{rot}$ can be calculated from the function parameters. 

%Alternatively, taking the natural logarithm of Eq. \ref{rot} leads to:

%\begin{equation}
%ln \frac{N_u}{g_{up}} = ln \frac{N_T}{Q(T_{rot})} - \frac{E_{up}}{T_{rot}}. 
%\label{rotlin}
%\end{equation}

%If the measured $N_u/g_{up}$ ratio for the different transitions are semi-log plotted against their upper level energies $E_{up}$, the resulting population diagram can be fitted with a linear least squares regression, where $T_{rot}$ is the inverse of the slope, and $N_T$ can be calculated from the intercept of the fit. 

Since $N_u$ depends on the rotational temperature through the optical depth correction factor C$_\tau$ (Eq. \ref{nucorr}), an iterative process is needed in order to derive the $T_{rot}$ and $N_{tot}$ values from Eq. \ref{rot}. 
In this work we have used the affine-invariant MCMC package \texttt{emcee} \citep{foreman13} to find the COM excitation temperatures and column densities that best fit the observations(i.e., the 50th percentile from the posteriors). Additional information on the MCMC fitting can be found in \citep{ryan18} and \citet{jenny19}.

\begin{deluxetable*}{cccc}[ht!]
\caption{COM column densities and rotational temperatures in the hot corino of Ser-emb 11 W. Values are estimated from a Gaussian fit of the observed spectrum using the MADCUBA software assuming a circular source with size $\Omega_{source}$ = 0.14$\arcsec$.\label{abundances_MADCUBA}}
\tablehead{
\colhead{Molecule} & \colhead{T$_{rot}$} & \colhead{$N_T$} & \colhead{$N_T$/$N_T$(CH$_3$OH)}\\
& \colhead{(K)} & \colhead{(cm$^{-2}$)} & \colhead{(\%)}}
\startdata
CH$_3$OH & 208 $\pm$ 4 & (1.82 $\pm$ 0.05) $\times$ 10$^{18}$ & 100 \\ %% T 20%, N/2
CH$_2$DOH & 115 $\pm$ 11 & (2.2 $\pm$ 0.1) $\times$ 10$^{17}$ & 13\\ %% T 39%, N/2
C$_2$H$_5$OH & \nodata & \nodata &\nodata\\ 
CH$_3$OCH$_3$ & 197 $\pm$ 17 &(4.6 $\pm$ 0.4) $\times$ 10$^{17}$ & 26\\ %% T 15%, N 15%
CH$_3$OCHO & 177 $\pm$ 6 & (2.7 $\pm$ 0.2) $\times$ 10$^{17}$ & 15\\ %% T 60%, N/3
CH$_3$COCH$_3$ & \nodata & \nodata &\nodata\\ 
\hline
NH$_2$CHO & \nodata & \nodata &\nodata\\ 
CH$_2$DCN & 172 $\pm$ 46 & (2.3 $\pm$ 0.4) $\times$ 10$^{15}$ & 0.13 \\ %% T within errors, N within errors
$^{13}$CH$_3$CN & \nodata & \nodata &\nodata\\
CH$_3$C$^{15}$N & 167 $\pm$ 83 & (7 $\pm$ 2) $\times$ 10$^{14}$ & 0.04 \\ %% T within errors, N within errors
C$_2$H$_5$CN & 202 $\pm$ 88 & (4.5 $\pm$ 0.6) $\times$ 10$^{15}$ & 0.25 \\ %% T within errors, N/2
\enddata
\tablecomments{%These column densities are calculated assuming a circular source with size $\Omega_{source}$ = 0.14$\arcsec$, which is the average of the major and minor axis of the 2D Gaussian fit of the CH$_3$OH 10$_{3,7}$ $-$ 11$_{2,9}$ E moment 0 map (see the text).
The \texttt{Autofit} function of the MADCUBA software did not converge for CH$_3$COCH$_3$, NH$_2$CHO, $^{13}$CH$_3$CN, and C$_2$H$_5$CN.%, and CH$_3$C$^{15}$N, but the simulated spectrum using the rotational temperatures and column densities reported in Table \ref{abundances} was consistent with the observations.
}
\end{deluxetable*}

\section{Rotational temperatures and column densities estimated with the MADCUBA pacakge}\label{app-madcuba}

The \texttt{Autofit} function of the SLIM tool within the MADCUBA package was used to perform a non-linear least-squares fit of a synthetic LTE spectrum to the observed spectrum \citep{madcuba}.
We used the values derived from the population diagram analysis (Table \ref{abundances}) as the initial parameters of the fit, 
%EXCEPT FOR CH3OCHO, FOR WHICH I USED THE SLIGHTLY DIFFERENT ALTERNATIVE VALUES
with a fixed line FWHM  according to the values reported in Sect. \ref{sec:results_col}, and a fixed source size of 0.14$\arcsec$. 
The best-fit rotational temperatures and column densities are listed in Table \ref{abundances_MADCUBA}  and were used to cross-check those presented in Table \ref{abundances}. 
We note that even though the fit did not converge for C$_2$H$_5$OH, CH$_3$COCH$_3$, NH$_2$CHO, and $^{13}$CH$_3$CN, the simulated spectrum assuming the values in Table \ref{abundances} was consistent with the observations.
Similar excitation temperatures were found, within a 35\% difference, for CH$_3$OH, CH$_2$DOH, and CH$_3$OCH$_3$, while they were the same, within errors, for the N-bearing COMs. 
In the cases of CH$_3$OCH$_3$ and CH$_3$OCHO, the estimated temperatures were consistent, within errors, with the temperatures found for CH$_3$OH and CH$_2$DOH through the MCMC fitting of the population diagrams. 
In Sect. \ref{sec:results_col} we explain that the CH$_3$OCH$_3$ rotational temperature calculated with the population diagram analysis could have been overestimated. %while the estimated CH$_3$OCHO excitation temperature was lower than that found for the rest of O-bearing COMs. 
The derived column densities for the O-bearing COMs were within a factor of 2 compared to those reported in Table \ref{abundances}, and the resulting column density ratios were near identical except for CH$_3$OCH$_3$, with a 35\% higher abundance with respect to CH$_3$OH. 
On the other hand, the derived column densities for CH$_2$DCN, CH$_3$C$^{15}$N, and C$_2$H$_5$CN were the same, within errors, as those estimated through the population diagram analysis.

\section{Outflow signature in Ser-emb 11 W}\label{app-co-cont}

\begin{figure}
    \centering
    \includegraphics[width=8cm]{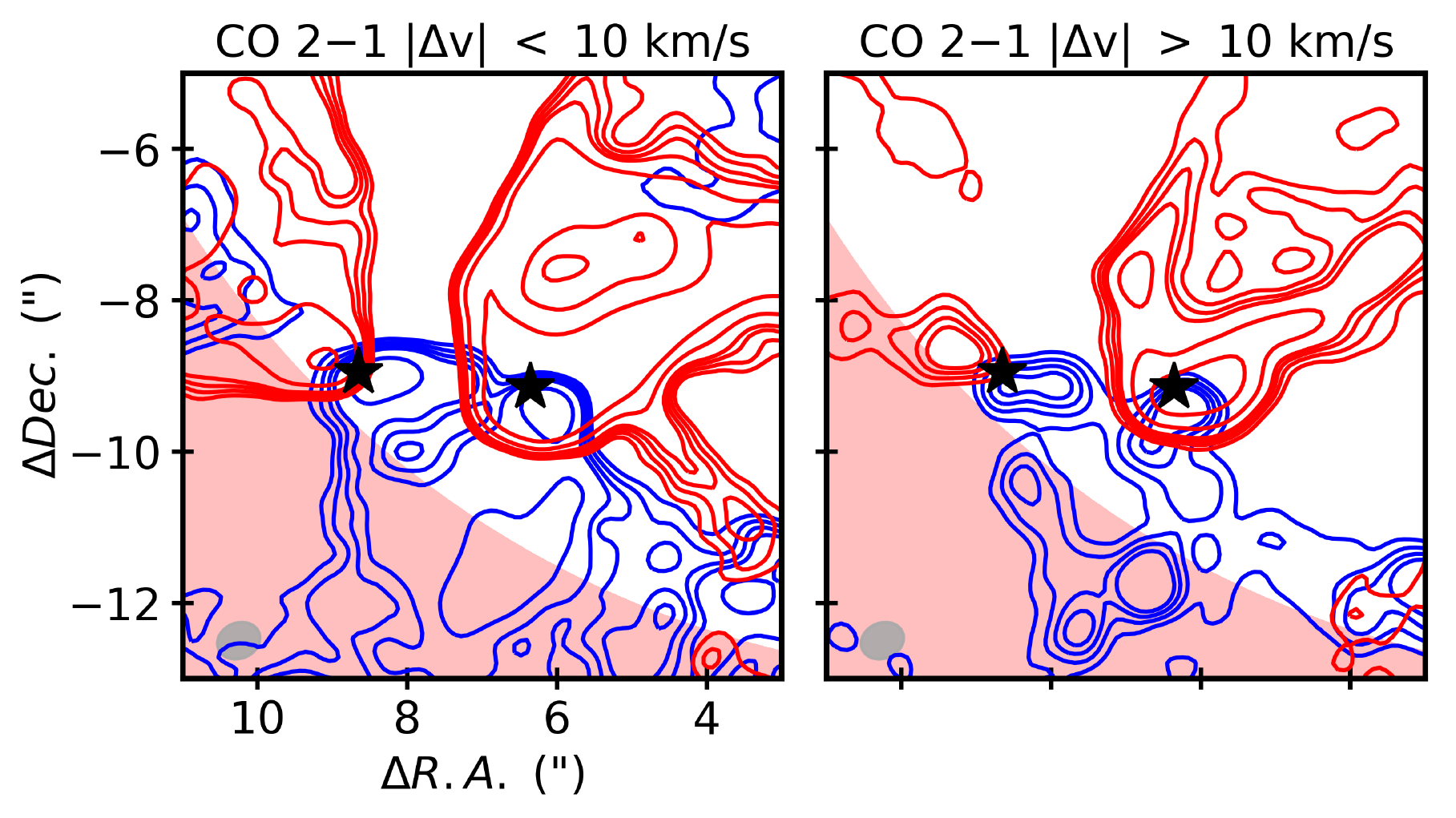}
    \caption{6, 12, 18, 24, 48, and 96$\sigma$ contours ($\sigma$ $\sim$ 18 mJy) of the low velocity ($\Delta$v $<$ 10 km/s, left panel) and high velocity ($\Delta$v $>$ 10 km/s, right panel) integrated  redshifted and blueshifted CO 2$-1$ emission toward Ser-emb 11. 
    The position of the continuum emission peaks is marked with a star symbol in each panel. 
    The size of the synthesized beam 
    is shown on the lower left corner of each panel. 
    The region of the field of view observed outside the half power beam width of the primary beam is colored in red.}
    \label{fig:co_cont}
\end{figure}

Fig. \ref{fig:co_cont} presents the integrated redshifted and blueshifted CO J = 2$-$1 emission decomposed into the low velocity ($\Delta$V $<$ 10 km/s) and high-velocity ($\Delta$V $>$ 10 km/s) components, in order to disentangle the higher velocity emission, more likely corresponding to the v-shaped outflow. We have used a velocity source of $V_{LSR}$ = 8.2 km/s, as measured for the C$^{18}$O J = 2$-$1 emission in a 2$\arcsec$ diameter region around the continuum peak of Ser-emb 11 W. 

\section{Reported COM column densities in Class 0 and Class I hot corinos}\label{app-hot_corinos}

Tables \ref{abundances_0} and \ref{abundances_1} list the column densities of a series of O- and N-bearing COMs in Class 0 and Class I hot corino sources reported in the literature, along with those found toward Ser-emb 11 W in this work. 
For space-saving purposes, only two significant figures are provided for most of the column density values.  
%(except for the B335, %CH3COCH3 needed 3 significant figures with the current exponential
%B1-bS, %NH2CHO needed more significant figures with the current exponential
%Ser-emb 1, and Ser-emb 8) %the error in CH3OH was lower than that
%and the uncertainties in some sources are not presented. 
For the exact reported values and uncertainties, please check the corresponding references. 

\begin{deluxetable*}{cccccccccccccc}
\tabletypesize{\scriptsize}
\caption{Column densities (cm$^{-2}$) of selected O- and N-bearing COMs in Class 0 hot corinos reported in the literature
\label{abundances_0}}
\tablehead{
Species & 16293A$^a$ & 16293B$^b$ & 4A2$^c$ & 2A$^d$  & HH212$^e$ & B335$^f$ & L483$^g$ & B1-bS$^h$ & Ser-emb 1$^i$ & Ser-emb 8$^j$ & BHR71$^k$ & B1-c$^l$  }
\startdata
CH$_2$CO & 9.1$\pm$2.7 & 48 & 7.0 & \nodata & 3.8$\pm$2.0& \nodata&\nodata & 0.13$\pm$0.03& \nodata & 1.0$\pm$0.1&  \nodata&1.3$\pm$0.1 & $\times$10$^{15}$
\\
CH$_3$OH  & 13$\pm$4 & 10 & 10 & 1.2$\pm$0.4 &1.3$\pm$0.7 & \nodata & 17 & 0.50$\pm$0.06 & 0.12$\pm$0.04 & 1.4$\pm$0.6 & 2.4 & 1.9$\pm$0.6&$\times$10$^{18}$\\ 
CH$_2$DOH  & 11$\pm$3 & 7.1 & \nodata & \nodata&1.6$\pm$0.7 & \nodata &4.0 &\nodata & 0.48$\pm$0.18  &0.60$\pm$0.07 & 0.7 & 1.6$\pm$0.1&$\times$10$^{17}$ \\
$^{13}$CH$_3$OH & 20$\pm$7 & \nodata& \nodata& 4.8$\pm$1.3& 2.5$\pm$1.4& \nodata& 25 & 0.15$\pm$0.05 & \nodata& 1.0$\pm$0.1& 1.3 & 1.8$\pm$0.2 &$\times$10$^{16}$ \\
CH$_3$CHO & 0.4$\pm$0.1 & 12 & 1.9 & \nodata& 1.5$\pm$0.6& 0.14$\pm$0.02 & 8.0&0.06$\pm$0.01 & \nodata& 0.10$\pm$0.01& \nodata&0.46$\pm$0.10 &$\times$10$^{16}$ \\
C$_2$H$_5$OH & 8.0$\pm$2.4 & 23 & 4.9 & 5.1$\pm$2.2 & 2.7$\pm$1.2& 0.21$\pm$0.03 &  10& \nodata& \nodata& 0.30$\pm$0.02 &1.3 & 1.5$\pm$0.3 &$\times$10$^{16}$ \\
CH$_3$OCH$_3$ & 52$\pm$16 & 24 & 4.5 & 4.1$\pm$1.6& \nodata& 0.19$\pm$0.02 & 8.0&  0.67$\pm$0.04 & 9.2$\pm$3.8 & 1.2$\pm$0.1 & 1.0 & 2.4$\pm$0.1& $\times$10$^{16}$ \\
CH$_3$OCHO & 27$\pm$8 & 26 & 3.5 & 6.4$\pm$1.9 & 3.2$\pm$1.6&  0.26$\pm$0.03 & 13 & 0.38$\pm$0.02 & 9.1$\pm$3.6 & 1.6$\pm$0.2& 8.1 & 1.9$\pm$0.1&$\times$10$^{16}$ \\
CH$_3$COCH$_3$ & 2.4$\pm$0.7 & \nodata & 0.6 &\nodata& \nodata&0.05$\pm$0.05 &\nodata & 0.05$\pm$0.01 &\nodata &0.53$\pm$0.03 & 1.0 & 0.80$\pm$0.09& $\times$10$^{16}$ \\
\hline
NH$_2$CHO & 1.9$\pm$0.6 & 10 & 2.0 & 12 & 1.6$\pm$0.9& 0.24$\pm$0.02& 10 & \nodata & \nodata& 0.27$\pm$0.03 &\nodata &0.51$\pm$0.24 &$\times$10$^{15}$\\
CH$_3$CN  & 8.0$\pm$1.0 & 4.0$\pm$1.0 & 4.8 & 1.0$\pm$0.2 & \nodata& \nodata& \nodata& \nodata& \nodata& 0.29$\pm$0.10& \nodata& 0.74$\pm$0.22&$\times$10$^{16}$  \\
CH$_2$DCN & 3.5$\pm$0.5 & 1.4$\pm$0.2 & \nodata& \nodata& 0.4$\pm$0.2 & \nodata& \nodata& \nodata& \nodata& 0.10$\pm$0.02& \nodata& 0.26$\pm$0.02&$\times$10$^{15}$\\
$^{13}$CH$_3$CN & 10$\pm$2 & 6.0$\pm$0.2 & \nodata& \nodata& \nodata& \nodata& \nodata& \nodata& \nodata& \nodata& \nodata& \nodata& $\times$10$^{14}$ \\
CH$_3$C$^{15}$N & 3.4$\pm$0.2 & 1.6$\pm$0.2 & \nodata& \nodata& \nodata& \nodata& \nodata& \nodata& \nodata& \nodata& \nodata& \nodata &$\times$10$^{14}$\\
C$_2$H$_5$CN & 9.0$\pm$0.2 & 3.6$\pm$0.2& 1.5 & 1.2 & \nodata&0.10 & \nodata& \nodata& \nodata& 1.1$\pm$0.6& \nodata&1.1$\pm$0.3 &$\times$10$^{15}$\\
\enddata
\tablecomments{
$^a$O-bearing COMs from \citet{manigand20}. 
N-bearing COMs from \citet{calcutt18}. 
$^b$O-bearing COMs from \citet{jorgensen18}. 
CH$_3$OH was derived from CH$_3^{18}$OH assuming a $^{16}$O/$^{18}$O abundance ratio of 560 as in \citet{wilson94}. 
%except for CH$_3$COCH$_3$, extracted from \citet{lykke17}. 
20\% uncertainties were assumed. 
N-bearing COMs from \citet{calcutt18}. 
$^c$From \citet{lopezsepulcre16} (assuming T$_{rot}$ = 200 K, except for CH$_3$OCH$_3$ and CH$_3$OCHO).  
%(an uncertainty of $\sim$3 $\times$10$^{15}$ cm$^{-2}$ was reported for these two species). 
CH$_3$OH derived from centimeter observations with the Very Large Array \citep[VLA,][]{desimone20}. 
$^d$From \citet{taquet15} (assuming T$_{rot}$ = 179 K for NH$_2$CHO and C$_2$H$_5$CN). 
$^e$From \citet{lee19}. 
CH$_3$OH was derived from $^{13}$CH$_3$O assuming a $^{12}$C/$^{13}$C ratio of 50 for the Orion complex as in \citet{kahane18}. 
$^f$From \citet{imai16} (assuming T$_{rot}$ = 100 K). C$_2$H$_5$OH, CH$_3$OCH$_3$, CH$_3$OCH$_3$,  and C$_2$H$_5$CN detections are tentative. 
$^g$From \citet{jacobsen19}. 
%$^h$From \citet{marcelino18} (assuming T$_{rot}$ = 200 K). 
$^h$From \citet{vanGelder20} using Band 6 observations only. 
CH$_3$OH was derived from CH$_3^{18}$OH assuming a $^{16}$O/$^{18}$O abundance ratio of 560 as in \citet{wilson94}. 
CH$_2$CO detection was considered tentative. 
$^i$From \citet{martin19}. 
$^j$O-bearing COMs from \citet{vanGelder20} using Band 6 observations only. 
CH$_3$OH was derived from CH$_3^{18}$OH assuming a $^{16}$O/$^{18}$O abundance ratio of 560 as in \citet{wilson94}. 
CH$_2$CO detection was considered tentative.  
N-bearing COMs from \citet{nazari21}. 
%There are large discrepancies between the column densities listed in this Table and those reported in \citet{jenny19}. In particular, CH$_3$OH column density is 3 times higher, while the rest of column densities are one order of magnitude lower. 
$^k$From \citet{yang20}. %assuming T$_{rot}$ = 100 K.
$^l$O-bearing COMs from \citet{vanGelder20} using Band 6 observations only. 
CH$_3$OH was derived from CH$_3^{18}$OH assuming a $^{16}$O/$^{18}$O abundance ratio of 560 as in \citet{wilson94}. 
CH$_2$CO detection was considered tentative.  
N-bearing COMs from \citet{nazari21}. 
%$^m$Estimated from the reported $^{13}$CH$_3$OH column density assuming a $^{12}$C/$^{13}$C ratio of 81.3 \citep{botelho20}.
}
\end{deluxetable*}

\begin{deluxetable*}{cccccc}
\caption{Column densities (cm$^{-2}$) of selected O- and N-bearing COMs in Class I hot corinos reported in the literature
\label{abundances_1}}
\tablehead{
Species & SVS13-A$^a$ &  L1551-IRS5$^b$  & Ser-emb 17$^c$ & Ser-emb 11$^d$ }
\startdata
CH$_2$CO & \nodata& \nodata& 3.1$^{+4.3}_{-2.1}$ & \nodata& $\times$10$^{15}$\\
CH$_3$OH  & 0.81$^{+0.50}_{-0.31}$ & $\ge$10 & 0.28$^{+0.01}_{-0.01}$ &  3.7$^{+0.4}_{-0.3}$ &  $\times$10$^{18}$ \\ 
CH$_2$DOH  & \nodata& $\ge$5 & \nodata& 4.4$^{+0.4}_{-0.4}$ &  $\times$10$^{17}$\\
%$^{13}$CH$_3$OH & & &  & & $\times$10$^{16}$\\
CH$_3$CHO & 2.5$^{+1.7}_{-1.4}$ & \nodata& \nodata& \nodata& $\times$10$^{16}$\\
C$_2$H$_5$OH & 9.6$^{+1.8}_{-3.1}$ & 14.9$\pm$1.3 & \nodata& 11.3$^{+0.6}_{-0.6}$ & $\times$10$^{16}$\\
CH$_3$OCH$_3$ & 11$^{+2}_{-1}$ & \nodata& 12$^{+18}_{-8}$ & 54$^{+1}_{-1}$ & $\times$10$^{16}$\\
CH$_3$OCHO  & 33$^{+12}_{-6}$ & 33$\pm$2 & 17$^{+35}_{-8}$ & 79$^{+6}_{-8}$ & $\times$10$^{16}$\\
CH$_3$COCH$_3$ & 6.3$^{+3.7}_{-4.4}$ &\nodata & \nodata & 7.1$^{+0.8}_{-0.8}$ &  $\times$10$^{16}$\\
\hline
NH$_2$CHO & 1.9$^{+0.6}_{-0.7}$ &\nodata & 1.9$^{+2.7}_{-1.2}$ & 3.9$^{+0.9}_{-0.8}$ & $\times$10$^{15}$\\
CH$_3$CN & 0.59$^{+0.18}_{-0.13}$ & \nodata& \nodata&\nodata & $\times$10$^{16}$\\
CH$_2$DCN & 1.9$^{+1.0}_{-0.8}$ &\nodata &\nodata & 2.8$^{+0.2}_{-0.1}$ &  $\times$10$^{15}$\\
$^{13}$CH$_3$CN & \nodata&\nodata & \nodata& 1.23$^{+0.08}_{-0.08}$ & $\times$10$^{15}$\\
CH$_3$C$^{15}$N & \nodata& \nodata& \nodata& 8.2$^{+0.5}_{-0.5}$ & $\times$10$^{14}$\\
C$_2$H$_5$CN & 2.8$^{+0.9}_{-0.6}$ & \nodata& \nodata& 8.1$^{+1.2}_{-1.0}$ & $\times$10$^{15}$\\
\enddata
\tablecomments{$^a$From \citet{yang21}. 
$^b$From \citet{bianchi20}. 
$^c$From \citet{jenny19}. 
$^d$This work.}
\end{deluxetable*}

%%The C$_2$H$_5$OH column density in B1-bS takes into account only the trans transitions \citep{marcelino18} while the value reported in \citet{yang20} for BHR 71 is measured from the gauche transitions only. 
%%The CH$_3$CHO column density in B1-bS takes into account the ground and first excited vibrational states \citep{marcelino18}. 
%%The CH$_3$OCHO column density in B1-bS takes into account the ground and the two first excited vibrational states \citep{marcelino18}. 
%%The CH$_3$CHO column density in BHR 71 takes into account the ground and first excited vibrational states \citep{yang20}. 
%We refer the reader to the corresponding references for further information about the estimation process of the different column densities used in Fig. \ref{fig:comp_ME}.
%%The specifics of the estimation of the column densities for the different species in the different sources should not significantly affect the following discussion, since any assumption made (such as the employed analysis technique or the size of the emitting region) would affect all column densities measured toward a particular source in a similar way, and would thus cancel out when comparing abundance ratios. 
%%In any case, \citet{jenny19} suggested that some of the observed variations between sources could be an artifact of the different angular resolutions and analysis techniques used to derive the column densities.

\end{document}